\documentclass[useAMS,usenatbib]{mn2e}

\usepackage{graphicx}
\usepackage{rotating}
\usepackage{amsmath}
\usepackage{amssymb}
\usepackage{lscape}
\usepackage{natbib}
\usepackage{keyval}
\usepackage{lastpage}

\newcommand{\tbgscold}{\hbox{$T_\mathrm{C}^{\,\mathrm{ISM}}$}}
\newcommand{\fmu}{\hbox{$f_\mu$}}
\newcommand{\ldust}{\hbox{$\rm{L}_{\mathrm{d}}^{\,\mathrm{tot}}$}}
\newcommand{\tauv}{\hbox{$\hat{\tau}_{V}$}}
\newcommand{\tauvISM}{\hbox{$\hat{\tau}_{V}^{ISM}$}}
\newcommand{\ssfr}{\hbox{$\psi_{\mathrm S}$}}
\newcommand{\sfr}{\hbox{$\psi$}}
\newcommand{\mdms}{\hbox{$\rm{M}_{\mathrm d}/\rm{M}_\ast$}}

\def\arcsec{\hbox{$^{\prime\prime}$}}
\def\arcmin{$^{\prime}$}
\def\deg{\hbox{$^\circ$}}

\title[H-ATLAS/GAMA: Dusty early-type galaxies and passive spirals]
{{\it Herschel}\thanks{{\it Herschel} is an ESA space observatory with science instruments provided by European-led Principal Investigator consortia 
and with important participation from NASA.}-ATLAS/GAMA: Dusty early-type galaxies and passive spirals}

\author[K. Rowlands et al.]
{K. Rowlands$^{1}$\thanks{E-mail:ppxkr@nottingham.ac.uk}, ~L. Dunne$^{1}$, S. Maddox$^{1}$, N. Bourne$^{1}$, H.~L. Gomez$^{2}$, S. Kaviraj$^{3}$, 
\newauthor S.P.~Bamford$^{1}$, S. Brough$^{4}$, S. Charlot$^{5}$, E. da Cunha$^{6}$, S.P Driver$^{7,8}$, S.A. Eales$^{2}$, 
\newauthor A. M. Hopkins$^{4}$, L. Kelvin$^{7,8}$, R.C.~Nichol$^{9}$, A.E. Sansom$^{10}$, R. Sharp$^{11}$,
\newauthor D.J.B. Smith$^{1,12}$, P. Temi$^{13}$, P. van der Werf$^{14}$, M. Baes$^{15}$, A. Cava$^{16}$, A. Cooray$^{17}$, 
\newauthor S.M.~Croom$^{18}$, A. Dariush$^{3}$, G. De Zotti$^{19,20}$, S. Dye$^{1,2}$, J. Fritz$^{15}$, R. Hopwood$^{3}$, 
\newauthor E. Ibar$^{21}$, R.J. Ivison$^{21,22}$, J.~Liske$^{23}$, J.~Loveday$^{24}$, B. Madore$^{25}$,
\newauthor P.~Norberg$^{22}$, C.C.~Popescu$^{10}$, E.E. Rigby$^{1,22}$, A. Robotham$^{8}$, G. Rodighiero$^{19}$,
\newauthor M. Seibert$^{25}$, R.J.~Tuffs$^{26}$\\
  $^{1}$School of Physics \&\ Astronomy, The University of Nottingham, University Park Campus, Nottingham, NG7 2RD, UK \\ 
  $^{2}$School of Physics \&\ Astronomy, Cardiff University, Queens Buildings, The Parade, Cardiff, CF24 3AA, UK \\
  $^{3}$Physics Department, Imperial College London, Prince Consort Road, London SW7 2AZ\\
  $^{4}$Australian Astronomical Observatory, PO Box 296, Epping, NSW 1710, Australia \\
  $^{5}$Institut d'Astrophysique de Paris, CNRS, Universit\'e Pierre \& Marie Curie, UMR 7095, 98bis bd Arago, 75014 Paris, France \\
  $^{6}$Max Planck Institute for Astronomy, Konigstuhl 17, 69117, Heidelberg, Germany\\
  $^{7}$International Centre for Radio Astronomy (ICRAR), University of Western Australia, Crawley, WA6009, Australia \\
  $^{8}$(SUPA) School of Physics \& Astronomy, University of St Andrews, North Haugh, St Andrews, KY16 9SS, UK \\
  $^{9}$Institute of Cosmology and Gravitation (ICG), Dennis Sciama Building, Burnaby Road, Portsmouth, PO1 3FX, UK \\ 
  $^{10}$Jeremiah Horrocks Institute, University of Central Lancashire, Preston, PR1 2HE, UK\\
  $^{11}$Research School of Astronomy \& Astrophysics, Mount Stromlo Observatory, Cotter Road, Weston Creek, ACT 2611, Australia\\
  $^{12}$Centre for Astrophysics, Science \& Technology Research Institute, University of Hertfordshire, Hatfield, Herts, AL10 9AB, UK\\
  $^{13}$Astrophysics Branch, NASA Ames Research Center, Mail Stop 2456, Moffett Field, CA 94035, USA\\
  $^{14}$Leiden University, P.O. Box 9500, 2300 RA Leiden, The Netherlands\\
  $^{15}$Sterrenkundig Observatorium, Universiteit Gent, Krijgslaan 281 S9, B-9000 Gent, Belgium\\
  $^{16}$Departamento de Astrof\'{\i}sica, Facultad de CC. F\'{\i}sicas, Universidad Complutense de Madrid, E-28040 Madrid, Spain\\
  $^{17}$Department of Physics and Astronomy, University of California, Irvine, CA 92697, USA \\
  $^{18}$Sydney Institute for Astronomy, School of Physics, University of Sydney, NSW 2006, Australia \\
  $^{19}$INAF -- University of Padova, Department of Astronomy, Vicolo Osservatorio 3, I-35122, Padova, Italy\\
  $^{20}$SISSA, Via Bonomea 265, I-34136 Trieste, Italy\\
  $^{21}$UK Astronomy Technology Centre, Royal Observatory, Edinburgh, EH9 3HJ, UK\\
  $^{22}$Institute for Astronomy, University of Edinburgh, Royal Observatory, Blackford Hill, Edinburgh EH9 3HJ, UK\\
  $^{23}$European Southern Observatory, Karl-Schwarzschild-Str. 2, 85748 Garching, Germany\\
  $^{24}$Astronomy Centre, University of Sussex, Falmer, Brighton BN1 9QH, UK\\
  $^{25}$Observatories of the Carnegie Institution of Washington, 813 Santa Barbera St., Pasadena, CA 91101, USA\\
  $^{26}$Max Planck Institute for Nuclear Physics (MPIK), Saupfercheckweg 1, 69117 Heidelberg, Germany\\
}

\begin{document}

\date{}

\pagerange{\pageref{firstpage}--\pageref{lastpage}} \pubyear{2011}

\maketitle

\label{firstpage}

\clearpage

\begin{abstract}
We present the dust properties and star-formation histories of local 
submillimetre-selected galaxies, classified by optical morphology. Most 
of the galaxies are late types and very few are early types. The 
early-type galaxies that are detected contain as much dust as typical 
spirals, and form a unique sample that has been blindly selected at 
submillimetre wavelengths. Additionally, we investigate the properties 
of the most passive, dusty spirals.

We morphologically classify 1087 galaxies detected in the \emph{Herschel}-ATLAS 
Science Demonstration Phase data. Comparing to a control sample of 
optically selected galaxies, we find 5.5\% of luminous early-type 
galaxies are detected in H-ATLAS. The H-ATLAS early-type galaxies 
contain a significant mass of cold dust: the mean dust mass is 
$5.5\times{10}^{7}$M$_\odot$, with individual galaxies ranging 
from $9 \times 10^5 - 4 \times 10^8 $M$_\odot$. 
This is comparable to that of spiral galaxies in our sample, and is an 
order of magnitude more dust than that found for the control 
early-types, which have a median dust mass inferred from stacking of 
$(0.8-4.0)\times10^{6}{\rm{M}}_\odot$ for a cold dust temperature of 25-15 K. The 
early-types detected in H-ATLAS tend to have bluer $NUV-r$ colours, 
higher specific star-formation rates and younger stellar populations 
than early-types which are optically selected, and may be transitioning 
from the blue cloud to the red sequence. We also find that H-ATLAS and 
control early-types inhabit similar low-density environments. We 
investigate whether the observed dust in H-ATLAS early-types is from 
evolved stars, or has been acquired from external sources through 
interactions and mergers. We conclude that 
the dust in H-ATLAS and control ETGs cannot be solely from stellar 
sources, and a large contribution from dust formed in the ISM or 
external sources is required. Alternatively, dust destruction may not be 
as efficient as predicted.

We also explore the properties of the most passive spiral galaxies in 
our sample with SSFR\,$<10^{-11}\rm{yr^{-1}}$. We find these passive spirals have 
lower dust-to-stellar mass ratios, higher stellar masses and older stellar population 
ages than normal spirals. The passive spirals inhabit low density 
environments similar to those of the normal spiral galaxies in our 
sample. This shows that the processes which turn spirals passive do not 
occur solely in the intermediate density environments of group and 
cluster outskirts.
\end{abstract}


\section{Introduction}
\label{Intro}
It has long been known that there is a relationship between galaxy optical colour
and morphology. Galaxies can be split into a red sequence and blue
cloud \citep{Tresse99, Strateva01, Blanton03, Baldry04, Bell04}. Red galaxies are generally
passive early-type galaxies (ETGs, those that have elliptical or S0
morphology), but with $\sim25\%$ being spirals which are red either due to 
dust or because they are passive \citep{Driver06_MGC}. Blue galaxies are actively
star-forming and mostly of spiral or irregular morphology. The colour
bimodality of galaxies is linked to their star-formation history (SFH), with
the stellar population of galaxies transitioning from blue to red as
their star-formation ceases due to the removal or consumption of cold gas
\citep[e.g.][]{Faber07, Hughes_Cortese09}. The injection of
gas and dust via mergers may temporarily rejuvenate star-formation,
and so this evolution of colour can be reversed
\citep{Cortese_Hughes09, Kannappan09, Wei10}. Such ``rejuvenators''  may have had substantially
different star-formation histories from those which make up the majority
of their morphological type, and may provide insight into the
evolutionary processes that shape galaxies today.

Massive ($>10^{10}\rm{M}_\odot$) ETGs are traditionally thought to be ``red and dead''
\citep*[e.g.][]{Temi09a}, having formed most of their stellar mass at early epochs over a relatively short period of time
\citep[e.g.][]{Cimatti04, Thomas05} and then evolved passively to
their present state. Their optical light is dominated by old stellar
populations, however, recent ultra-violet (UV) studies of large samples of ETGs have shown
that many of these galaxies exhibit low to moderate levels of
star-formation \citep{Yi05, Schawinski07a, Kaviraj07, Kaviraj08, Kaviraj10, Kaviraj10_Stripe82}. 
The UV-optical colours suggest that at least $\sim$30\% of
UV-selected early-type galaxies at $z<0.11$ have evidence of recent
star-formation within the last 1\,Gyr \citep{Kaviraj07}, however, it is difficult to
determine the contribution of UV flux from old stars. 

Mergers are likely to trigger star-formation, since a high incidence of ETGs with
disturbed morphologies (18\%) has been observed \citep{Kaviraj10_Stripe82}, 
and these disturbed ETGs also have bluer $NUV-r$ colours than normal ETGs. 
The major merger rate at low and intermediate redshifts is thought to be too
low to account for the number of galaxies which have disturbed
morphologies \citep[e.g.][]{DePropris07, DePropris10, Lotz08a}, therefore 
\citet{Kaviraj10} conclude that minor mergers are the most likely 
trigger of recent star-formation in ETGs.

Although there is evidence for limited quantities of dust in ETGs, these galaxies are generally 
thought to be gas- and dust-poor, which gives an insight into their evolutionary state. 
UV starlight is preferentially absorbed and re-emitted by dust in the
far-infrared (FIR)-submillimetre, so the presence of dust emission is often 
viewed as evidence for ongoing star-formation \citep{Kennicutt98}, although dust can also be heated 
by the radiation field of an old stellar population. Evidence for dust in ETGs was
first found in the optical \citep[e.g.][]{Hawarden81,SG85, vDF95}, yet it is difficult to estimate the total dust mass purely from optical observations. Warm dust ($>30$K) was detected
in 12\% of local ETGs by \emph{IRAS} \citep{Bregman98}, but \emph{IRAS} was less 
sensitive to the cold dust component which dominates the dust mass in local
galaxies \citep*{DE01, SLUGS05, Smith11b}. There have been few studies of ETGs at 
FIR-submillimetre wavelengths to date, since surveys conducted at
these wavelengths have been limited in areal coverage. Consequently, studies of
ETGs have been targeted observations of relatively small, and often
biased, samples. Cold dust has been detected in ETGs
through observations with \emph{ISO}, SCUBA, {\em Spitzer\/}, and SHARC II
(\citealt[e.g.][]{Temi04}; \citealt{Leeuw04, SLUGS05}; \citealt*{Temi07, Stickel07}; \citealt{Leeuw08}; \citealt*{Savoy09}). Cold dust has also
been observed by \emph{Herschel} in 10 nearby ETGs \citep{Skibba11}, and in the Virgo cluster elliptical
galaxy M86, which contains dust stripped from the nearby spiral NGC
4438 \citep{Gomez10, Cortese10}.

Conversely, spiral galaxies are generally rich in dust and gas, and make up the
majority of the star-forming population. Their blue optical colours indicate
young stellar populations, yet for some time optically red spirals with no
spectroscopic evidence of star-formation have been known to exist in
the outskirts of clusters \citep{vdBergh76, Poggianti99, Poggianti04, 
Goto03}. These spirals can be red due to dust obscuration, or because of an
ageing stellar population \citep{Wolf09}. It is generally believed
that passive red spirals have had their star-formation quenched due to
environmental effects, since they are found to mostly reside in
intermediate density environments \citep{Skibba09, Bamford09, Masters10}. The
star-formation rate (SFR) was found to be lower for red spirals than blue spirals
in all environments, which indicates that factors other than
environment can truncate star-formation in red spirals \citep{Bamford09, Masters10}.
The same authors also find that a large fraction of red spirals are massive 
($>10^{10}$M$_\odot$).

We can now get an unprecedented view of dust in local galaxies from the
the {\em Herschel\/}-ATLAS survey (H-ATLAS, \citealt{Eales_ATLAS10}). 
The telescope observes at FIR-submillimetre 
wavelengths across the peak of the dust emission, 
making it an unbiased tracer of the dust mass in galaxies.
In this paper we examine the properties of galaxies detected in 
the H-ATLAS Science Demonstration Phase (SDP) field
as a function of morphological type, and highlight
interesting populations which do not conform to the usual trend of
colour and morphology. In particular, we focus our analysis on the properties 
of H-ATLAS ETGs and how these galaxies are different to optically selected ETGs, 
in addition to studying a population of dusty, passive spirals. We present the detection of the
very dustiest ETGs in a large area blind submillimetre survey with
\emph{Herschel}, where the lack of pre-selection in other bands makes
it the first unbiased survey for cold dust in ETGs. 
In Section \ref{sec:obs} we describe the survey, observations
and morphological classifications, and present the spectral energy
distribution (SED) fitting method to explore the properties and SFHs of
galaxies in our sample; the results of which are presented in Section
\ref{sec:results}. The properties of
a population of passive spirals are examined in Section
\ref{sec:Passive spirals}. The AGN fraction of dusty ETGs is
explored in Section \ref{AGN}, and we investigate whether environment may
be an influential factor in the properties of our galaxies in Section
\ref{sec:Environment}. The submillimetre detected ETGs are compared 
to a control sample of optically selected ETGs in Section \ref{sec:control}, 
and we discuss the origin of the dust in Section \ref{sec:chem_ev}. 
We adopt a cosmology with $\Omega_m=0.27,\,\Omega_{\Lambda}=0.73$ and $H_o=71\,
\rm{km\,s^{-1}\,Mpc^{-1}}$.

\section{Observations and sample selection}
\label{sec:obs}
The H-ATLAS \citep{Eales_ATLAS10} is a $\sim$570 deg$^2$
survey undertaken by the \emph{Herschel Space Observatory}
\citep{Pilbratt10} at 100, 160, 250, 350 and 500$\mu$m to provide an
unbiased view of the submillimetre universe. Observations are carried
out in parallel mode using the PACS \citep{PACS10} and SPIRE
\citep{Griffin10} instruments simultaneously. In this paper, we use
observations in the SDP field, with an area of $\sim$14 deg$^2$
centered on $\alpha$=09$^{h}$05$^{m}$30.0$^{s}$,
$\delta=$00\deg30\arcmin00.0\arcsec{} (J2000). Details of the map making
can be found in \citet{Pascale11, Ibar10}. A catalogue of
$\geq$5$\sigma$ detections in any of the 250, 350 and 500$\mu$m bands
was produced \citep{Rigby11} using the MAD-X algorithm (Maddox et
al. in prep) and contains 6876 sources. The $5\sigma$ noise levels are
132, 126, 32, 36 and 45mJy per beam at 100, 160, 250, 350 and
500$\mu$m, respectively; the beam sizes are $\sim9$, $\sim13$, 18, 25 and 35
arcsec in these bands. 

The H-ATLAS SDP field overlaps with that of the 
Galaxy And Mass Assembly (GAMA) survey \citep{GAMA_Driver11, GAMA_Hill11, GAMA_Robotham10, GAMA_Baldry10}, 
which will provide $\sim350\,000$ spectra
for galaxies at low redshifts over 6 regions, covering
$\sim300$ square degrees. The GAMA data comprise $r$-band defined
aperture matched photometry as described in \citet{GAMA_Hill11} from UV \emph{GALEX} 
\citep[][Seibert et al. in prep.]{GALEX_Martin05, GALEXDR3}, optical $ugriz$ 
SDSS DR6 \citep{AM09} and near-infrared $YJHK$ UKIDSS-LAS \citep{UKIDSS-LAS} imaging. 
Spectroscopic redshifts and spectra from the AAOmega spectrograph are provided for $r_\mathrm{petro}<19.8$ or $(K_\mathrm{Kron}<17.6$ and $r_\mathrm{modelmag}<20.5$) or $(z_\mathrm{Kron}<18.2$ and $r_\mathrm{modelmag}<20.5$)\footnote{$r_{\rm{petro}}$ is the $r$-band Petrosian magnitude, which is measured using a circular aperture of twice the Petrosian radius, which is defined using the light profile of the galaxy. $r_\mathrm{modelmag}$ is the SDSS $r$-band model magnitude, which is determined from the best fit of an exponential or de Vaucouleurs profile; further details are presented in \citet{GAMA_Baldry10}.} in the G12 field, 
and $r_\mathrm{petro}<19.4$ or $(K_\mathrm{Kron}<17.6$ and $r_\mathrm{modelmag}<20.5$) or $(z_\mathrm{Kron}<18.2$ and $r_\mathrm{modelmag}<20.5$) in G15 and G09 which includes the H-ATLAS SDP field.

A likelihood-ratio analysis \citep{SuthSaund92} 
is performed to match 250$\mu$m sources to SDSS DR7
\citep{SDSS_DR7} sources with $r<22.4$ within a 10\arcsec{} radius \citep{Smith11a}, and accounts 
for the possibility of the true counterpart being below the optical magnitude limit. 
The reliability of an association is defined as the probability that an
optical source is associated with the submillimetre source.
SDSS sources with reliability$\geq0.8$ are considered to be likely matches
to submillimetre sources, these are matched to GAMA survey data to provide spectra when
available. There are 2423 reliable optical counterparts to H-ATLAS sources, with either
photometric or spectroscopic redshifts. Around two-thirds of the objects without reliable optical counterparts are unidentified because their counterparts lie below the optical magnitude limit. These sources mostly reside at $z>0.5$ (see \citealt{Dunne11}). The remaining unidentified sources are believed to have a counterpart in the SDSS catalogue but we are unable to unambiguously  identify the correct counterpart in all cases due to near neighbours and the non-negligible probability of a background galaxy of the same magnitude being found at this distance.
\citet{Smith11a} estimate the completeness of the H-ATLAS sample as a function of redshift by calculating the total number sources that we would expect to have a counterpart above the SDSS magnitude limit in H-ATLAS; we refer the reader to \citet{Smith11a, Dunne11} for further details. \citet{Smith11b} find that at $z<0.35$ the $r$-band selection does not bias our sample towards less obscured sources. Since the majority of our spirals and ETGs lie at redshifts less than this, our sample should be representative of the low-redshift galaxy population.
Matches are also made to the\emph{IRAS} \citep*{IRAS_FSC_1992} and FIRST radio catalogues \citep*{FIRST_Becker95} as described in \citet{Smith11a}.

\subsection{Morphology}
Morphological classification of sources was performed by eye using SDSS standard
depth $gri$ composite images, and objects were assigned one of four categories: early-type, late-type, merger and unknown. 
The classification fractions are shown in Table \ref{table:control_morphs}. ETGs were identified by looking for a dominant bulge
and a complete lack of spiral arms, and late-types were identified by 
the presence of spiral arms. Due to the shallow depth of the SDSS
images, we do not discriminate between E and S0 types, however, it is
possible that these populations may have different properties
\citep*[e.g.][]{Temi09b}.
The merger category contains systems of galaxies that are clearly interacting.
Galaxies were classified as `unknown' if it was impossible to assign a morphology, usually because
the galaxy was too faint or small. This situation becomes more common as 
spatial resolution and signal-to-noise decrease at higher redshifts. 
It is possible that at low redshifts some of the unknown classifications 
are irregulars, which tend to have small angular size and are therefore difficult to identify. Additionally, very few H-ATLAS galaxies are low stellar mass objects, which is due to the flux limit in the submillimetre. Therefore the dearth of irregulars is likely to be a real effect and not an inability to classify them.
Given the sample size, visual inspection is the
preferred method to classify our galaxies into broad morphological
classes. It has been shown that visual inspection is superior in
identifying contaminants in samples of ETGs (e.g. face on spirals
which have a dominant bulge but have weak spiral arms) than automated
classification methods \citep{Kaviraj07, Schawinski07a, Lintott08, Bamford09}.
Since we are interested in selecting spheroids, inclination is not an
issue. It is possible that at higher redshifts Sa type galaxies
with faint spiral arms not visible in the shallow imaging could 
be classified as ETGs.

\begin{table*}
\caption{Morphologies obtained by visual classification of 1087 H-ATLAS sources and 1052 control sample galaxies. 
The control sample galaxies are selected to have the same $r$-band magnitude
and redshift distribution as those detected in H-ATLAS. The estimated detection fraction of galaxies in each 
morphological class is shown in the last row. These are estimated as explained in Section~\ref{sec:control_sample}.}
\begin{tabular}{ c | c | c | c | c }
\hline
 & Early-type & Late-type & Merger & Unknown \\
\hline
All (detected) 1087     & 44 & 496 & 23 & 524 \\
                        & 4.1\% & 45.6\% & 2.1\% & 48.2\% \\
\hline
All (non-detected) 1052 & 233 & 378 & 22 & 419 \\
                        & 22.1\% & 35.8\% & 2.1\% & 39.8\% \\
\hline
H-ATLAS detected fraction & 5.5\% & 28.2\% & 25.0\% & 20.6\% \\
\hline
\label{table:control_morphs}
\end{tabular}
\end{table*}

\subsubsection{H-ATLAS sample}

\begin{figure*}
\centering
\includegraphics[scale=0.7]{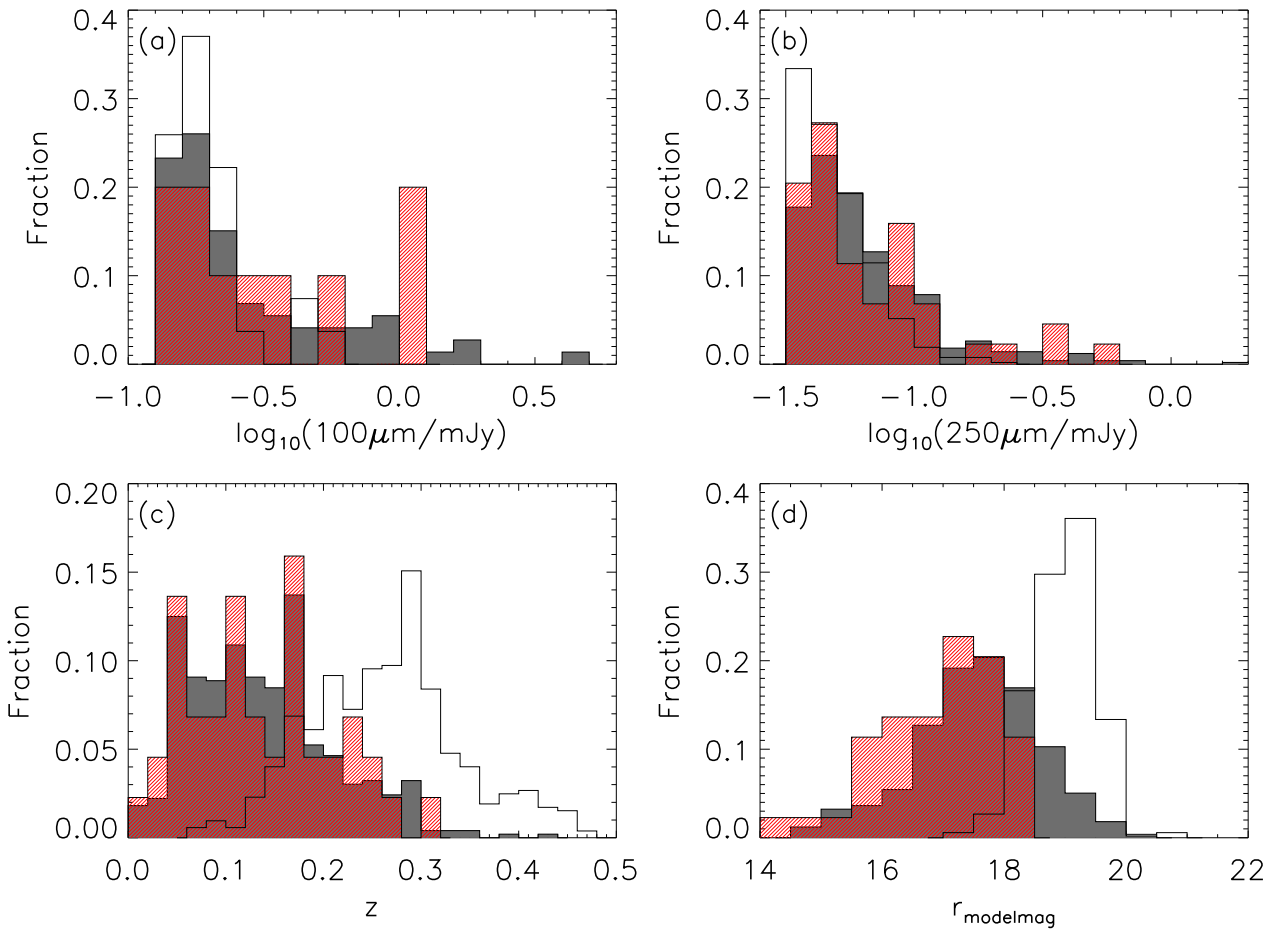}
\caption{The distribution of $\geq5\sigma$ 100$\mu$m (a) and 250$\mu$m point-source fluxes (b) are shown for the H-ATLAS morphologically classified sample. Spirals are shown as grey/filled, ETGs as red/hatched and unknown morphologies as an open histogram. It can be seen that the flux distributions are similar for each type of morphology. (c): Redshift distribution of our H-ATLAS sample for each morphological type. The ETGs and spirals have similar redshift distributions, but galaxies classified as unknown lie at much higher redshift on average. (d): The distribution of SDSS $r$-band model magnitudes for the same morphological classifications. Sources with unknown morphologies are fainter on average than those which are classified.}
\label{fig:z-dist}
\end{figure*}

We morphologically classify 1087 H-ATLAS sources which have reliability$\geq$0.8 
of being associated with an SDSS source, and which have good quality spectroscopic
redshifts (flagged with {\sc z\_quality} $(nQ) \geq 3$). Additionally, we require that sources are 
at a redshift of $z<0.5$; above this redshift only a very small number of galaxies have spectroscopic redshifts and will be difficult to classify.
Two sources with stellar or QSO IDs were removed from the sample, 
as were the five sources identified as being lensed in \citet{Negrello10}.
We calculate the number of false IDs expected in the classified sample from the sum of the probabilities of a false ID as 
$\sum(1-R)$, where $R$ is the reliability. This indicates that 21 galaxies (2\%) in our 
sample are likely to be false IDs.
There are 115 and 199 sources for which we have PACS 100 and 160$\mu$m point-source detections at $\geq5\sigma$, respectively. The selection effects arising from the PACS detections are discussed in \citet{Smith11b}, who found that the SED results in Section~\ref{sec:results} are not significantly influenced by the inclusion of upper limits for PACS data in the majority of the sample. All sources are detected $\geq5\sigma$ at 250$\mu$m (which is a requirement for our sample selection), 272 sources have a $\geq5\sigma$ detection at 350$\mu$m, and 138 sources have a $\geq3\sigma$ detection at 500$\mu$m. The distribution of 100$\mu$m PACS and 250$\mu$m SPIRE detections is shown for each morphological type in Figures~\ref{fig:z-dist} (a) and (b).

We visually classify 44 galaxies as early-type (E or S0), with $0.01<z<0.32$. 
It can be seen from Table \ref{table:control_morphs} that there are few
ETGs in our sample compared to spirals, so it is evident that H-ATLAS
preferentially selects spiral galaxies over ETGs. This is as expected 
since ETGs are generally passive and have little dust content.
The late-type category in principle encompasses both spirals and irregular
galaxies, however no irregular galaxies are found in our sample. This
may be because these objects are difficult to classify at all but the
very lowest redshifts, but H-ATLAS also does not detect many
low optical luminosity (and therefore low mass) sources in the SDP field \citep{Dunne11, Dariush11}.
The number of mergers in our classified sample is underestimated because the
reliability$\geq0.8$ criteria inherently assumes a 1:1 correspondence
between optical and submillimetre sources \citep{SuthSaund92, Smith11a}. In the case of
mergers there can be two optical sources close to the SPIRE position
which both have a high likelihood of association but the probability
(reliability) is split between the sources, sometimes reducing the
reliability below our threshold of 0.8. The median redshifts of the ETGs
and spirals in our sample are both $\sim0.13$, and the redshift
distribution is shown in Figure~\ref{fig:z-dist} (c).
As galaxies become faint and small with increasing redshift, classification
becomes difficult, and the unknown fraction increases significantly
for $r_{\rm{petro}}>18.5$ (see Figure~\ref{fig:z-dist} (d)). It also seems easier to classify spirals than ETGs at fainter $r$-magnitudes. 
We observe morphological disturbances in 13/44
($30^{+8}_{-6}\%$) ETGs and 22/496 ($4\pm1\%$) spirals\footnote{The errors are $1\sigma$ confidence intervals on a binomial population using a beta distribution, which is appropriate for small population numbers \citep{Cameron11}.}. These galaxies show evidence of dust or tidal features and may be signs of a merger remnant, however this is a lower limit on the number which may be disturbed since faint
features may not be visible in standard depth SDSS
images. Morphologically disturbed sources occupy a range of redshifts
up to $z\sim0.26$. We find a higher fraction of morphological disturbance in ETGs
compared to \citet{Kaviraj10_Stripe82} who find 18\% for a sample of optically selected ETGs (with $r<16.5$ and $z<0.05$).

To check our ETG classifications, we compare to those in the Galaxy Zoo sample \citep{Lintott08, Lintott11}, in which galaxies were visually classified by over 100,000 volunteers. Only the brighter members of our H-ATLAS sample ($r<17.77$ and $z<0.25$) overlap with Galaxy Zoo. Galaxies were classified as either elliptical\footnote{The `elliptical' classification also contains the majority of S0 galaxies, as shown in \citet{Bamford09}.}, spiral, merger or `don't know'. We assign a galaxy one of these classifications if it has $>50$ percent of the vote fraction. Debiased votes are used to account for the tendency for Galaxy Zoo classifiers to assign small or faint galaxies (usually at higher redshift due to a lack of resolution) to the `elliptical' category. The debiasing procedure is described fully in \citet{Bamford09, Lintott11}. There are 22 of our ETGs which have a match in Galaxy Zoo, 17/22 are classified as elliptical, 3/22 are classified as spiral, and 2 are ambiguous. The ETGs which are classified as spirals in Galaxy Zoo either have evidence of disturbed morphology which could have been mistaken for spiral structure, or have evidence of a disk yet no spiral arms. The majority our H-ATLAS ETGs which match with the Galaxy Zoo sample are classified as `elliptical', and so our morphological classifications agree well with overlapping studies.

\begin{figure}
\centering
\includegraphics[scale=0.7]{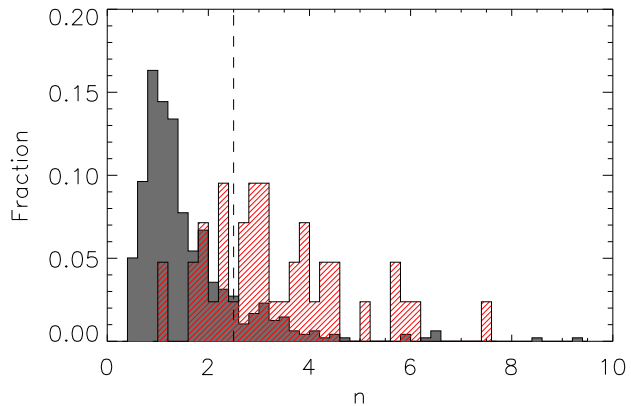}
\caption{S\'{e}rsic index (n) distribution of the spirals (grey), and ETGs (red/hatched) in our H-ATLAS sample. The dashed line at $n=2.5$ denotes the traditional cut between `early-type' ($n>2.5$) and `late-type' ($n<2.5$), and gives an indication of the contamination that can occur in samples selected on S\'{e}rsic index.}
\label{fig:Sersic}
\end{figure}

We also examine the S\'{e}rsic index ($n$) of our ETGs and spirals in Figure~\ref{fig:Sersic}, to check that our morphological classifications are broadly consistent with what is expected from automated galaxy classification. This is accomplished by fitting single component S\'{e}rsic models to the light profile of the galaxy (Kelvin et al. in prep.). Generally, late-type galaxies have an exponential profile ($n=1$), and ETGs have a de Vaucouleurs profile ($n=4$). As expected, our visually classified spirals have a very strong peak at $n=1.3$, whereas the ETGs have a variety of Sersic indicies, but have a higher average $n$ of 3.1. The wide range of S\'{e}rsic indicies is because we include S0s in our early-type classification, which may have a substantial disk component. We note that although $n$ broadly agrees with our visual morphologies,  S\'{e}rsic index is not the ideal classification method, because a spiral with a bright nucleus may appear to have a high value of $n$ and would be mis-classified as an ETG \citep[e.g.][]{Bamford09}.

\subsubsection{Control sample}
\label{sec:control_sample}
In order to understand how the H-ATLAS and optically selected 
ETG populations differ, we obtain visual morphological classifications of a control
sample drawn from the GAMA galaxy catalogue which overlaps with the H-ATLAS SDP field.
Galaxies are required to be undetected in H-ATLAS and have good quality spectroscopic 
redshifts, and were chosen to have the same redshift and $r_{\rm{petro}}$ -magnitude 
distribution $n(r,z)$ as our H-ATLAS detected, morphologically classified sample. 
This was accomplished by splitting the H-ATLAS sample into $(r,z)$ bins, and randomly picking approximately 
the same number of galaxies in each bin from the GAMA catalogue, so that the control sample comprises 1052 galaxies.
By selecting a control sample of galaxies which are matched in redshift to the H-ATLAS sample, we 
avoid selection effects.

The morphologies of the control sample are summarised in Table
\ref{table:control_morphs}. It can be seen that there are many more
ETGs compared to spirals in the optically selected sample than 
in the 250$\mu$m selected sample. To estimate the fraction of galaxies which are detected at the depth of H-ATLAS as a function of morphology, we pick a random sample of 1076 galaxies\footnote{The random sample is chosen to be approximately the same size as the H-ATLAS detected sample, but 11 H-ATLAS galaxies are not in the GAMA survey region and lack $r_{\rm{petro}}$-magnitude information. Therefore the size of the random sample is smaller than the H-ATLAS sample, but this should not affect any of our conclusions.} from the GAMA catalogue in the SDP field, disregarding whether they are detected by \emph{Herschel}. We make sure the selected galaxies follow the same $n(r,z)$ as our H-ATLAS detected and control samples. We repeat the random sampling 1000 times to estimate the average number of H-ATLAS detected and undetected galaxies. On average, 225 galaxies are in the H-ATLAS detected sample, and 11 of these are ETGs. There are 851 undetected galaxies, and from the control sample fractions we expect 22\% (188) of these to be ETGs. Consequently, there are 199 ETGs in total in the random sample, so we estimate 5.5\% of ETGs are detected in H-ATLAS compared to the total number of ETGs in the SDP field, for this $n(r,z)$. The detected fractions of other morphological types are presented in Table \ref{table:control_morphs}. We cannot reliably extrapolate the control sample fractions to the entire SDP field, since morphology is a function of both $r$ and $z$, and we have not probed the full $(r,z)$ parameter space in this work.

\subsubsection{Classification bias}

\begin{figure*}
\centering
\includegraphics{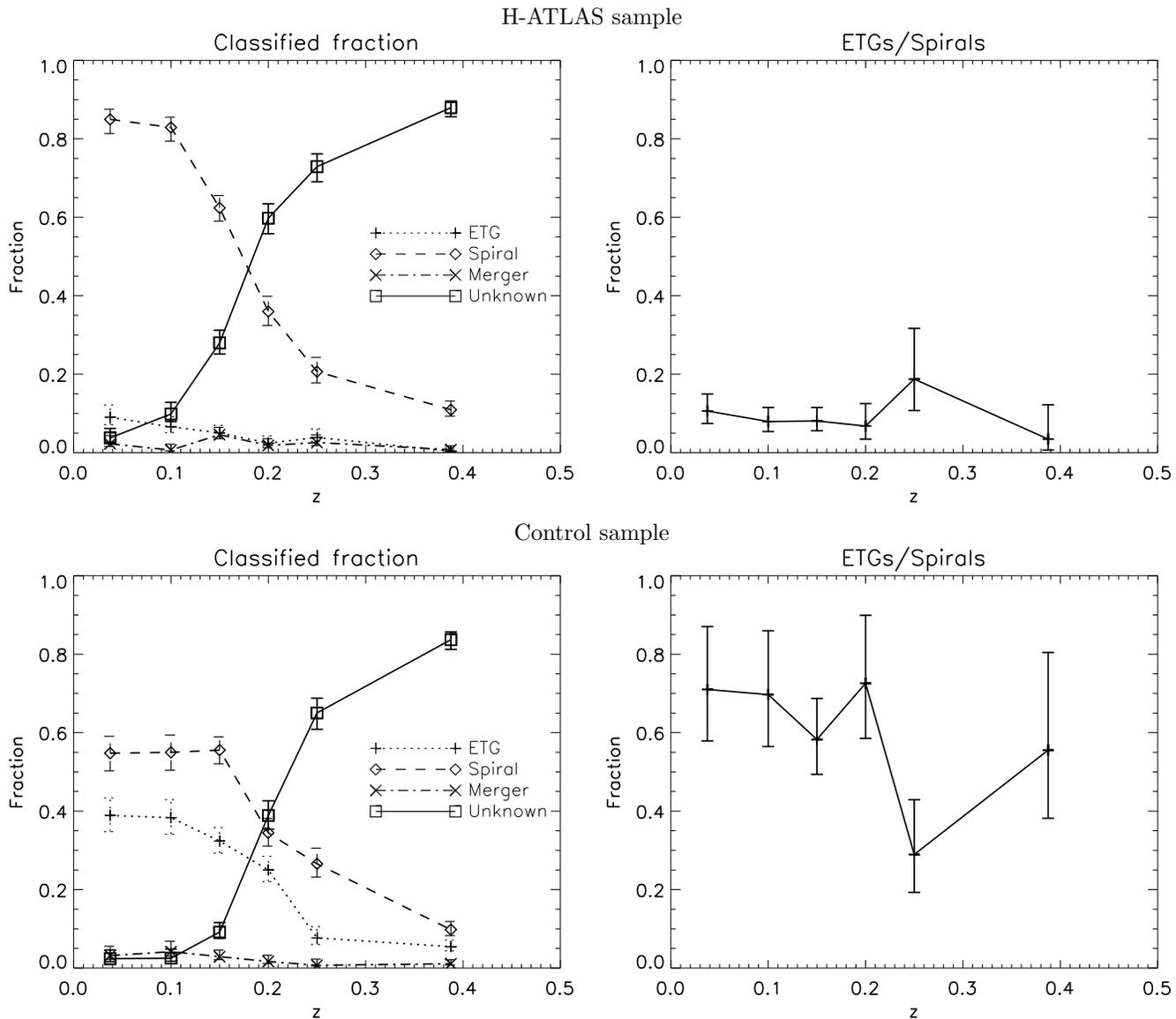}
\caption{Morphological classification fractions as a function of redshift for the H-ATLAS detected sample (top) and the control sample (bottom). Errors bars are the $1\sigma$ confidence intervals for a binomial population, derived from a beta distribution (see \citealt{Cameron11}). Also shown is the ETG-to-spiral fraction for both samples, where error bars are the $1\sigma$ confidence intervals for a binomial distribution, using the approximation of \citet{Gehrels86}. The ETG-to-spiral fraction does not increase with redshift, therefore we do not observe a bias towards classifying more ETGs as they become smaller and fainter.}
\label{fig:morph_frac}
\end{figure*}

\citet{Bamford09} showed that in Galaxy Zoo the fraction 
of galaxies classified as `elliptical' increases with redshift compared to 
spirals. This is because the spatial resolution and signal-to-noise decreases with redshift, 
so features such as spiral arms become invisible. Also, in Galaxy Zoo, images are presented to the 
classifier without any indication of angular scale, so distant, unresolved galaxies 
could have been classified as elliptical. We should therefore check if this bias is present in our 
classifications. We show the classification fractions of our H-ATLAS sample in Figure
\ref{fig:morph_frac}, and there is no trend that we classify more 
ETGs with increasing redshift. Indeed, we see that we classify fewer.
This may be because unlike Galaxy Zoo volunteers, our expert classifier recognises 
the limitations of the resolution of the image, and will classify an object 
as unknown instead of as an ETG.

\subsection{SED fitting}
\label{sec:SED_fitting}
\citet{Smith11b} fit the UV-submillimetre SEDs of 1404 H-ATLAS galaxies with reliability $>0.8$ of being
associated with an optical counterpart in the SDSS $r$-band catalogue,
and which have available multiwavelength photometry.
Using the physically motivated method of \citet*[][hereafter DCE08]{DCE08} 
allows us to recover the physical properties of these galaxies.
In this method the energy from UV-optical radiation emitted by stellar populations is
absorbed by dust, and this is matched to that re-radiated in the FIR. Spectral
libraries of 25000 optical models with stochastic star-formation histories, and 50000
infrared models, are produced at the redshift of each galaxy in our
sample, containing model parameters and synthetic photometry from the
UV to the submillimetre. The optical libraries are produced using the
spectral evolution of stellar populations using a \citet{Chabrier03}
Galactic-disk Initial Mass Function (IMF), calculated from the latest version of the
population synthesis code of \citet{BC03}, which includes a revised prescription 
for thermally-pulsing asymptotic giant branch (TP-AGB)
stars (Bruzual and Charlot, in prep). These libraries contain model spectra with a wide range of star-formation
histories, metallicities and dust attenuations. The two-component dust
model of \citet{CF00} is used to calculate the attenuation of
starlight by dust, which accounts for the increased attenuation of
stars in birth clouds compared to old stars in the ambient interstellar medium (ISM).
The model assumes angle averaged spectral properties and so does not include 
any spatial or dynamical information.

The infrared libraries
contain SEDs with different temperature dust components, which include
polycyclic aromatic hydrocarbons (PAHs), hot dust (stochastically heated small grains, $130-250$ K), warm dust in birth clouds
($30-60$ K) and cold dust grains ($15-25$ K) in thermal equilibrium in the diffuse ISM, from
which the dust mass ($\rm{M}_\mathrm{d}$) is calculated. A dust emissivity
index $\beta=1.5$ is assumed for warm dust, and $\beta=2.0$ for cold
dust, as described in DCE08. 

The attenuated stellar emission and dust emission models in the two spectral libraries are combined using a simple energy balance argument: that the energy absorbed by dust in stellar birth clouds and the diffuse ISM are re-emitted by dust in the infrared. In practise, this means that each model in the optical library is matched to models in the infrared library which have the same value of \fmu (within a tolerance of 0.15) and are scaled to the total dust luminosity\footnote{Integrated between $3-1000\mu$m.} \ldust.
We derive statistical constraints on the various parameters of the model using the Bayesian approach described in DCE08. We compare each observed galaxy SED to the library of stochastic models which encompasses all plausible parameter combinations. For each galaxy, we build the marginalised likelihood distribution of any physical parameter by evaluating how well each model in the library can account for the observed properties of the galaxy (by computing the $\chi^{2}$ goodness of fit). This method ensures that possible degeneracies between model parameters are included in the final probability density function (PDF) of each parameter. The effects of individual wavebands on the derived parameters are explored in DCE08, and \citet{Smith11b}, but we emphasise the importance of using the H-ATLAS FIR-submillimetre data to sample the peak of the dust emission and the Rayleigh-Jeans slope in order to get reliable constraints on the dust mass.

\begin{figure*}
\centering
\includegraphics[]{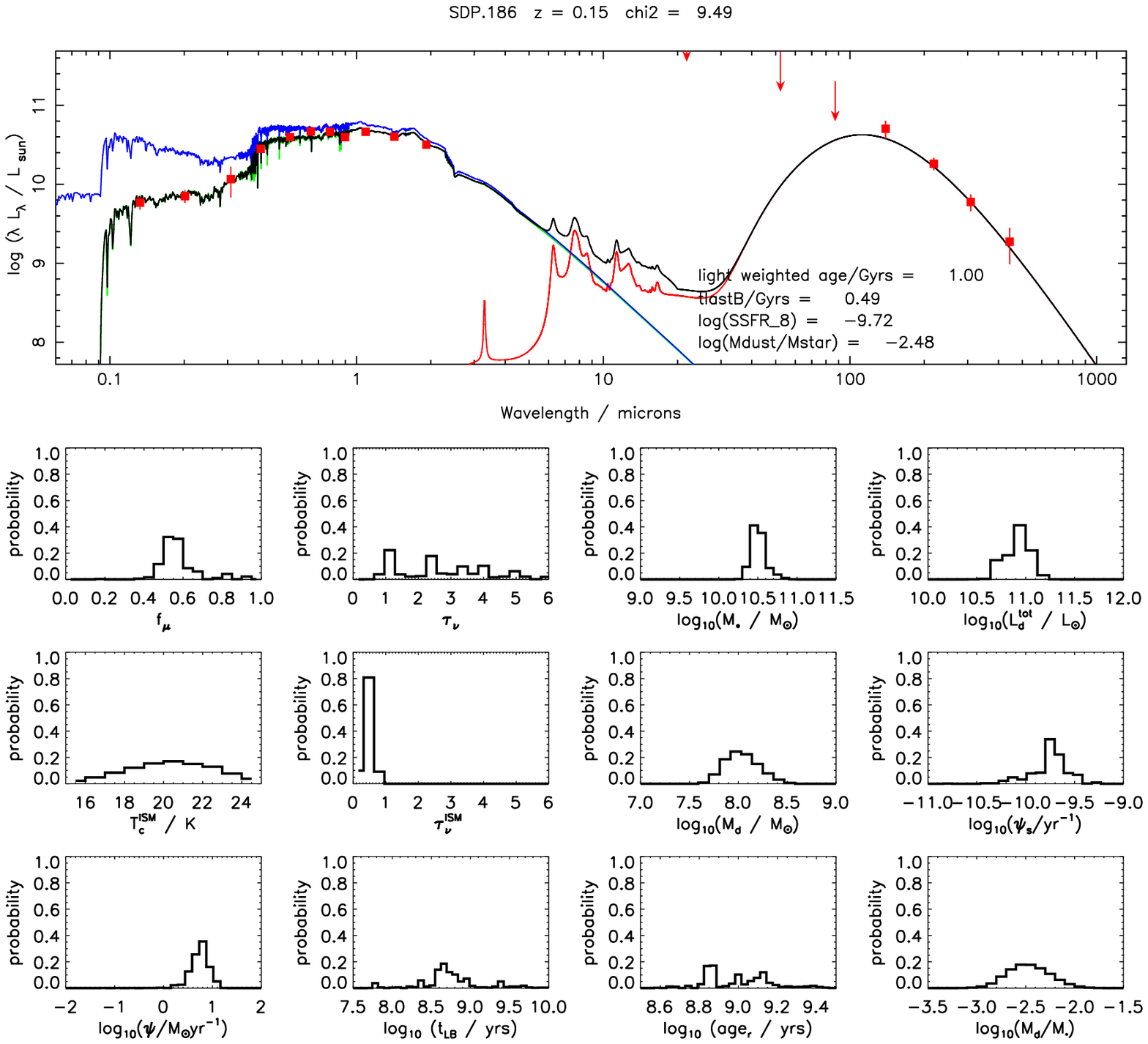}
\caption{{\bf Top}: Example best-fit rest-frame SED of an ETG, with observed photometry points from the UV to the submillimetre (red squares). 5$\sigma$ upper limits are shown as arrows. Errors on the photometry are described in \citet{Smith11b}. The black line is the best fit model, the green line is the attenuated optical model, the blue line is the unattenuated optical model, and the red line is the infrared model. {\bf Bottom}: probability density functions of parameters for this ETG.}
\label{fig:SED_example}
\end{figure*}

An example best-fit SED and set of PDFs are shown in Figure
\ref{fig:SED_example}. The parameters we compute are \fmu, the
fraction of total dust luminosity contributed by the diffuse ISM;
\tauv, total effective V-band optical depth seen by stars in birth
clouds; $\rm{M}_\ast/\rm{M}_\odot$, stellar mass; \ldust/$\rm{L}_\odot$, dust luminosity; \tbgscold/K, temperature of
the cold diffuse ISM dust component; \tauvISM, the V-band optical depth in the
ambient ISM; $\rm{M}_\mathrm{d}/\rm{M}_\odot$, dust mass; \ssfr/$yr^{-1}$,
specific star-formation rate (SSFR); \sfr/$\rm{M}_\odot yr^{-1}$, SFR; $\rm{t_{LB}}$, time
of last burst; $\rm{age_r}$, $r$-band light-weighted age and \mdms, dust to stellar mass
ratio. For more details of the method we refer the reader to DCE08.

\section{Properties of ETGs compared to spirals}
\label{sec:results}
Here we explore the multiwavelength properties of our sample of
morphologically classified spirals and ETGs detected in H-ATLAS. We
present parameters derived from the SED fitting method as described in
Section \ref{sec:SED_fitting} for 42 of the 44 ETGs, and as a
comparison we also explore the properties of 450 out of the 496 spiral
galaxies in our sample. We present the SDSS images, best-fit SEDs
and optical spectra of these ETGs in Figure~\ref{fig:ETGs}.
The galaxies which are excluded from our analysis do
not have available aperture-matched GAMA photometry (2 ETGs, 17 spirals); additionally we reject 
29 galaxies from our analysis which have poor quality SED fits with $\chi^2>30$. Our sample 
covers a range of redshifts, but since the median redshifts of the ETGs
and spirals are approximately the same, differences between the samples 
due to evolution in the redshift range are likely to be small. Additionally, 
we have checked that the following trends are present if we look at galaxies 
at $z<0.13$, and $z>0.13$. We also observe similar results if we separate our 
successfully classified sample into `early-type' ($n>2.5$) and `late-type' ($n<2.5$) 
using S\'{e}rsic index.

\subsection{SED parameters}
\label{sec:SED parameters}

\begin{figure*}
\centering
\includegraphics[width=18cm]{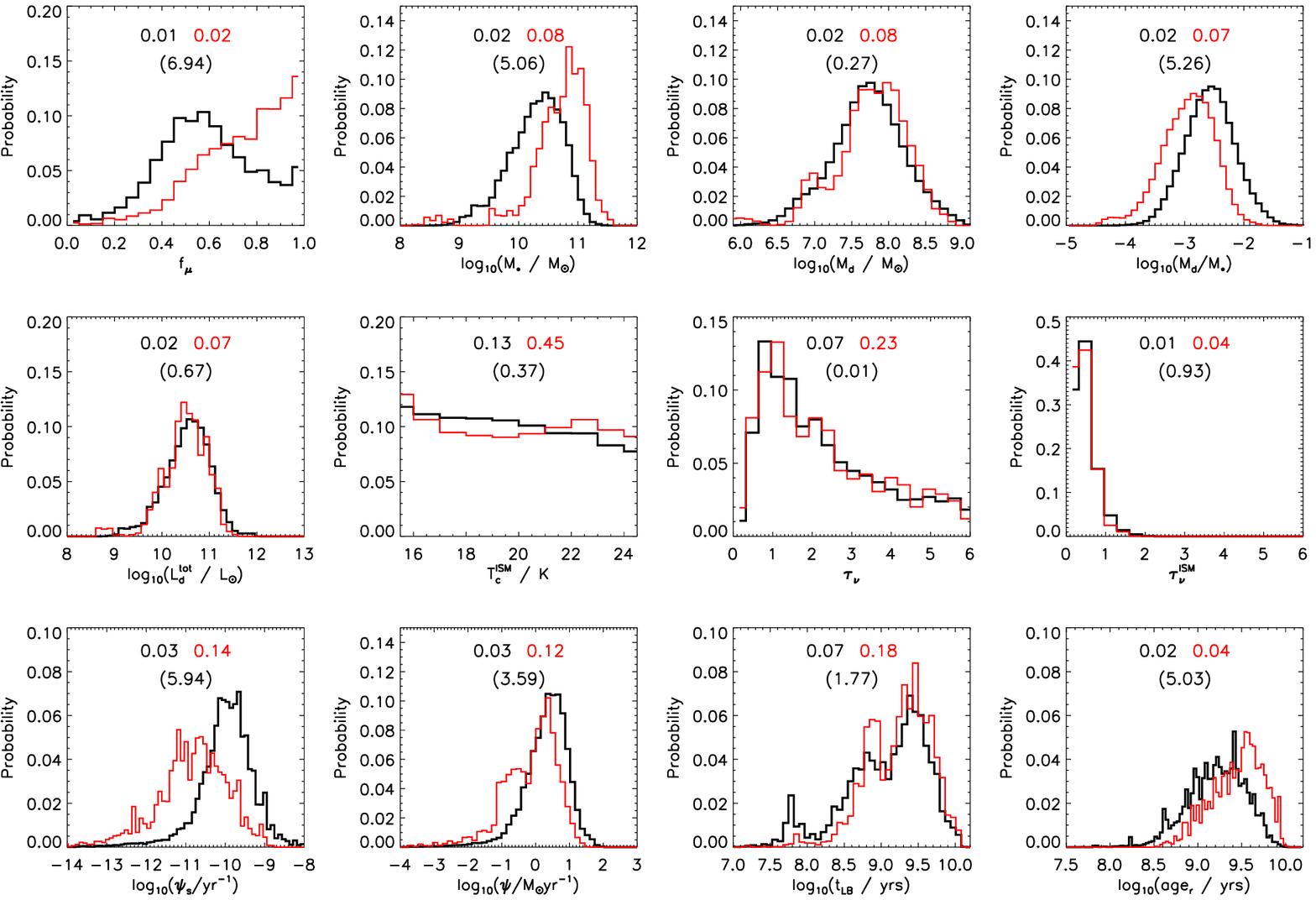}
\caption{Average PDFs of the SED parameters of 42 ETGs
  (red line) compared to 450 spirals (black line). The parameters are (from left to
  right): \fmu, the fraction of total dust luminosity contributed by
  the diffuse ISM; $\rm{M}_\ast$, stellar mass; $\rm{M}_\mathrm{d}$, dust mass;
  \mdms, dust to stellar mass ratio; \ldust, dust luminosity;
  \tbgscold, temperature of the cold ISM dust component; \tauv, total
  effective V-band optical depth seen by stars in birth clouds;
  \tauvISM, the V-band optical depth in the ambient ISM; SSFR and SFR 
  averaged over the last ${10}^{8}$ years; $\rm{t_{LB}}$, time of last burst; 
  and $\rm{age_r}$, the $r$-band light-weighted age of the stellar population. 
  The uncertainty on each distribution for ETGs and spirals is given by 
  the error on the mean and is shown at the top of each histogram with corresponding colours, and 
  the significance of the difference in the means in brackets. The errors for logarithmic parameters are in dex.}
\label{fig:SED_params}
\end{figure*}

In order to compare physical parameters for ETGs and spirals in our sample, we compute 
the average probability density function of parameters derived from our SED fitting.
The average PDFs of ETGs (red) and spirals (black) are shown in Figure~\ref{fig:SED_params}, 
and the mean values and errors are summarised in Table \ref{tab:summary_properties}.
For each parameter, we use the first moment of the average PDF to estimate the mean of the population. 
We can estimate the variance on the population mean as the second
moment of the average PDF minus the mean squared, divided by the number 
of galaxies in the sample. The error on the mean is simply the square
root of the population variance.
The significance of the difference in the means is shown in brackets in Figure~\ref{fig:SED_params} 
and uses the quadrature sum of the errors on the mean for the two populations.

The ETGs have a mean \fmu{} of $0.74\pm0.02$, which is significantly higher than that of
spirals which have a mean \fmu{} of $0.59\pm0.01$. This means that most of the
FIR luminosity in ETGs is from dust in the diffuse ISM, which is mostly 
heated by old stellar populations
(stars older than $10^7$ years). Some ETGs have lower values
of \fmu, indicating that more of the FIR luminosity comes from dust in birth clouds,
which is heated by young stars and implies
ongoing star-formation. The ETGs in our sample are more massive than
spirals, with ETGs having a mean stellar mass ($\rm{M}_\ast$) of
$(4.9^{+1.0}_{-0.8})\times{10^{10}}\,$M$_\odot$ compared to $\rm{M}_\ast$ of
$(1.9\pm0.1)\times{10^{10}}$\,M$_\odot$ for spirals. ETGs have approximately the
same mean dust mass ($\rm{M}_\mathrm{d}$) and dust luminosity (\ldust) as spirals, 
although the mean ratio of dust to stellar mass (\mdms) for ETGs is
lower than that for spirals in our sample by 0.38 dex, meaning that
ETGs are dust deficient for their stellar mass compared to spirals. 
Our median \mdms{} value for ETGs is $(1.6\pm0.1)\times10^{-3}$, which is larger than 
the average found by \citet{Skibba11} of $1.7\times{10}^{-4}$ for 10 ETGs.
In order to contain enough dust to be detected in H-ATLAS, galaxies which
have low \mdms{} need to be more massive in general, which may explain
why ETGs have a higher $\rm{M}_\ast$ on average than spirals in our
sample. We find a mean dust mass of
$(5.5^{+1.1}_{-0.9})\times{10}^{7}$\,M$_\odot$ for the sample of ETGs, which is larger
than the highest dust masses found in some previous studies of ETGs
e.g. \citet{Temi04} found ${10}^{5}-{10}^{7}$\,M$_\odot$, which is similar to the dust masses found in ETGs with optical dust lanes \citep{Kaviraj11} (although these are likely to be underestimated by the use of \emph{IRAS} data - this issue will be addressed in future work.) Our mean dust mass is
consistent with \citet{SLUGS05} who found dust masses greater than
${10}^{7}$\,M$_\odot$ for 6 elliptical galaxies from an optically
selected sample observed with SCUBA. The dust mass inferred for the 
SCUBA ellipticals may include contamination from synchrotron radiation, 
but for the sample of ETGs studied here, we find that synchrotron radiation is negligible compared to thermal emission from dust (see Section \ref{sec:radio}).
We find no significant difference in the
\tbgscold of the spirals and ETGs, and find a wide range of values for the dust temperature.
The total effective V-band optical depth seen by stars (\tauv, \tauvISM) is approximately the same for ETGs and spirals. This shows that ETGs have
approximately the same attenuation as spiral galaxies
(though with rather large uncertainties).

\subsection{Star-formation histories}
\label{sec:SFH}
We investigate the SFH of our galaxies by examining the SFR, (\sfr) 
and SSFR, (\ssfr, defined as \sfr/$\rm{M}_\ast$) averaged over the last ${10}^{8}$
years. These parameters are derived from the SED fitting as described
in Section \ref{sec:SED_fitting}. The model SFHs are described by a
continuous exponentially decreasing star-formation rate,
with superimposed randomly distributed bursts of star-formation
\citep{Kauffmann03_SFH} lasting between $3 \times 10^{7}$ and $3
\times 10^{8}$ years. These bursts occur with equal probability
throughout the lifetime of the galaxy. The probability is set such that 50 percent of the galaxies in the library
have undergone a burst of star-formation in the
last 2\,Gyr. The amplitude of the burst (ratio of mass formed in the burst
to mass formed in continuous star-formation over the lifetime of the galaxy) is distributed
logarithmically between 0.03 and 4.0. For further details of the models, and the effects 
of model assumptions on derived parameters we refer the reader to \citet{Kauffmann03_SFH} and DCE08.

The mean SFR for ETGs is $0.7\pm0.2\,\rm{M}_\odot$yr$^{-1}$, with a range of
$0.04-12.4$\,M$_\odot$yr$^{-1}$. It is interesting to note that distribution of ETG 
SFRs in Figure~\ref{fig:SED_params} shows signs of bimodality. Our range of SFRs is
comparable to that found for optically blue ETGs by \citet{Schawinski09}, who
find SFRs of $0.5-50$\,$\rm{M}_{\odot}$yr$^{-1}$ using a range of indicators
(H$\alpha$ luminosity, $u$-band light, infrared luminosity from \emph{IRAS}). 
H-ATLAS ETG SFRs are also larger than those found in recent studies of ETGs in the
SAURON sample, which is a representative sample of local ETGs, 
which are located in both clusters and the field \citep{deZeeuw02}. 
\citet{Temi09b} find that the SFR for SAURON S0s as estimated from 24$\mu$m luminosity is
$0.02-0.2\,\rm{M}_\odot$yr$^{-1}$, and \citet{Shapiro10} 
calculated the SFR in the SAURON sample from non-stellar
8$\mu$m emission, and this was found to be $<0.4$\,M$_\odot$yr$^{-1}$. These findings of low level
star-formation in the SAURON galaxies can possibly be explained by the optical selection, 
which is not biased towards highly star-forming galaxies. 
This is in contrast to the H-ATLAS sample which selects the dustiest ETGs, and therefore the highest SFRs.
Additionally, the SAURON measurements only give the obscured SFR, and may not be representative of the total SFR of the galaxy.

SSFR is defined as the star-formation rate per unit stellar mass 
and measures the star-formation efficiency of a galaxy. Figure
\ref{fig:SED_params} shows that the mean SSFR averaged over the last $10^8$
years for ETGs ($1.4^{+0.5}_{-0.3}\times 10^{-11}\rm{yr}^{-1}$) is lower than that
of spirals ($1.0\pm0.7\times 10^{-10}\rm{yr}^{-1}$). This trend is 
insensitive to changes in the timescale over which the SSFR is
averaged. There is, however, a wide range of SSFR and 17 percent of
ETGs have a SSFR greater than the mean of the spiral sample.

\begin{figure}
\centering
\includegraphics[clip=true, scale=0.85]{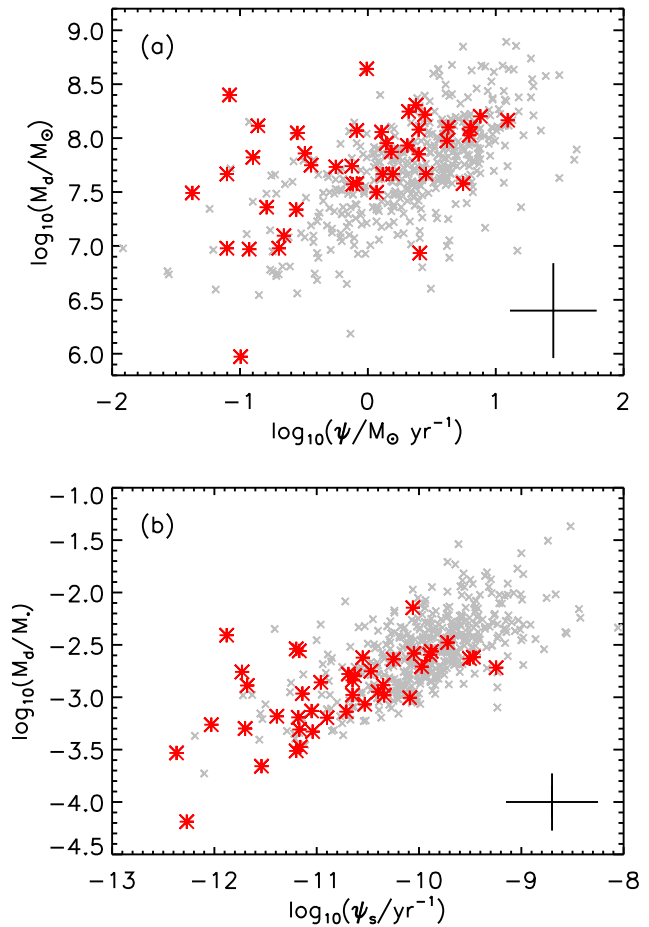}
\caption{{\bf (a):} $\rm{M}_\mathrm{d}$ vs. SFR of the ETGs (large red stars) compared to the spiral galaxies (small grey crosses).
{\bf (b):} \mdms{} vs. SSFR of the ETGs compared to the spiral galaxies. The error bars indicate the median $1\sigma$ uncertainty on a data point.}
\label{fig:SSFR_Md_mstar}
\end{figure}

In Figure~\ref{fig:SSFR_Md_mstar} (a), we show a plot of dust mass versus SFR for spirals and ETGs in our sample. 
It can be seen that galaxies with the highest dust mass also have a high SFR.
This trend is expected since both dust mass and SFR will depend on the total stellar mass of a galaxy.
We can remove this trend by dividing by stellar mass and so we plot \mdms{} vs. SSFR 
in Figure~\ref{fig:SSFR_Md_mstar} (b). As was found in \citet{dC10}, 
there is a strong correlation between these two parameters. It can be 
seen that typically the ETGs have lower SSFR and \mdms{} than spirals.
There are some spirals with very low SSFR and \mdms, which are discussed in Section \ref{sec:Passive spirals}.

We can use the results of our SED fitting to see if star-formation is dominated by a recent burst or
continuous star-formation using the model parameter $\rm{t_{LB}}$. Although 
there is a large uncertainty on this parameter, our results are still useful for a statistical 
comparison of two populations. As shown in Figure~\ref{fig:SED_params}, $\sim 76$
percent of our ETGs have not had a burst of star-formation in the last
${10}^{9}$ years, and have therefore not formed a substantial fraction
of their mass in recent bursts. It seems that most of our
sample have residual star-formation left over from the
last major burst. \citet{Kauffmann03_AGN} find
that galaxies with $\rm{M}_\ast>10^{10}\,\rm{M}_\odot$ typically have not had
recent bursts of star formation, which may explain why our generally high mass ETG sample
shows few recent bursts. The time since the last burst can also be characterised by
the age of the young stellar population, parametrised in our models by
the $r$-band light-weighted age ($\rm{age}_{r}$). It is found that the mean stellar population age of the ETGs is $2.8\pm0.3\,$Gyr, 
which is older than that found for the spirals of $1.6\pm0.1\,$Gyr. This is consistent with the general picture that ETGs are older than spirals.  
We note that 3/10 ETGs with bursts of
star-formation in the last 1\,Gyr show disturbed morphologies, so
galaxy interactions may be the cause of the burst.  It is possible that more ETGs in this sample
are disturbed at a level which is not detected in the shallow imaging that we have available. 
Without deeper imaging, conclusions cannot be drawn about
whether there is a correlation between morphological disturbance and
recent star-formation in this sample.

\subsection{Comparison of broadband photometric and spectroscopic star-formation parameters}
SFH parameters are traditionally measured using spectroscopic
information, whereas we used broadband data, so
there may be a large uncertainty on some parameters.
\citet{Walcher08} explored degeneracies in the SFH parameters from broadband photometry using
similar stellar population models to those in this work, 
and found that $\rm{M}_\ast$, $\rm{age}_{r}$ and \ssfr\ are well
determined. \citet{Wild09} classify galaxies into star-forming
galaxies and quiescent galaxies using broadband and spectroscopic data, and found a good agreement between 
these two classification methods. They also find the time of last burst derived
from broadband SED fitting agrees with that derived from spectroscopy. 

\begin{figure*}
\includegraphics{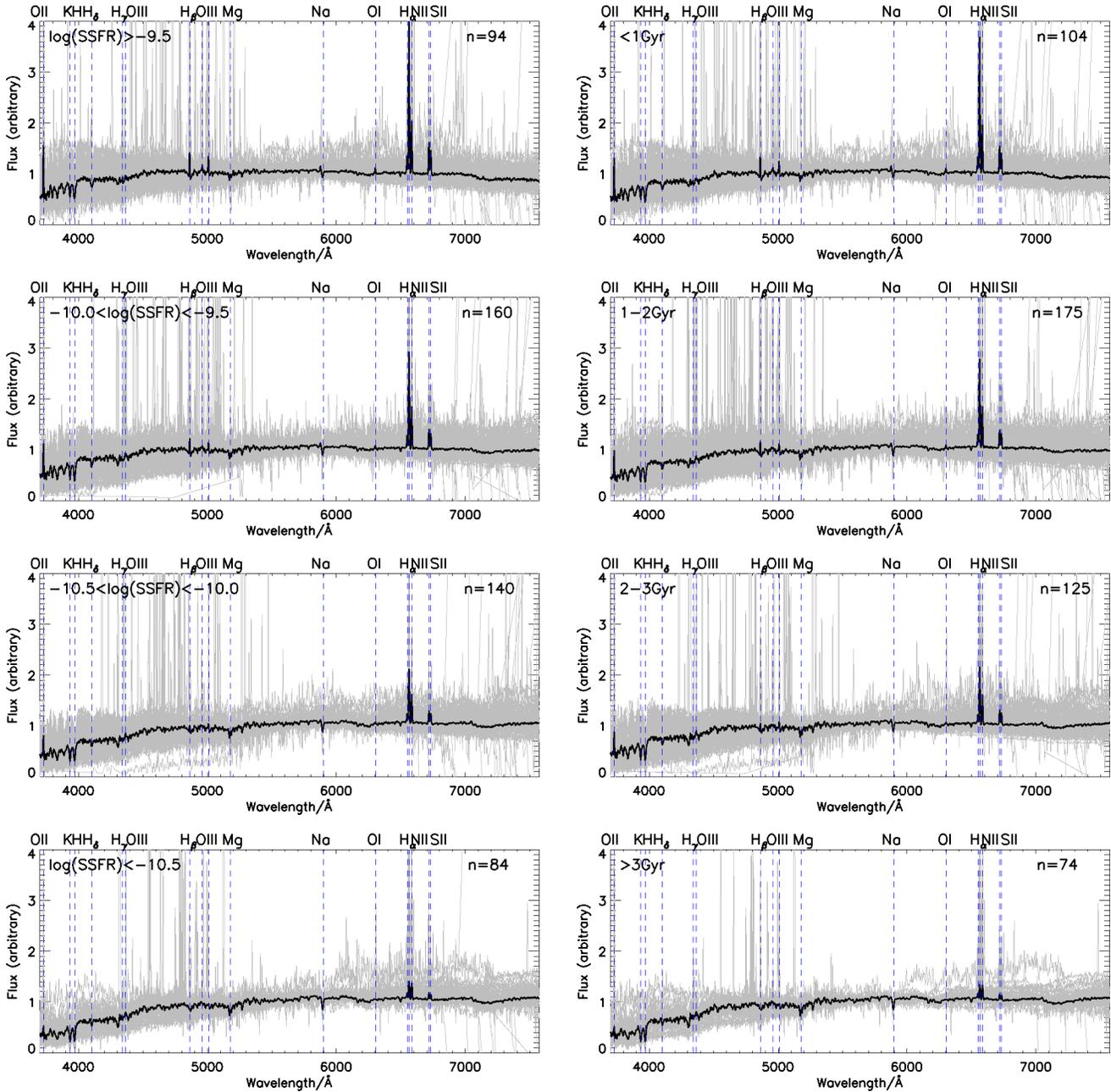}
\caption{Stacked spectra in bins of SSFR (left) and stellar population age (right). The bins are arranged 
from high SSFR/young stellar population (top) to low SSFR/old stellar population (bottom). 
The grey area shows the normalised, individual spectra, and the thick black line is the median of these spectra, smoothed by a boxcar of 5\AA\/. Prominent emission 
lines are shown by the blue dashed lines. The number of spectra in each stack is indicated in each panel.}
\label{fig:SSFR_stacking}
\end{figure*}

To investigate whether fitting SEDs to broadband photometry can accurately
describe the SFHs of our galaxies, we stack spectra of ETGs and
spirals together in bins of SSFR and $r$-band weighted age to look for trends in spectral
features. The spectra are shifted to rest wavelength and resampled
onto a common wavelength array.  The spectra are normalised to the
median of the spectrum, and then combined using the median of the
spectra in each bin. Spectra which show signs of AGN (see Section
\ref{AGN}), or have anomalous effects such as bad sky subtraction or
fibre fringing \citep{Colless01_2dF} have been removed.
It can be seen from Figure~\ref{fig:SSFR_stacking} that as expected,
the galaxies with the highest SSFR show signatures of star-formation
such as strong H$\alpha$ and [OIII] and [OII] emission lines. Going from high 
to low SSFR, the strength of the emission lines decrease;
the same trends are found for age, with
older stellar populations showing minimal signs of star-formation. 

\subsection{UV--Optical Colours}
\label{Colours}
\begin{figure}
\centering
\includegraphics[scale=0.85]{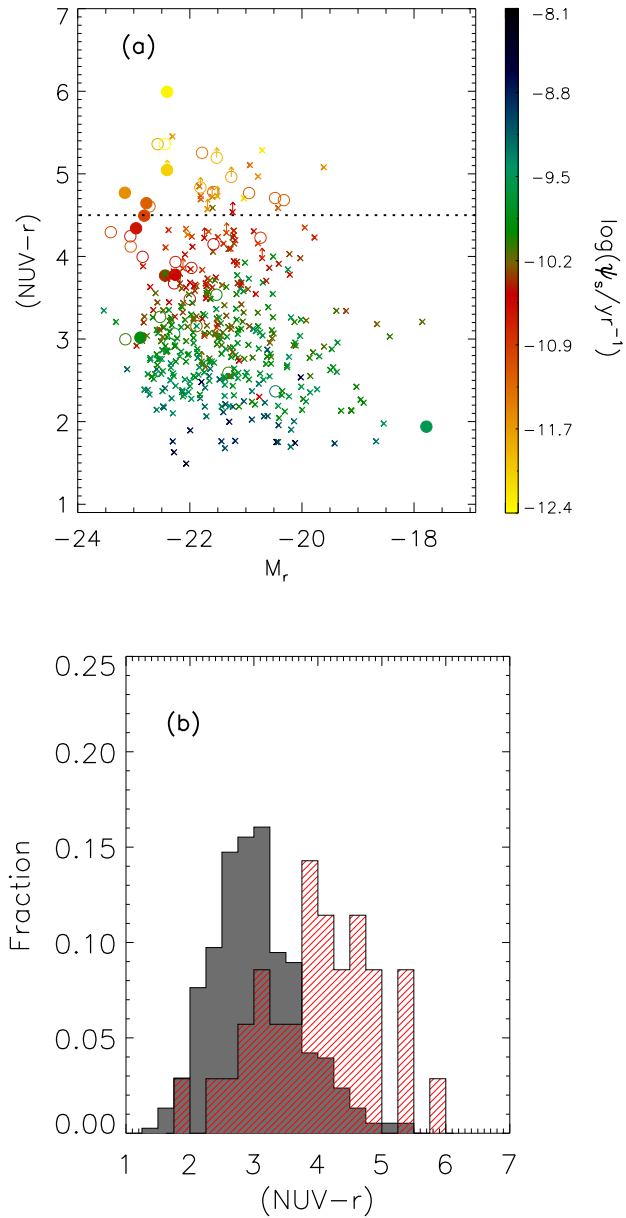}
\caption{{\bf (a)}: UV-optical colour magnitude diagram, colour coded according to SSFR. 
Circles are ETGs, crosses are spirals, filled circles indicate that the ETG is morphologically disturbed.
The dashed line shows the separation between `blue' and`red' as defined in \citet{Dariush11}.
Lower limits are shown for galaxies which do not have a $\geq5\sigma$ NUV detection.
{\bf (b)}: The distribution of ($NUV-r$) colours for the ETGs (red/hatched) and spirals (grey) which have a $\geq5\sigma$ NUV detection.}
\label{fig:(NUV-r)}
\end{figure}

Galaxy colour is often used as a proxy for the age of a stellar
population, with red galaxies assumed to be old due to a lack of UV emission from young stars. 
This simple interpretation can become complicated, with young star-forming galaxies appearing 
red due to dust obscuration, and old galaxies appearing blue due to contamination of the UV 
light by horizontal branch stars\footnote{We note that UV contamination from old stars is unlikely to be a concern, since our sample does not contain giant elliptical galaxies \citep*{Yi97}, and UV flux from old stars is likely to be swamped by that produced by young stars \citep{Kaviraj10}.} \citep{OConnell99, Yi05, Kaviraj09}.
\citet{Dariush11} separate red and blue galaxies in the H-ATLAS sample at $NUV-r=4.5$ by
fitting double Gaussians to the colour distribution. They found that \emph{Herschel} preferentially selects
blue galaxies, and that 90 percent of H-ATLAS sources with red colours are not old/passive\footnote{\citet{Dariush11} define `passive' systems as galaxies which have red ($NUV-r>4.5$) colours, after correcting for dust obscuration.}
but have their light attenuated by dust.
We examine the $NUV-r$ colours of our morphologically selected galaxies using aperture matched \emph{GALEX} UV and GAMA optical photometry in
Figure~\ref{fig:(NUV-r)}. Rest-frame photometry is calculated
using {\sc k-correct.v4.2} \citep{Kcorrect}, and is corrected for galactic extinction using the reddening data of \citet*{SFD98}.
Overall, 93 percent of ETGs have available NUV photometry. For sources which have a
$<5\sigma$ NUV detection \footnote{corresponding to $NUV>23.0$ in the AB magnitude system after galactic extinction correction.}, we compute lower limits for the colours. The mean error in the $NUV-r$ colour is 0.08 magnitudes.

Using the colour cut of \citet{Dariush11} at $NUV-r=4.5$ in Figure~\ref{fig:(NUV-r)} (a),
we find the ETGs have a range of colours, with 24 `blue' and 15 `red'
ETGs. Many exist in the transition region between the red sequence
and blue cloud. The SSFR of each galaxy is represented by the colour of each point, and a
correlation with $NUV-r$ colour is observed. As expected, blue galaxies tend to
have a higher SSFR, and red galaxies a lower SSFR, although with some
exceptions. In Figure~\ref{fig:(NUV-r)} (b) there is a wide range in the colours of both morphological types, although the
median $NUV-r$ colour for the spirals is bluer than that of the ETGs.
This trend is expected since spirals have the bulk of their stellar population
dominated by young stars.

The red ETGs generally have low SSFR, but still contain an appreciable
amount of dust. These sources have high \fmu{} values which indicate the dust in these sources is predominantly heated by
an older stellar population, which gives rise to the
red colour of these galaxies. We are observing these objects at a time when their
star-formation has mostly ceased, either because they have used up all
their gas, or because star-formation has been quenched by some
process. Their dust has not yet been destroyed by sputtering and
shocks from type Ia SNe, and this is discussed in Section
\ref{sec:chem_ev}.

The ETGs which show signs of morphological disturbance
(denoted by filled circles in Figure~\ref{fig:(NUV-r)} (a)) span a range
of colours. A Kolmogorov-Smirnov (K-S) test gives a probability of 0.14 of the colours of disturbed and non-disturbed ETGs 
being drawn from the same distribution, however, this is not significantly different ($1.1\sigma$).
In contrast, \citet{Kaviraj10_Stripe82} find that peculiar ETGs have significantly
bluer $NUV-r$ colours than relaxed ETGs. There is also a small 
population of 15 spirals with $NUV-r>4.5$, and these are discussed in the following section.

\section{Passive and Red Spirals}
\label{sec:Passive spirals}
There has been much discussion in the literature about whether the red
colour of some spirals is due to dust extinction or an old stellar
population (\citealt*{WGM05}; \citealt{Wolf09, Masters10}). \citet{Wolf09} find optically red
spirals have a lower SFR than blue spirals, but also contain large amounts of
dust which obscures star-formation. This may be due to the
inclusion of edge-on spirals in their sample, which would inherently
have a higher dust extinction because the central dust lane is oriented
along our line-of-sight.

Of the 15 red ($NUV-r > 4.5$) spirals in our sample, only two have moderate
levels of star-formation with SSFR$\geq10^{-11}\rm{yr^{-1}}$. The majority of the red spirals have SSFR much lower than this.
By selecting spirals with SSFR\,$<10^{-11}\rm{yr^{-1}}$ we explore the properties of the 19 ($\sim5$\%)
most passive galaxies in our spiral sample. 
We note that this is different from the `passive' definition used by \citet{Dariush11}, 
which was based on dust-corrected UV-optical colour. 
The error on the SSFR for some passive spirals is large (up to 2.1 dex), meaning that some passive spirals could plausibly have SSFR\,$>10^{-11}\rm{yr^{-1}}$, however, the mean of the average SSFR PDF is $(2.6^{+1.3}_{-0.9})\times10^{-12}\rm{yr^{-1}}$. As a population, we can regard the average SSFR of passive spirals as being significantly ($9.3\sigma$) different from those of normal spirals (which have a mean of $(1.2\pm0.1)\times10^{-10}\rm{yr^{-1}}$).
SDSS images, best-fit SEDs and
optical spectra of the passive spirals are presented in Figure~\ref{fig:Red_spirals}. 

These spirals have $NUV-r$ colours ranging from
4.3 to 5.5, although there are 2/19 spirals for which NUV magnitudes are not measured due to the source being in close proximity to a bright star. 
We find 13/17 passive spirals are `red', and 3/17 are `blue', with one passive spiral having ambiguous colour due to an upper limit on the $NUV$ magnitude.
The majority of the passive spirals are not found at the extremes of the colour distribution, and lie in the green valley.

\subsection{Properties of passive spirals}
\begin{figure}
\centering
\includegraphics[clip=true, scale=0.48]{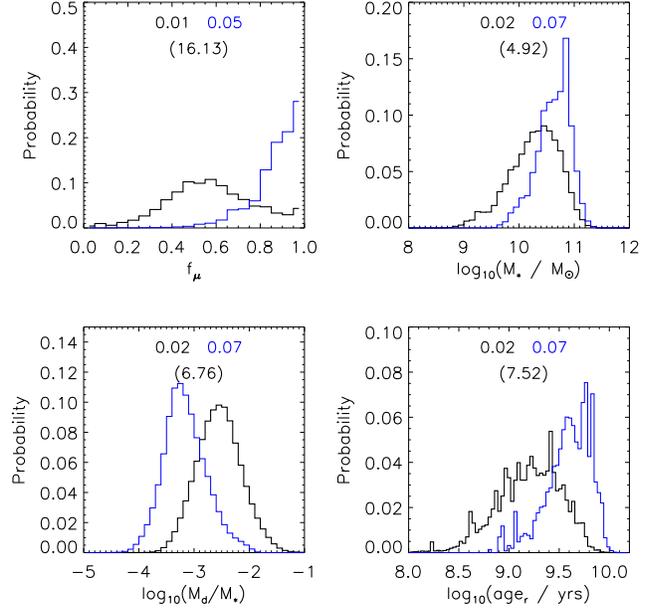} 
\caption{Average PDFs of the SED parameters of 19 passive spirals (SSFR$<10^{-11}\rm{yr^{-1}}$)
  (blue line) compared to 431 normal spirals with SSFR$\geq10^{-11}\rm{yr^{-1}}$ (black line).  
  The parameters are (from left to
  right): \fmu, the fraction of total dust luminosity contributed by
  the diffuse ISM; $\rm{M}_\ast$, stellar mass; \mdms, dust to stellar mass ratio;
  and $\rm{age_r}$, the $r$-band light-weighted age of the stellar population. 
  The uncertainty on each distribution for ETGs and spirals is given by 
  the error on the mean and is shown at the top of each histogram with corresponding colours, and 
  the significance of the difference in the means in brackets. The errors for logarithmic parameters are in dex.}
\label{fig:passive_spirals_mdms}
\end{figure}

A comparison of average PDFs derived from the SED fitting for 19 
passive and 431 normal (SSFR\,$>10^{-11}\rm{yr}^{-1}$) spirals 
is shown in Figure~\ref{fig:passive_spirals_mdms}, and the mean values of the parameters
with errors are summarised in Table \ref{tab:summary_properties}. The passive spirals have a
high mean \fmu{} of $0.87\pm0.02$, indicating that the majority of the dust luminosity is produced 
in the diffuse ISM, and powered mostly by old stellar populations. 
The distribution of V-band optical
depths in the passive and normal spirals is similar, which argues
against the passive spirals being red due to higher dust obscuration.
The differences found in opacity between our passive spirals and
the \citet{Wolf09} red spirals (which have twice the dust extinction
of blue spirals) may be because we only examine passive spirals, and
they select their sample of red spirals on the basis of optical colour
alone. As we have shown in Section \ref{Colours}, red colour does not
necessarily mean that galaxies are passive.

The mean $\rm{M}_\ast$ of the passive
spirals is $(4.2^{+0.7}_{-0.6})\times{10}^{10}\,\rm{M}_\odot$, in comparison to that of
the normal spiral population which has a mean of
$(1.9\pm0.1)\times{10}^{10}\,\mathrm{M}_\odot$. We find 95 percent of the passive spirals are massive with
$\rm{M}_\ast>{10}^{10}\,\rm{M}_\odot$. This could be a selection
bias in that we can only detect dust in the most massive passive
spirals as their \mdms{} ratios are much lower than
the normal spiral population.
Alternatively, \citet{Masters10} found that in their sample almost all 
red spirals were massive ($\rm{M}_\ast>{10}^{10}\,\rm{M}_\odot$).
Figure~\ref{fig:passive_spirals_mdms} shows that the
passive spirals in our sample have much older stellar populations than
the normal spiral population. This is consistent with \citet{Masters10}, 
who found red, face-on spirals have older stellar populations than blue spirals, and are
not post-starburst objects. This suggests our spirals
have not stopped forming stars recently, and may have low SSFR because
they have used up most of their gas. This implies that, under some 
circumstances, spirals can retain their spiral appearance for a few Gyr 
following the cessation of their star-formation \citep*[e.g.][]{Bekki02}. 
This interpretation is supported by the time of last burst, for which 
we find a mean of $1.8^{+0.5}_{-0.4}\,$Gyr for our passive spirals.

\subsection{Inclination effects}
\label{sec:inclination}
Our sample of passive spirals is separated from the other morphologically-classified 
spirals on the basis of our SED fitting results, which uses an `angle-averaged' approach.
Results may be biased for sources with high inclinations \citep{dC10}, so we
calculate the inclination of our passive spirals to check that there is not a high
fraction of edge-on galaxies in our sample. The minor to major observed axis ratio
$b/a$ of the SDSS $g$-band isophote at 25 mag $\rm{arcsec}^{-2}$ can
be used to determine inclination. A ratio of $b/a$ of $\sim$1
indicates that a galaxy is face-on, $b/a$ decreases as the galaxy
inclination becomes edge-on. The observed axial ratio $b/a$ can be
converted into an inclination using the relation \citep[as used in][]{Masters10a}

\begin{equation}
{\rm{cos}}^2 i = \frac{(b/a)^2-q^2}{1-q^2}
\label{inclination}
\end{equation}

\noindent
where $q$ is the intrinsic axial ratio that would be measured for an
edge-on galaxy ($i=90^{\circ}$). An estimate of $q$ can be obtained from the observed 
distribution of axial ratios for SDSS galaxies with different values of 
the parameter $f_{Dev}$. This SDSS parameter describes 
the fraction of the galaxy light which is fit by a de
Vaucouleurs profile (the other fraction of the luminosity is fit by an
exponential profile), and gives information about the bulge-to-disk
ratio. We adopt the relation found in \citet{Masters10a} 
$q = 0.12+0.10\times f_{Dev}$, and use the $g$-band defined $f_{Dev}$. 
The inclinations are listed in Table \ref{inclinations}. Assuming that
galaxies appear approximately edge-on for $i>75^{\circ}$, then a
random sample of inclinations would lead to 17 percent of galaxies appearing
edge-on. We find that 5/19 of our passive spirals have an edge-on
inclination, and so within $2\sigma$ binomial errors our sample is consistent with a random
distribution of inclinations. \citet{dC10} show that the SSFR derived
from SED fitting may be biased low for high inclinations ($b/a<0.4$, 
corresponding to $i>67^{\circ}$). However, the SSFR of these passive 
spirals are sufficiently low that after accounting for this small bias 
we would still regard the majority of these galaxies as being passive.

\begin{table}
\begin{center}
\caption{Inclinations ($i$) in degrees of the 19 passive spirals in our sample. The \emph{Herschel} SDP ID is given in column 1, $b/a$ is the minor to major observed axis ratio of the SDSS $g$-band isophote at 25 mag $\rm{arcsec}^{-2}$, $f_{Dev}$ is an SDSS parameter which is the fraction of the galaxy fit by a de Vaucouleurs profile, $q$ is the intrinsic axial ratio that would be measured for $i=90^{\circ}$.}
\begin{tabular}{|l|l|l|l|l|}
\hline
  \multicolumn{1}{|c|}{SDP ID} &
  \multicolumn{1}{c|}{$b/a$} &
  \multicolumn{1}{c|}{$f_{Dev}$} &
  \multicolumn{1}{c|}{$q$} &
  \multicolumn{1}{c|}{$i$} \\
\hline
  SDP.30   & 0.87 & 0.7 & 0.19 & 29.9\\
  SDP.77   & 0.95 & 0.82 & 0.2 & 17.9\\
  SDP.143  & 0.33 & 1.0 & 0.22 & 75.1\\
  SDP.271  & 0.24 & 0.27 & 0.15 & 79.2\\
  SDP.372  & 0.32 & 0.98 & 0.22 & 76.1\\
  SDP.1544 & 0.78 & 1.0 & 0.22 & 40.0\\
  SDP.1773 & 0.32 & 0.94 & 0.21 & 75.9\\
  SDP.1888 & 0.51 & 0.4 & 0.16 & 60.3\\
  SDP.2547 & 0.43 & 0.01 & 0.12 & 65.6\\
  SDP.2612 & 0.31 & 0.0 & 0.12 & 73.1\\
  SDP.3578 & 0.28 & 0.56 & 0.18 & 77.4\\
  SDP.3935 & 0.58 & 0.98 & 0.22 & 56.4\\
  SDP.4548 & 0.36 & 0.51 & 0.17 & 71.4\\
  SDP.4639 & 0.73 & 0.02 & 0.12 & 43.9\\
  SDP.4859 & 0.43 & 0.59 & 0.18 & 66.8\\
  SDP.4964 & 0.64 & 0.35 & 0.16 & 51.3\\
  SDP.5108 & 0.49 & 0.47 & 0.17 & 61.8\\
  SDP.5226 & 0.62 & 0.95 & 0.22 & 53.4\\
  SDP.7324 & 0.38 & 1.0 & 0.22 & 71.3\\
\hline\end{tabular}
\label{inclinations}
\end{center}
\end{table}

We conclude that most of the `passive' spirals are red because they 
harbour old stellar populations, not because of increased
amounts of dust which obscures star-formation. This agrees with the
findings of \citet{Masters10}, who find that red spirals have similar
dust content (measured from Balmer decrements) to blue spirals at the
same stellar mass.

\section{Star-formation and AGN fractions}
\label{AGN}
\subsection{Emission line diagnostics}
We use optical emission line ratios plotted on a BPT diagram \citep*{BPT81} to
characterise the AGN activity in our H-ATLAS ETGs and spirals. Line ratios and equivalent widths (EWs) are derived from
the SDSS MPA-JHU catalogue\footnote{http://www.mpa-garching.mpg.de/SDSS/DR7/}
\citep{Tremonti04} and the GAMA survey \citep{GAMA_Driver11}.
We regard a line detection as $>3\sigma$ above the
continuum, but lines affected by sky emission or fibre fringing are not
used. For line fluxes derived from the SDSS sample, corrections are
made for stellar continuum absorption by subtracting a stellar population
model from the spectrum, and measuring emission lines from the residual \citep{Tremonti04}. 
Where line fluxes are derived from GAMA
measurements, a correction of 1.3\AA\ for stellar absorption is
applied to the EW of the H$\alpha$ and H$\beta$ emission lines \citep{Hopkins03, Gunawardhana11, Wijesinghe11}.
\citet{Gunawardhana11} found for H$\alpha$ lines with 
log(H$\alpha$ EW)$<0.9$ there was a difference of more than 5\% in 
EW when a range of absorption corrections from $0.7-1.3$\AA\ was applied. 
Some of our sources are below log(H$\alpha$ EW)$<0.9$, but our results 
are unchanged if this range of absorption corrections are used.
In the cases where there are multiple measurements of the same galaxy, 
we take the signal--to--noise weighted mean of the line fluxes. 

\begin{figure*}
\centering
\includegraphics[width=18.0cm]{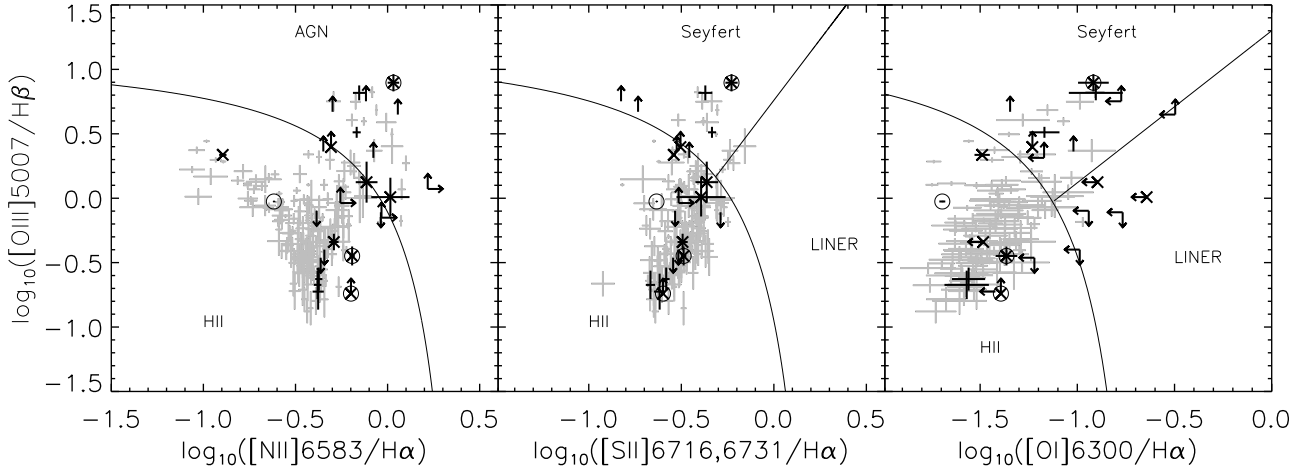}
\caption{BPT diagram showing H-ATLAS spirals (grey) and ETGs (black) with all four lines in each diagram detected at $>3\sigma$, with error bars shown. 
Those below the curved line are classified as star-forming, those above the curved line are classified as AGN. 
Upper limits are shown for the ETGs where at least two emission lines are detected. 
ETGs which have radio emission (Section \ref{sec:radio}) are marked with a circle, and disturbed morphologies with a diagonal cross.}
\label{fig:BPT}
\end{figure*}

We plot the [OIII]/H$\beta$ line ratio as a function of
the [NII]/H$\alpha$, [SII]/H$\alpha$ and [OI]/H$\alpha$ line ratios in
Figure~\ref{fig:BPT} for spirals and ETGs. We classify objects as AGN or
star-forming first from the [OI]/H$\alpha$ diagram, since [OI] is the
most sensitive to the presence of an AGN. If the galaxy is not
present in the [OI] diagram, we use the [SII]/H$\alpha$ diagram, and finally the [NII]/H$\alpha$ diagram.  
On all diagrams, galaxies that lie above the curved
line are classified as AGN \citep{Kewley01}, and galaxies below the line are star-forming. Low signal-to-noise [OIII] 
and H$\beta$ lines mean that some sources cannot be located on the BPT diagram. 
In these cases, a source is classified as an AGN if [NII]/H$\alpha\geq0.2$.
In many of our ETGs we do not detect all four required emission lines, so 
we derive upper limits and locate the galaxy on the BPT diagram where at least two lines are
present\footnote{In the case where there is H$\beta$ absorption, the H$\beta$ 
flux is not measured in the GAMA spectra, so we derive 3$\sigma$ upper limits. 
Assuming a flat continuum, we estimate the area under a Gaussian line in pixels (Npix) 
with FWHM equal to the instrumental resolution of 3.5\AA\/, and estimate the error 
on this line given the mean noise in the spectrum ($\sigma$) as $\sqrt{\rm{Npix}} \times \sigma$.}.
We present the classification fractions of ETGs in Table \ref{table:AGN}; 
more than half of ETGs are star-forming, but 45 percent of ETGs cannot be classified due to
their weak emission lines. For comparison, spiral galaxies are plotted on 
Figure~\ref{fig:BPT} in grey, and lie mostly in the star-forming region.

\begin{table}
\begin{center}
 \caption{Emission line classifications of H-ATLAS ETGs which can be unambiguously classified on the BPT diagram, with Poisson errors. These fractions do not include galaxies which we cannot classify into either category, which comprises 45 percent of the sample. Ambiguous classifications result from one or more weak emission lines not detected at $>3\sigma$, or measurements affected by skylines. The errors are $1\sigma$ confidence intervals on a binomial population using a beta distribution, which is appropriate for small population numbers \citep{Cameron11}.}
 \begin{tabular}{ l | c | c }
 \hline
 Classification & Number & ETG Percentage  \\
 \hline
   \vspace{0.1cm} 
   ETGs & 23 & 100\% \\ 
   \vspace{0.1cm} 
   Star-forming & $13\pm2$ & $57^{+9}_{-10}\%$ \\
   AGN & $10\pm2$ & $43^{+10}_{-9}\%$ \\
 \hline
 \label{table:AGN}
 \end{tabular}
 \end{center}
 \end{table}

In a sample of optically selected ETGs, \citet{Schawinski07a} 
found 61\% are star-forming, and 39\% are AGN dominated, which is similar to the fractions
in our H-ATLAS sample. Since our AGN fraction is consistent with that from an optically selected sample, 
this would suggest there is no link between the presence of AGN and dust emission,
although it is interesting to note that we detect few LINERs\footnote{Low-ionization nuclear emission-line region.} 
in our sample (although some galaxies with upper limits may fall into this category). 
The lack of LINERs in our sample may be because they are dust poor
\citep{Kauffmann03_AGN, Kewley06}, and therefore we may potentially be biased against
detecting LINERS in H-ATLAS, although we need a larger sample of galaxies to confirm this. 

We do not account for AGN emission in the SED fitting, so we may expect the galaxies with AGN to be 
poorly fit by the models. For the ETGs which host AGN, their SEDs look
similar in the optical to those which are classified as
star-forming. \citet{Kauffmann03_AGN} find that the optical spectra of
type-2 AGN have a small fraction of their optical light from
non-stellar sources, and are very similar to spectra of non-AGN host
galaxies, except for emission lines. Since our physical properties are
determined from broadband fitting and not from line strengths,
properties from optical data should not be affected by the presence of
a type-2 AGN. Since \citet{Hatziminaoglou10} find no difference between the FIR/submillimetre 
colours of star-forming and AGN galaxies; the FIR is insensitive to presence of 
AGN and therefore will not produce a bias in SED parameters.

\subsection{H$\alpha$ equivalent widths}
\begin{figure}
\centering
\includegraphics[scale=0.65]{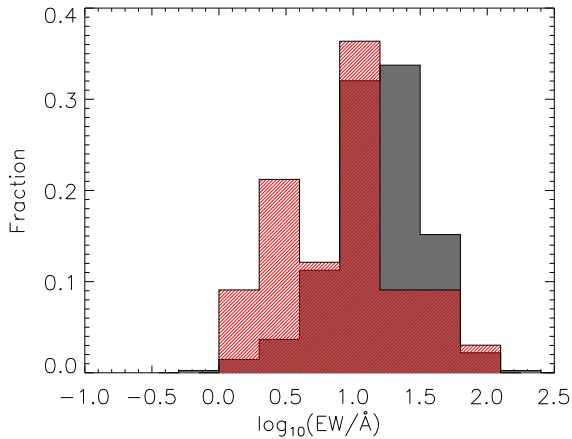} 
\caption{H$\alpha$ EW (corrected for stellar absorption) of spiral galaxies (grey) and ETGs (red/hatched) which have $\geq3\sigma$ H$\alpha$ detections. 
A K-S test shows that the ETGs have a probability of $1.7 \times 10^{-4}$ of being drawn from the same distribution as the spirals.}
\label{fig:Halpha_EW}
\end{figure}

We present the H$\alpha$ EW distribution of our ETGs in comparison to spirals in
Figure~\ref{fig:Halpha_EW}. For the ETGs there is a range in EW from
0-109\AA\, with a median of 8.7\AA\/, which is lower than the median
for the spirals in our sample (16.4\AA\/). The median value for
the spirals is similar to that found for field galaxies by
\citet{Tresse99}. It is not unsurprising that the EW of ETGs is less than
that of spirals, but nonetheless some EWs are substantial and indicate
ongoing star-formation (consistent with the broad-band SED
fitting). The range of H$\alpha$ EWs in ETGs are comparable to those found by \citet{Schawinski09},
who found EWs up to 85\AA\ in their blue ETG
sample. \citet{Fukugita04} found that visually classified ETGs (with
$r<15.9$ and $z\lesssim 0.12$) have a similar H$\alpha$ EW range as
our sample, with 19 out of 420 E/S0s with H$\alpha$ EW $>10$\AA\
(which represents star-forming galaxies). 
In our sample, we find that a much larger fraction (31
percent) of our ETGs have H$\alpha$ EW $>10$\AA, which is 
unsurprising given our submillimetre selection.

\subsection{Radio detections}
\label{sec:radio}
Another indicator of star-formation and AGN activity is radio
emission. \citet{Smith11a} computed the statistical probability of a
chance alignment between radio and H-ATLAS sources using the
frequentist technique of \citet{Downes86}, which used 
a method to determine the most likely radio counterpart by choosing the source with
the lowest probability $P$ of being a chance
alignment. We cross-match our ETGs with the FIRST radio
catalogue and find that 5/42 ETGs have radio
counterparts with $P<0.2$, and so are considered to be likely associations. 
The radio emission may indicate the presence of an AGN and/or star-formation, so we compute the ratio 
of the bolometric IR flux to the 1.4\,GHz radio flux ($q_{\rm{IR}}$) 
using the method of \citet*{Helou85, Bell03}. $q_{\rm{IR}}$ is
defined as

\begin{equation}
q_{\rm IR} = \log_{10} \left( \frac{\rm TIR}{3.75 \times 10^{12} 
	{\rm W\,m^{-2}}} \right) - \log_{10} \left( 
	\frac{S_{\rm 1.4\,GHz}}{\rm W\,m^{-2}\,Hz^{-1}} \right) ,
\end{equation}

\noindent
where $S_{\rm 1.4\,GHz}$ is the rest-frame 1.4\,GHz $k$-corrected flux density and ${\rm TIR}$ is the total infrared luminosity (\ldust), 
which is integrated between $3-1000\mu$m.

The $q_{\rm{IR}}$ values for the ETGs are presented in Table \ref{qIR}.
Three ETGs have $q_{IR}$ values consistent with that found for 162 star-forming galaxies in \citet{Bell03}, with a median $q_{IR}=2.64\pm0.02$. 
We find two ETGs have $q_{\rm{IR}}$ values which are significantly
lower than that for star-formation, which suggests the presence of a radio-loud AGN in these galaxies. To
rule out synchrotron contamination of the 500$\mu$m flux, we
extrapolate the 1.4\,GHz radio flux to 500$\mu$m using a power law with a spectral
slope, $\alpha$. Assuming $\alpha=-0.8$ we find that the synchrotron
emission at this wavelength is negligible compared to the dust
emission measured at 500$\mu$m.

\begin{table}
\begin{center}
\caption{$q_{IR}$ values for the 5/42 \emph{Herschel} ETGs with reliable radio counterparts and SED fits. Errors are propagated from the 1$\sigma$ error on TIR and the local noise estimate at the source position measured in mJy. $\rm{F_{int}}$ is the integrated flux density at 1.4\,GHz in mJy. $P$ is the probability of a chance alignment of the 
submillimetre and the radio source as computed in \citet{Smith11a}.}
\begin{tabular}{|l|l|c|c|c|}
\hline
  \multicolumn{1}{|c|}{Name} &
  \multicolumn{1}{c|}{SDP ID} &
  \multicolumn{1}{c|}{$P$} &
  \multicolumn{1}{c|}{$\rm{F_{int}}$} &
  \multicolumn{1}{c|}{$q_{IR}$} \\
\hline
  J091205.8+002656 & SDP.15 & 0.014 & 4.25 & $2.53\pm0.15$ \\
  J090352.0-005353 & SDP.45 & 0.048 & 0.98 & $2.59\pm0.17$ \\
  J090718.9-005210 & SDP.350 & 0.148 & 1.18 & $2.58\pm0.18$ \\
  J090752.3+012945 & SDP.1027 & 0.083 & 2.14 & $1.68\pm0.41$ \\
  J085947.9-002143 & SDP.6427 & 0.077 & 8.62 & $0.86\pm0.56$ \\
\hline\end{tabular}
\label{qIR}
\end{center}
\end{table}

It is interesting to note the classifications of ETGs using emission
line ratios are consistent with those from radio emission. The three
ETGs with radio emission from recent star-formation also have
some of the bluest optical colours and largest H$\alpha$ EWs, consistent
with recent star-formation. For the ETGs which are classified as AGN using
radio emission, one (SDP.6427) is classified as an AGN using emission
lines. The other (SDP.1027) is likely to be an AGN from its line emission, although it has
insufficient signal--to--noise to confirm this.

\subsection{Passive spirals}
In most cases the spectra of the passive spirals show little or no H$\alpha$ emission and 
a strong 4000\AA\ break (see Figure~\ref{fig:Red_spirals}), indicating low SFR 
and an old stellar population. Strong sodium and
magnesium absorption is often observed in the spectra, which can
indicate the presence of an old stellar population, or high metallicity. 
Only 4/19 passive spirals have sufficiently strong emission lines such that they can be located 
on a BPT diagram, and all of these are classified as AGN. This may be because AGN are more 
common in massive galaxies \citep{Kauffmann03_AGN}, although \citet{Masters10} found that red, 
face-on spirals have a higher AGN fraction than blue, face-on spirals. 
The lack of emission lines in the majority of the sample
is consistent with their being selected as passive in terms of star-formation, and also
indicates a lack of AGN activity. This agrees with radio data,
as there are no matches for these sources in the FIRST radio catalogue. 

\section{Environment of Herschel detected sources}
\label{sec:Environment}
We examine the environment of ETGs and spirals by computing the local density around each one (Brough et al. in prep). 
To define the local density, we use a volume limited sample of galaxies with $\rm{M}_r<-20$ and $z<0.18$. 
The density $\Sigma_{N}$ in $\rm{Mpc}^{-2}$ is computed as
 
 \begin{equation}
 \Sigma_{N} = \frac{N}{\pi{d_N^2}}
 \label{density}
 \end{equation}

\noindent
where $d_N$ is the projected comoving distance to the $N$th nearest
neighbour within $\pm1000\rm{kms}^{-1}$, and $N=5$.
Densities are computed for all H-ATLAS galaxies which have
$r_{petro}\leq19.4$, and have good quality spectroscopic redshifts
with $0.01<z<0.18$, which is the limit defined by the absolute
magnitude limit of the sample.

\subsection{H-ATLAS ETGs}
\begin{figure}
  \centering \includegraphics[scale=0.7]{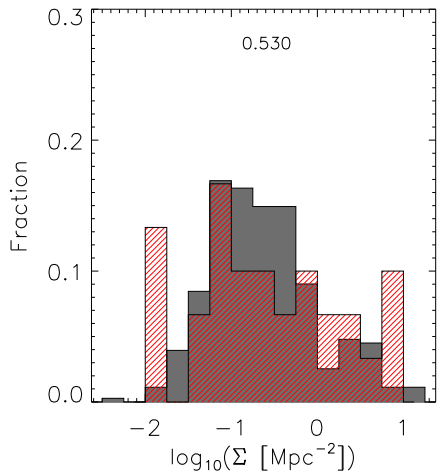}
\caption{Comparison of densities for H-ATLAS spirals (grey) and ETGs (red/hatched). 
A K-S test (shown at top of histogram) shows that we can't statistically rule out 
the null hypothesis of the samples being drawn from the same distribution.}
\label{fig:Environment}
\end{figure}

Using these criteria we are able to measure densities for 30 ETGs and
354 spirals detected in H-ATLAS, which are compared in Figure~\ref{fig:Environment}. 
The densities for ETGs and spirals both range from void to 
group environments \citep{Baldry06}, with most galaxies residing
in field environments. There are few H-ATLAS galaxies in group/cluster
environments, so our galaxies do not sample the full range 
of densities in the SDP field, which range from ($\sim0.01-100$) galaxies $\rm{Mpc}^{-2}$. 
From the morphology-density relation
\citep{Dressler80}, spirals are more numerous in low density
environments, and ETGs generally reside in high density environments,
however, a K-S test reveals there is no significant difference between
the densities of spirals and ETGs detected in H-ATLAS.
This is consistent with the findings of \citet{Dariush11}, who
found that the detection rate of H-ATLAS galaxies split into blue and
red colours does not depend on environment. \citet{Young11} observed 
a volume-limited sample of ETGs and found a statistically weak dependence of 
molecular CO (which is often associated with star-formation) on local 
galaxy density, where CO detections were only marginally lower in the cluster 
environment compared to the field. Conversely, blue ETGs have been found in 
lower density environments than red ETGs \citep{Schawinski07b, Bamford09}, 
although these studies sampled both field and cluster environments. 
\citet{Kannappan09} found that intermediate mass ETGs are common in low density environments, 
and suggest that they may be undergoing disk re-growth.

It is possible that we are not sampling a large enough range of environments 
to see a significant difference in the densities of H-ATLAS sources as a 
function of morphology. The full H-ATLAS data set will encompass the Coma cluster and many
other rich Abell clusters and will allow a more in depth investigation
of environmental effects. Since some of the H-ATLAS galaxies are in low density regions, 
it is possible that our measure of environment does not always reflect 
the true local density, and instead traces inter-halo distances.

\subsection{Passive spirals}
\begin{figure}
\centering
\includegraphics[scale=0.7]{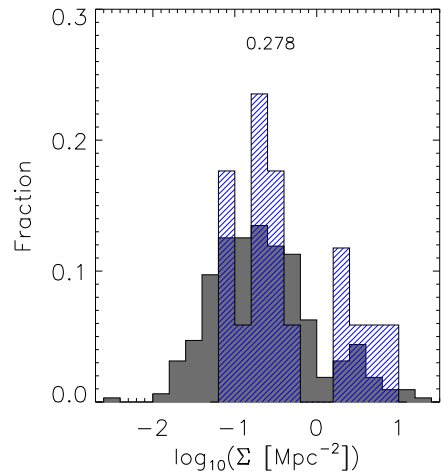} 
\caption{Environment of passive spirals with SSFR\,$<10^{-11}\rm{yr}^{-1}$ detected in H-ATLAS (blue), compared to normal spirals with SSFR\,$>10^{-11}\rm{yr}^{-1}$ (grey). 
A K-S test shows a high probability (shown at top of histogram) that both samples are drawn from the same distribution.}
\label{fig:Environment_red_spirals}
\end{figure}

It is thought that passive spirals have had their star-formation
quenched as a result of galaxy interactions with the intra-cluster medium. 
This can remove gas from the outer halo, which stops the supply of fuel for star-formation in the disk
\citep[e.g.][]{Bekki02, Wolf09}. We use the local density estimates
to test for any environmental differences between our passive and
normal spirals, where densities can be measured for 17/19 passive spirals. 
Figure~\ref{fig:Environment_red_spirals} shows that
passive spirals in our sample mostly inhabit low density environments
with a median density of 0.28 galaxies $\rm{Mpc}^{-2}$, which is slightly higher 
than the median density of normal spirals (0.19 galaxies $\rm{Mpc}^{-2}$). 
A K-S test shows that the distributions of densities of passive and normal spirals
are not significantly different, although this may be due to our small sample size.
Our median density is different from \citet{Masters10} who found the red, face-on spiral fraction peaks at
$1\,\rm{Mpc^{-2}}$, and \citet{Bamford09} who found that the density of red
spirals peaks at $6\,\rm{Mpc^{-2}}$. While 19\% of their red spirals are
found at densities $<1\,\rm{Mpc^{-2}}$, we find that 71\% of
our passive spirals lie at densities lower than this. The differences
in these fractions may be due to selection effects, since the Bamford
and Masters samples are selected to be `red', rather than `passive', 
and also because our H-ATLAS spirals are in low density environments. 
Our sample shows that it is possible to have passive spirals at low
densities. We can conclude that environment is not the only factor 
influencing whether galaxies are passive, and the processes which 
turn spirals passive occur at both high and low densities.

\section{Properties of non-detected ETGs}
\label{sec:control}
We have identified a population of ETGs with substantial dust masses,
some of which are actively star-forming. In order to understand how
this population is different from optically selected ETGs, we compare to a control
sample of morphologically classified galaxies in the SDP field with have the 
same $n(r,z)$ as the H-ATLAS sample, and which are not
detected in H-ATLAS. The selection method for this sample is described 
in Section \ref{sec:control_sample}.

\subsection{Dust masses}
\label{sec:stacking}
Our control sample is comprised of galaxies which are not detected 
in the submillimetre, however, we can investigate the
average dust mass of optically selected ETGs with stacking techniques.
The stacking is performed on background-subtracted, unfiltered
SPIRE maps. All detected SPIRE
sources are subtracted from the map, so we do not contaminate the stack.
We stack at the positions of the ETGs in the control sample,
using the same method as \citet{Bourne11b}. We assume all
galaxies are unresolved point sources, and for each source we convolve a cut-out
of the map with a point spread function (PSF) centred on the optical position,
which is interpolated to the same pixel grid as the data map. We
share out the flux of blended sources as described
in the appendix of \citet{Bourne11b},
but find the effect to be negligible in this sparsely distributed sample, so
double-counting of flux does not affect the stacked values.
This method is effectively similar to stacking in a PSF-filtered
map. The background level is estimated by stacking at random positions
and this is subtracted from the stacked flux. We use the
median value in the stack in order to avoid bias from outliers. 
Following the same method as \citet{Bourne11b}, we
estimate the $1\sigma$ error on the median from the distribution
of values in the stack as described by \citet{Gott01}.
This error estimate automatically takes into account both
the measurement error, which reduces as the square root
of the number of objects stacked, and the intrinsic spread of fluxes
within the stack. By stacking on the positions of 233 ETGs
in the control sample, we find median fluxes of $2.9\pm0.5$mJy
at $250\mu$m ($5.8\sigma$), $0.8\pm0.6$mJy at $350\mu$m and $-0.6\pm0.6$mJy at $500\mu$m.
The $250\mu$m flux is consistent with the typical fluxes of the 
optically red galaxies in \citet{Bourne11b}.

To obtain the median stacked dust mass, we calculate the dust mass of
each object in the sample from its measured flux in Jy and its
redshift, using equation \ref{dust_mass}.
Again the error is calculated from the distribution of dust mass values in the stack
using the \citet{Gott01} method.

\begin{equation}
\mathrm{{M}_{d}} = \frac{S_{250}\,D_L^2 K}{\kappa_d(\nu)\,B(\nu,\, T_{\rm{d}})(1+z)}
\label{dust_mass}
\end{equation}

\noindent
$S_{250}$ is the observed 250$\mu$m flux, $D_L$ is the luminosity
distance at redshift $z$, $B(\nu,\, T_{\rm{d}})$ is the value of the
Planck function at 250$\mu$m for a dust temperature $T_{\rm{d}}$, the
dust mass opacity coefficient $\kappa_d(\nu)$ is
0.89$\mathrm{m}^{2}\mathrm{kg}^{-1}$ (following \citealt{Dunne11}). 
$K$ is the $k$-correction, which is given by
 
 \begin{equation}
 K = \left(\frac{\nu_{\rm{o}}}{\nu_{\rm{e}}}\right)^{3+\beta} \frac{e^{h\nu_{\rm{e}}/kT_{\rm{iso}}}-1}{e^{h\nu_{\rm{o}}/kT_{\rm iso}}-1}
 \label{dust_Kcorr}
 \end{equation}
 
\noindent
where $\nu_{\rm{o}}$ is the observed frequency at 250$\mu$m,
$\nu_{\rm{e}}$ is the emitted frequency and $\rm{T_{iso}}$ is
the isothermal temperature of a greybody model normalised 
to recover the stacked flux at 250$\mu$m.
We assume a dust emissivity index $\beta=2.0$ and $\rm{T_{iso}}=18.5$ K, 
which adequately describes the SEDs of optically selected galaxies 
\citep{Bourne11b}.

Assuming a realistic range of temperatures of 25-15\,K\footnote{Higher dust temperatures have been found in some studies of ETGs \citep[e.g.][]{Savoy09, Skibba11}, but these used $\beta=1.5$ which results in a higher dust temperature ($\sim3-4$\,K) being calculated \citep{Bendo03}. Accounting for this difference in $\beta$, these studies yield dust temperatures which are consistent with our range of adopted values.} \citep[][Smith, M. et al. in prep.]{Temi04, Leeuw04}, we find median
dust masses ranging from $(0.8-4.0)\times10^{6}\,{\rm{M}}_\odot$. 
This is more than an order of magnitude less than
the dust masses of the H-ATLAS ETGs, indicating that the $250\mu$m selected 
ETGs are indeed much dustier than the average optically selected ETG. 

\subsection{Star formation histories and optical colours}
\begin{figure}
\centering
\includegraphics[scale=0.48]{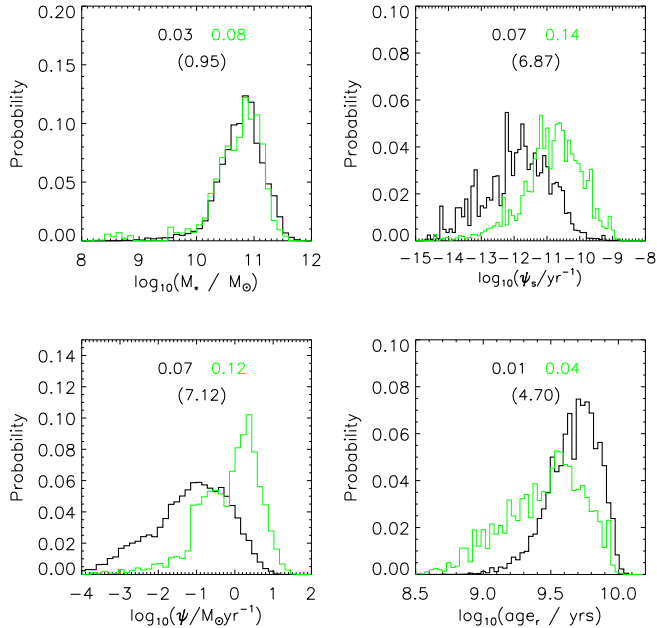} 
\caption{Average PDFs of the SED parameters of 42 detected ETGs 
  (green line) compared to 222 control ETGs (black line).  
  The parameters are (from left to
  right): $\rm{M}_\ast$, stellar mass; \ssfr/$yr^{-1}$, SSFR; \sfr/$\rm{M}_\odot yr^{-1}$, SFR;
  and $\rm{age_r}$, the $r$-band light-weighted age of the stellar population. 
  The uncertainty on each distribution for ETGs and spirals is given by 
  the error on the mean and is shown at the top of each histogram with corresponding colours, and 
  the significance of the difference in the means in brackets. The errors for logarithmic parameters are in dex.}
\label{fig:detected_control_ETGs}
\end{figure}

We use the same technique as described in Section~\ref{sec:SED_fitting} to
fit the multiwavelength SEDs of the control sample galaxies, using
$5\sigma$ upper limits for the FIR-submillimetre fluxes. We reject 27 ETGs and spirals 
which have poor quality SED fits with $\chi^2>30$. Although
the parameters derived from the FIR-submillimetre region of the SED are 
only constrained by the UV--NIR data, we can put similar
constraints on SFH parameters as for the $250\mu$m selected
sample, as most of the constraint for SFH parameters comes from the UV--NIR photometry.
A summary of the parameters derived from the mean PDFs is provided in Table \ref{tab:summary_properties}.
Stellar mass is one of the main drivers of galaxy properties, so it 
is important to check that the $\rm{M}_\ast$ distributions are the same for the 
H-ATLAS detected and control ETGs, so that we can compare physical 
properties without a dependence on galaxy mass.
Figure~\ref{fig:detected_control_ETGs} shows the stacked PDFs of the stellar mass distributions for H-ATLAS and control ETGs 
are not significantly different, since the control sample is selected to have the same $r,z$ distribution. 
The range of \mdms{} for the control ETGs is $(1.4-6.8)\times{10}^{-5}$ for 25-15\,K dust, and
on average, the mean SSFR of the control ETGs is 1.1 dex lower than that of H-ATLAS ETGs. 
A similar trend is found when comparing the mean SFR of the ETGs.
For our control ETGs the mean $r$-band light-weighted age
of the stellar population is $4.6\pm0.1$\,Gyr, which is 1.8\,Gyr older
than the H-ATLAS sample of ETGs.

\begin{figure}
\centering
\includegraphics[scale=0.85]{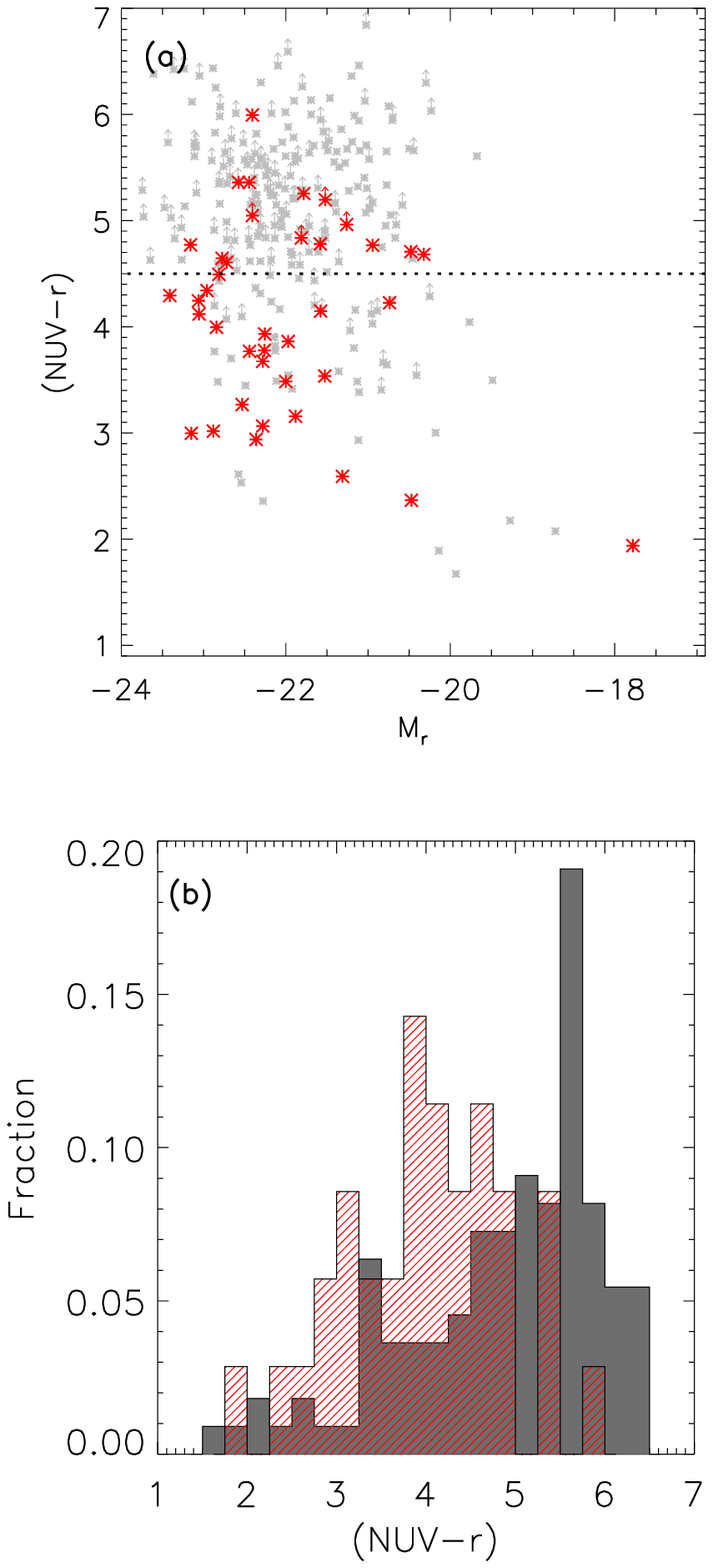} 
\caption{{\bf (a)}: UV-optical colour magnitude diagram for H-ATLAS detected (red stars) and control ETGs (grey crosses).
Lower limits are shown for galaxies which do not have a $\geq5\sigma$ NUV detection.
{\bf (b)}: Comparison of the $NUV-r$ colours for the detected ETGs (red/hatched) and control ETGs (grey/filled) } which have a $\geq5\sigma$ NUV detection.
\label{fig:non-d_ETGs_colours}
\end{figure}

The $NUV-r$ colours of the control ETGs are computed as in Section \ref{Colours}, and are compared to the H-ATLAS ETGs on a 
colour--magnitude diagram in Figure~\ref{fig:non-d_ETGs_colours} (a). 
These cover approximately the same range in ${\rm{M}}_r$ by design.
The distribution of colours are shown in Figure~\ref{fig:non-d_ETGs_colours} (b);
the control ETGs are on average 1.0 magnitude redder than
the H-ATLAS detected ETGs. Since the control ETGs are not detected in H-ATLAS 
these galaxies are less obscured by dust, with colours dominated 
by stellar population age rather than obscuration. The colour 
difference between detected and control ETGs is therefore intrinsic. 
A handful of control ETGs have very blue
$NUV-r$ colours, but the dust masses of these star-forming galaxies may not be high 
enough to be detected by H-ATLAS.
Alternatively, there could have been a failure in matching the optical 
counterpart and submillimetre source, which is a possibility for 7 of the control ETGs,
(of which 3 are `blue'). These, however, have a very small reliability of association 
as determined in \citet{Smith11a}.

\subsection{Environments of Herschel non-detected sources}
\begin{figure}
\centering
\includegraphics[scale=0.65]{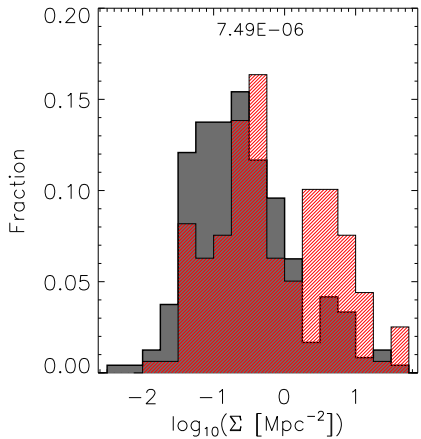} 
\caption{Comparison of densities for control sample ETGs (red/hatched) and spirals (grey). As expected, the spirals have a lower median density than the ETGs. 
A K-S test shows a low probability of the samples of ETGs and spirals being drawn from the same distribution.}
\label{fig:env_compare}
\end{figure}

We compare the environments of control sample ETGs and spirals in Figure
\ref{fig:env_compare}, with densities as calculated in
Section \ref{sec:Environment}. As expected, on average the median density of
control ETGs is higher than that of the spirals, and in contrast to the H-ATLAS ETGs and spirals, a K-S test 
shows a low probability of the control ETGs and spirals being 
drawn from the same distribution. 

\begin{figure}
\centering
\includegraphics[width=8.3cm]{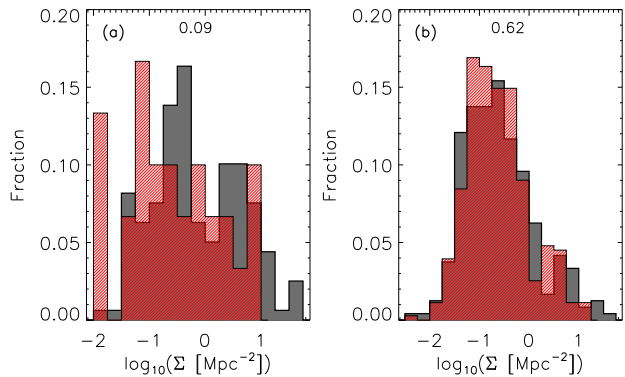} 
\caption{Comparison of environments between H-ATLAS detected (red/hatched) and control ETGs (a) and H-ATLAS and control spirals (b). 
For the detected and control ETGs a K-S test shows the two distributions have a low probability of being drawn from the same distribution, although this is only significant at the $1.8\sigma$ level. We cannot statistically rule out the null hypothesis that the control and H-ATLAS spirals are drawn from the same distribution.}
\label{fig:env_compare2}
\end{figure}

To see how the environments of the detected ETGs are different from those in the control sample, 
we compare the densities in Figure~\ref{fig:env_compare2} (a), and find they are different at only the $1.8\sigma$ level.
There is some indication that H-ATLAS ETGs are in lower density environments 
than optically selected ETGs, but a larger sample size is needed to confirm this.
A comparison of the detected and control spirals using a K-S test in Figure~\ref{fig:env_compare2} (b) 
shows that we cannot statistically rule out the null hypothesis that they are drawn from 
the same distribution. The similarity of the
distributions suggests that environment does not explain the
differences between the H-ATLAS detected and control sample ETGs, 
however, small sample statistics combined with a small range of 
environments currently limits the strength of our findings.

\section{The Origin of Dust Observed in ETGS}
\label{sec:chem_ev}
Interstellar dust in early-type galaxies can originate from either an
internal source (via mass loss from stars \citep[e.g.][]{GoudfrooijdeJong95} 
or an external source where gas and dust are accreted by minor mergers
of galaxies or galaxy-galaxy interactions \citep[e.g.][]{Temi07, Kaviraj09,Shapiro10}. 
Here we use a simple approach to investigate whether stellar mass 
loss could be responsible for the dust masses of the H-ATLAS ETGs.

Internal sources of interstellar dust in galaxies
are poorly understood, particularly the relative contributions from
supernovae (SNe) and the cool, stellar winds of low-intermediate
mass stars (LIMS) in their AGB phase. In spiral galaxies like the Milky Way
(MW), the major dust source is presumed to be LIMS \citep[e.g.][]{Whittet03}, 
which inject dust at a rate of $\sim 2 \times 10^{-3}\,\rm
M_{\odot}\,yr^{-1}$. Dust is destroyed via sputtering in SN driven shocks, with
theoretical models predicting that dust is destroyed on timescales of
$\rm \tau_{des} \sim 300-800\rm \,Myr$ in the MW \citep{Jones94, Tielens98}. 
Comparing total dust masses with destruction timescales
for both local \citep{Jones01, Matsuura09, Dunne11}
and high redshift \citep*{ME03, MWH10, Gall11} galaxies leads to a `{\em dust budget 
crisis\/}': the rates of dust injection required to maintain the
observed dust mass are around an order of magnitude higher than the
injection rates from LIMS.

Significant dust production in SNe would alleviate this dust budgetary
problem in both local and high-$z$ galaxies \citep{ME03}, yet there is 
still some controversy about whether SNe can produce the $\sim 1\, \rm{M_{\odot}}$
that would be required to alleviate the budget problem. Current estimates
suggest $\sim 10^{-3}-1\,\rm M_{\odot}$ of dust is formed per core-collapse SN
\citep[][Krause et al. in prep]{Rho08, Dunne09, Otsuka10, Barlow10, Matsuura11};
additionally the mass of dust created (if any) in Type Ia SNe is highly uncertain.

The ISM of ETGs differs from that of spirals due to the presence of a
hot X-ray emitting component at $10^6-10^7$ K. This presents a harsh
environment for dust, and indeed the destruction of dust in ETGs is
thought to be dominated by thermal sputtering due to immersion in this
hot gas \citep{Jones94}, in addition to a contribution
from Type 1a supernovae shock collisions. For thermal sputtering in the hot ISM, the
typical dust destruction timescale is $<50\,\rm Myr$. The destruction timescale is
increased in the warm phase of the ISM ($\sim 10^4\,\rm K$) with $\rm
\tau_{des} \sim 1\,\rm Gyr$ \citep{Barlow78}. Assuming the supernova 
rate in ETGs is dominated by Type Ia SNe, dust destruction due to Type Ia
shocks depends on the SFR and stellar mass \citep[e.g.][]{ScannapiecoBildsten05}. Estimates for our ETGs
(assuming the mean stellar mass of $4.9\times{10^{10}}$\,M$_\odot$) are
presented in Table~\ref{tab:dustsource}. The timescales for
destruction from SNe shocks are much longer than those from sputtering
in hot gas.

Applying all of these arguments to our sample, (see
Table~\ref{tab:dustsource} for details) the dust injection rate for an average 
H-ATLAS ETG would need to be 20--400 times higher than expected for
LIMS and SNe. Taking the full range of derived values for the SFR for the 
H-ATLAS ETG sample still produces the same discrepancy. This problem 
is however not confined only to our H-ATLAS ETGs, as
\citet{Dunne11} and Gomez et al. (in prep.) found difficulty in producing enough dust 
in chemical evolution models for all H-ATLAS galaxies at $z<0.5$.

The median dust mass for the control sample of ETGs is approximately
14-69 times lower than the detected sample ($\rm M_d =
(0.8-4.0)\times 10^6\,\rm M_{\odot}$), with a mean SFR of
$0.07\,\rm M_{\odot}\,yr^{-1}$ (Table~\ref{tab:dustsource}). If dust
is destroyed only by SNe shocks then the necessary injection rate is
comparable to that produced by stars, however if dust sputtering in
hot gas is important then we still observe more than 10 times the amount
of dust expected from stars. This suggests either sputtering is not as efficient 
as estimated, X-ray haloes are not as abundant or dust reforms quickly in the ISM.
It is not just the control ETGs which have these issues but also the H-ATLAS ETGs.

In summary, the dust in the ETGs studied here cannot have been replenished 
via stellar mass loss \citep*[see also][]{GoudfrooijdeJong95, Irwin01}. 
This result is not exclusive to ETGs since this shortfall is also seen in local 
and high redshift galaxies \citep{ME03} of all morphologies. Alternatively, the
dust destruction timescales in ETGs would need to be of order
$15\,\rm Gyrs$ (two orders of magnitude longer than currently
estimated). Although this is not likely, some massive ETGs are seen 
without a hot X-ray ISM \citep{Pellegrini99}, and X-ray luminosity is
found to correlate with the shape of the elliptical light profile:
those ETGs with power law profiles (as opposed to cores) are generally 
found to have lower X-ray luminosities \citep{Pellegrini05}. Post-merging 
systems also tend to be deficient in hot gas \citep*{ReadPonman98, OSullivan01, OSullivanPonman04, Brassington07}, 
and a lack of X-ray emission was also found in ETGs 
with stellar populations $<4$\,Gyr old \citep{Sansom06}. \citet{Brassington07} 
found that post-merger ETGs within 1\,Gyr of coalescence are very underluminous 
in X-rays, and find tentative evidence that over the period of a few Gyr 
galaxies regenerate their X-ray halo. It may be the case that the dusty ETGs 
are examples of systems with low hot gas content, possibly seen within $1-2$\,Gyr 
after a major star-formation episode which created the dust, however, 
X-ray observations are required to confirm this.

Longer destruction timescales may also be achieved if a significant
fraction of the star-dust is shielded from the hot gas in a cool,
dense phase of the ISM. Recent observations of ETGs have detected 
significant amounts of cold CO gas \citep{Young11}, with a
detection rate of 22 percent in a volume-limited sample of ETGs. It is possible 
that since our ETGs have significant dust mass they also have a large 
cold gas content, however observations of molecular gas are required 
to test this hypothesis.

Finally, multiple lines of evidence now favour dust growth in the cold
neutral phase ($n_H > 30 \,\rm{cm^{-3}}$) of the ISM on rapid
timescales of $\sim$ Myr \citep{DraineSalpeter79, DwekScalo80, Tielens98, Draine09}. 
Such growth can replenish dust mass lost through shocks and thermal 
sputtering (although seed nuclei must survive sputtering in order 
for this to occur). Given the shortfall of
dust from stars, one would then conclude that most
(i.e. $40-99$\% - Table~\ref{tab:dustsource}) of the dust in
the ISM of {\it Herschel} detected ETGs would need to be grown 
{\em in} the ISM. The dust yield of a galaxy would depend only on balancing
destruction timescales with the formation conditions in the ISM. 

The other scenario we can consider is that mergers with dust rich
galaxies could account for the dust content of the {\em Herschel\/}
ETGs. Such a case was demonstrated recently when \citet{Gomez10}
detected $\sim 10^6\,\rm M_{\odot}$ of dust associated with M86 which
originated from a recent tidal interaction with the nearby spiral
NGC~4438; this is of the same order as the dust mass in the stacked,
un-detected sample. However, to explain the discrepancy between the
observed total dust mass in the {\it Herschel}-detected ETGs and the
dust mass provided via stellar mass loss, we would require $>10^7\,\rm
M_{\odot}$ of dust to be accreted. This is akin to the average ETG in
this sample swallowing a large spiral galaxy with an equivalent dust
mass to the Milky Way in its recent history. As indicated by \citet{Maller06}, 
the major merger rate at $z=0.3$ is 0.018\,Gyr$^{-1}$ ($\mathrm{M_{gal}}<6.4\times10^{10}\,M_{\odot}$) 
to 0.054\,Gyr$^{-1}$ ($\mathrm{M_{gal}}>6.4\times10^{10}\,\mathrm{M}_{\odot}$) per galaxy. 
The small fraction of dusty ETGs (5.5\%) relative to similar optically 
selected ETGs could be produced assuming that the dusty ETGs are being 
observed within $0.5-1$\,Gyr after major morphological disturbance has 
subsided \citep{Lotz08a}, on the assumption that sputtering does not destroy the dust grains.
Observations of gas and stellar kinematics are required to test this hypothesis.
Lower levels of dust in other ETG samples are thought to be due to minor
mergers \citep[e.g.][]{Temi07}, although in the case of our sample this is less 
plausible since the rate of such mergers needs to be very high to create 
and sustain the dust mass.

Since only a small fraction of ETGs have dust masses as large as those
detected in {\em Herschel\/}-ATLAS, it could be that our sample
represents a short-lived phase in the evolution of some ETGs. Either we
are seeing them at a time when dust is present in a galaxy following a major merger, and the hot
X-ray component is also suppressed in this phase, allowing the dust to
survive. Alternatively, we have a sample of ETGs with sufficient residual ISM left
over from star formation to provide a haven for the dust grains to
grow and survive the hot X-ray gas (which may also be less abundant than
average in these galaxies). Current dust destruction timescales are 
inconsistent with the amount of dust observed in both H-ATLAS and control 
ETGs. We can only account for the dust observed in some H-ATLAS ETGs if we 
assume maximal supernova dust production (in Type I and II SNe),
and no destruction, which would produce $\sim 1 \times 10^{7}\,\rm{M}_{\odot}$ of dust in 1\,Gyr.

Further observations to study the kinematics, light profiles, gas
content and X-ray properties of this sample will be required to fully
answer the question of the origin of this dust. A more detailed
investigation of the evolution of dust in these galaxies using a
detailed chemical evolution model will be presented in Gomez et al (in
prep).

\begin{table*}
\begin{tabular}{lllll}  \\ \hline 
\multicolumn{1}{c}{} & 
\multicolumn{1}{l}{Milky Way} &
\multicolumn{1}{l}{LMC$^{a}$} & 
\multicolumn{1}{l}{ETGs (H-ATLAS)} &
\multicolumn{1}{l}{ETGs (control$^{b}$)} \\ \hline 
Average observed dust mass ($\rm M_{\odot}$) & $\sim 3\times 10^{7}$ &
$ 2\times 10^{6}$ & $5.5 \times 10^{7}$ & $(0.8-4.0) \times 10^{6}$ \\ 
Average star formation rate ($\rm M_{\odot}\,yr^{-1}$) & $\sim 1.0$ & $0.25$ & $0.7$ & $0.07$ \\ 
Destruction via SNe $\rm \tau_{des}$ (Gyr) & $0.5$ & $2.5$ & $1.2$ & $5.0$\\
Median dust injection rate $\rm M_d/\tau_{des}$ ($\rm M_{\odot}\,yr^{-1}$) & $0.06$ & $8\times 10^{-4}$ & $0.045$ & $(1.6-8.0)\times 10^{-4}$\\ 
& & & &  \\ 
Range observed dust masses ($\rm M_{\odot}$) & .. & .. & $(3.1-15.0) \times 10^{7}$ & .. \\ 
Range star formation rate ($\rm M_{\odot}\,yr^{-1}$) & .. & .. & $0.04-12.4$ & ..\\
Range destruction via SNe $\rm \tau_{des}$ (Gyr) & .. & .. & $21-0.07$ &..\\
Range dust injection rate $\rm M_d/\tau_{des}$ ($\rm M_{\odot}\,yr^{-1}$) & .. & .. & $1.5\times 10^{-4}- 2.2 $ & ..\\ 
& & & &  \\
Injection rate AGB stars $\dot{M}_{\rm in,AGB}$ ($\rm M_{\odot}\,yr^{-1}$)$^{c}$ & $2\times 10^{-3}$ & $5\times 10^{-5}$ & $(1-500)\times 10^{-4}$  & $1.4 \times 10^{-4}$\\ 
Injection rate SNe $\dot{M}_{\rm in,SNe}$ ($\rm M_{\odot}\,yr^{-1}$)$^{d}$ & $1\times 10^{-3}$ & $5\times 10^{-5}$ & $(0.5-250)\times 10^{-4}$ & $0.7 \times 10^{-4}$\\ 
& & & &  \\
Time to build up dust mass (Gyrs)$^e$ & $10$ & $19$ & $200 - 2$ & .. \\
Time to build up average dust mass (Gyrs)$^e$ &.. & .. & $400-1$ & $3.8-19.1$\\
Contribution of dust mass from stellar sources & $5\%$ &$13\%$ & $0.3-60\%$ & $100-17\%$\\ 
Extra dust per SNe needed ($\rm M_{\odot}$)$^f$ & $1.0$ & $0.6-0.9$ & $10-0.1$ & $0.0-1.01$\\ \hline
\end{tabular}
\caption{Summary of the dust sources and parameters in the H-ATLAS ETG sample with the Milky Way and Large Magellanic Cloud shown for comparison. $^a$ - \citep{Matsuura09}; $^b$ median dust mass of control sample for range of temperatures (15-25\,K). $^{c}$ - \citep[][and references therein]{Whittet03}; $^{d}$ - \citep{Rho08, Barlow10} for core-collapse SNe; $^{e}$ - calculated from $\rm M_d/(\dot{M}_{\rm in,AGB}+\dot{M}_{\rm in,SNe})$ assuming no destruction; $^f$ - assuming SNe are producing the dust required to match the dust injection rate.}
\label{tab:dustsource} 
\end{table*}

\section{Conclusions}
\label{sec:conclusions}
We present the properties of a $250\mu$m selected sample of galaxies
according to their morphology. Our sample consists of 44 early-type 
galaxies and 496 spiral galaxies in the 14 square degree
SDP field. Using an energy balance method of SED fitting we derive 
physical parameters, and use these to compare the properties of 
H-ATLAS galaxies as a function of morphological type. We also compare to a control 
sample of optical galaxies selected to have the same $n(r,z)$ as the 
H-ATLAS sample. Our main results are as follows.

\begin{itemize}

\item ETGs detected by \emph{Herschel} are atypical compared to optically
  selected ETGs. We detect significant dust masses in H-ATLAS ETGs, with a 
  mean of $5.5\times{10}^{7}$\,M$_\odot$. Through stacking we find that dust 
  masses are an order of magnitude lower in optically selected ETGs of a 
  similar stellar mass.

\item Only a small fraction of H-ATLAS ETGs (24 percent) have evidence 
  for a recent burst of star formation within the past Gyr. Some of these galaxies may have
  had star-formation triggered as a result of an interaction,
  indicated by disturbed morphologies in 31 percent of the sample,
  although not all disturbed sources show signs of a recent burst.
  The majority have residual low-level star-formation left over from
  the last burst a few Gyrs ago, and their optical colours suggest they exist in the
  transition region between the blue cloud and the red sequence.

\item We find that the control ETGs have lower SSFRs and older stellar 
  population ages than H-ATLAS ETGs, which is consistent with the red 
  UV-optical colours of the control ETGs. It is possible that the dust content may therefore be 
  related to the time of the last major star formation episode several Gyrs ago.

\item No significant difference is found in the environments of
  H-ATLAS and control ETGs, although this may be due to small sample 
  size. Environment does not seem to influence whether an ETG is dusty, 
  at the moderate-low densities probed in this study. Additionally, we 
  do not find any H-ATLAS ETGs in high density environments.

\item The 5.5\% of ETGs detected in H-ATLAS (compared to an optical sample of similar stellar mass) contain more dust 
  than can be accounted for by production in AGB stars, although 
  this problem also extends to the control ETGs (except at the very 
  lowest median dust mass). Most of the dust must be formed in the ISM, or 
  an external source of dust from major mergers is also a possibility. 
  It is also possible that in H-ATLAS and control ETGs the dust destruction 
  timescale is longer if they are deficient in X-ray gas. 
  Future studies of the kinematics of the gas and stars would be 
  beneficial in testing this hypothesis, in addition to X-ray and 
  CO observations.

\item We examine the properties of passive spirals in our sample which
  have low SSFR\,$<10^{-11}\rm{yr^{-1}}$, but still contain
  significant dust mass. They have larger $\rm{M}_{\ast}$ and lower
  \mdms{} than `normal' spirals, and are red in colour, which is due to an
  old stellar population, and not due to increased dust reddening.
  It is possible that these passive spirals have simply run out of
  gas to fuel star-formation, or their star-formation has been
  quenched by some process in the low density environment in which 
  they reside. 

\end{itemize}

Using \emph{Herschel} we can probe the dust content of different types
of galaxies over a wide range of redshifts. The full coverage of the
H-ATLAS survey will allow more investigation, with far larger numbers
of ETGs and passive spirals. This will improve our
understanding of objects which are transitioning between the blue and
red sequence, either through rejuvenated star-formation, or through
the cessation of star-formation as the supply of gas ends.

\section*{Acknowledgements}
We would like to thanks the anonymous referee for important suggestions 
which helped to improve the paper. We also thank Luca Cortese, Alfonso Arag\'{o}n-Salamanca and Andrew Baker for helpful comments.
The Herschel-ATLAS is a project with \emph{Herschel}, which is an ESA space
observatory with science instruments provided by European-led
Principal Investigator consortia and with important participation from
NASA. The H-ATLAS website is http://www.h-atlas.org/. \\ GAMA is a
joint European-Australasian project based around a spectroscopic
campaign using the Anglo-Australian Telescope. The GAMA input
catalogue is based on data taken from the Sloan Digital Sky Survey and
the UKIRT Infrared Deep Sky Survey. Complementary imaging of the GAMA
regions is being obtained by a number of independent survey programs
including \emph{GALEX} MIS, VST KIDS, VISTA VIKING, \emph{WISE}, \emph{Herschel}-ATLAS,
GMRT and ASKAP providing UV to radio coverage. GAMA is funded by the
STFC (UK), the ARC (Australia), the AAO, and the participating
institutions. The GAMA website is http://www.gama-survey.org/.
The Italian group acknowledges partial financial support from ASI/INAF agreement n. I/009/10/0.

\bibliographystyle{mn2e}
\bibliography{references}

\appendix
\section{Early-type galaxies}

\begin{landscape}
\begin{table}
\begin{center}
\caption{Properties of ETGs derived from SED fitting. The columns are (from left to right): ID, SDP ID, redshift, SDSS RA, SDSS DEC, $250\mu$m flux in Jy,
\fmu, the fraction of total dust luminosity contributed by the diffuse ISM; \tauv, total effective V-band optical depth seen by stars in birth clouds; 
$\rm{M}_\ast/\rm{M}_\odot$, log(stellar mass); \ldust/$\rm{L}_\odot$, log(dust luminosity); \tbgscold/K, temperature of the cold ISM 
dust component; \tauvISM, the V-band optical depth in the ambient ISM. $\rm{M}_\mathrm{d}/\rm{M}_\odot$, log(dust mass); \ssfr/$yr^{-1}$, log(SSFR); 
\sfr/$\rm{M}_\odot yr{-1}$, log(SFR), $\rm{t_{LB}}$, log(time of last burst); $\rm{age_r}$, log($r$-band light-weighted age of the stellar population), rest-frame NUV-$r$ colour (Section \ref{Colours}), density ($\Sigma$/galaxies $\rm{Mpc}^{-2}$, see Section \ref{sec:Environment}) H$\alpha$ EW/\AA (corrected for stellar absorption of 1.3\AA\/ if $>3\sigma$ detection). * indicates morphological disturbance.}
\begin{tabular}{|l|l|l|l|l|l|l|l|l|l|l|l|l|l|l|l|l|l|l|l|l|}
\hline
  \multicolumn{1}{|c|}{ID} &
  \multicolumn{1}{|c|}{SDP ID} &
  \multicolumn{1}{c|}{$z$} &
  \multicolumn{1}{|c|}{RA} &
  \multicolumn{1}{|c|}{DEC} &
  \multicolumn{1}{c|}{$\rm{F}_{250}$} &
  \multicolumn{1}{c|}{\fmu} &
  \multicolumn{1}{c|}{\tauv} &
  \multicolumn{1}{c|}{$\rm{M}_\ast$} &
  \multicolumn{1}{c|}{\ldust} &
  \multicolumn{1}{c|}{\tbgscold} &
  \multicolumn{1}{c|}{\tauvISM} &
  \multicolumn{1}{c|}{$\rm{M}_\mathrm{d}$} &
  \multicolumn{1}{c|}{\ssfr} &
  \multicolumn{1}{c|}{\sfr} &
  \multicolumn{1}{c|}{$\rm{t_{LB}}$} &
  \multicolumn{1}{c|}{$\rm{{age}_r}$} &
  \multicolumn{1}{c|}{$NUV-r$} &
  \multicolumn{1}{c|}{$\Sigma$} &
  \multicolumn{1}{c|}{H$\alpha$ EW} \\
  \hline
  J091205.8+002656 & 15* & 0.05   & 138.024 & 0.449 & 0.38 & 0.50 & 3.60 & 10.20 & 11.00 & 24.1 & 1.27 & 7.58 &  -9.47 & 0.74 & 9.04 & 9.08 & --   & 2.92 & 27.54\\
  J091448.7-003533 & 35* & 0.05   & 138.704 & -0.592& 0.25 & 0.73 & 2.31 & 10.45 & 10.44 & 21.5 & 0.69 & 7.50 & -10.40 & 0.07 & 9.38 & 9.52 & --   & 1.02 & 8.39\\
  J090352.0-005353 & 45* & 0.10   & 135.967 & -0.898& 0.20 & 0.73 & 2.20 & 10.96 & 11.00 & 22.2 & 0.64 & 7.97 & -10.34 & 0.62 & 9.25 & 9.41 & 3.77 & 0.06 & 14.87\\
  J091051.1+020121 & 128 & 0.05  & 137.714 & 2.022 & 0.11 & 0.58 & 1.71 &  9.87 & 10.00 & 15.7 & 0.46 & 7.74 &  -10.06 & -0.13 & 9.05 & 9.29 & 2.59  & 0.46 & 12.63\\
  J090234.3+012518 & 159 & 0.12  & 135.643 & 1.421 & 0.11 & 0.93 & 3.02 & 10.96 & 10.72 & 22.9 & 0.88 & 7.75 & -11.39 &-0.44 & 9.31 & 9.67 & 5.25 & 1.08 & 5.24\\
  J090647.7+011555 & 186 & 0.15  & 136.699 & 1.265 & 0.10 & 0.57 & 2.53 & 10.50 & 10.91 & 20.4 & 0.48 & 8.03 &  -9.72 & 0.79 & 8.69 & 9.00 & 3.07 & 0.08 & 24.35\\
  J090101.2-005541 & 273 & 0.09  & 135.256 & -0.929& 0.09 & 0.68 & 1.10 & 10.82 & 10.36 & 19.0 & 0.21 & 7.67 & -10.71 & 0.12 & 9.33 & 9.51 & 3.93 & 0.01 & 5.01\\
  J090238.7+013253 & 311 & 0.12  & 135.661 & 1.548 & 0.09 & 0.68 & 2.44 & 10.38 & 10.78 & 23.0 & 0.82 & 7.67 &  -9.97 & 0.46 & 8.61 & 9.11 & 3.54 & 0.63 & 39.30\\
  J090223.1+010709 & 328* & 0.20  & 135.597 & 1.120 & 0.09 & 0.83 & 2.08 & 11.02 & 11.04 & 20.8 & 0.60 & 8.22 & -10.63 & 0.44 & 8.90 & 9.31 & 4.34 &   -- & 8.87\\
  J090718.9-005210 & 350 & 0.06  & 136.829 & -0.869& 0.09 & 0.36 & 1.33 &  9.65 & 10.55 & 24.1 & 0.50 & 6.93 &  -9.25 & 0.41 & 8.80 & 8.92 & 2.37 & 0.01 & 109.07\\
  J091332.4+000631 & 366* & 0.23  & 138.386 & 0.108 & 0.08 & 0.93 & 3.65 & 10.88 & 11.13 & 23.7 & 0.84 & 8.07 & -10.96 &-0.09 & 8.78 & 9.00 & 4.49 &  --  & 14.76\\
  J091023.1+014023 & 370 & 0.14  & 137.596 & 1.673 & 0.08 & 0.95 & 3.68 & 10.82 & 10.46 & 15.7 & 0.71 & 8.40 & -11.88 &-1.08 & 9.48 & 9.80 & $>5.19$ & 0.07 & 0.48\\
  J090952.3-003019 & 451 & 0.05  & 137.468 & -0.505& 0.09 & 0.90 & 1.86 & 10.28 &  9.93 & 22.4 & 0.56 & 6.97 & -11.17 &-0.93 & 9.35 & 9.65 & 4.68 & 0.01 & 12.69\\
  J085915.7+002329 & 457* & 0.01  & 134.815 & 0.392 & 0.09 & 0.30 & 0.67 &  8.62 &  8.76 & 18.3 & 0.14 & 5.97 &  -9.51 &-1.00 & 9.16 & 9.28 & 1.94 &  --  & 45.87\\
  J090551.5+010752 & 628* & 0.05  & 136.465 & 1.131 & 0.07 & 0.92 & 1.08 & 11.18 &  9.94 & 21.7 & 0.12 & 6.98 & -12.27 &-1.10 & 9.57 & 9.92 & 5.99 & 0.17 & 1.70\\
  J090522.1-005925 & 786 & 0.10  & 136.343 & -0.991& 0.08 & 0.80 & 1.58 & 10.91 & 10.40 & 20.3 & 0.26 & 7.57 & -11.04 &-0.12 & 9.41 & 9.60 & --   & 0.08 & 1.59\\
  J090752.3+012945 & 1027 & 0.10 & 136.968 & 1.496 & 0.07 & 0.80 & 1.28 & 11.06 & 10.36 & 20.3 & 0.20 & 7.58 & -11.16 &-0.09 & 9.52 & 9.67 & 4.61 & 3.24 & 1.39\\
  J085852.1+010624 & 1278 & 0.12 & 134.718 & 1.106 & 0.06 & 0.63 & 1.14 & 10.60 & 10.35 & 17.4 & 0.20 & 7.87 & -10.47 & 0.18 & 8.86 & 9.24 & 3.67 & 0.50 & 2.21\\
  J091037.8+015654 & 1372* & 0.23 & 137.658 & 1.949 & 0.06 & 0.55 & 2.41 & 10.76 & 10.99 & 19.4 & 0.37 & 8.20 &  -9.88 & 0.88 & 8.77 & 9.05 & 3.02 &  --  & 32.32\\
  J090929.3+020327 & 1409 & 0.15 & 137.373 & 2.057 & 0.06 & 0.92 & 1.85 & 11.17 & 10.49 & 19.4 & 0.31 & 7.86 & -11.70 &-0.49 & 9.36 & 9.67 & 5.36 & 0.26 & 1.30\\
  J090618.0-002455 & 1955* & 0.17 & 136.575 & -0.415& 0.05 & 0.66 & 2.30 & 10.92 & 10.68 & 20.1 & 0.38 & 7.85 & -10.53 & 0.40 & 9.40 & 9.51 & 3.78 & 0.28 & 3.49\\
  J090259.5+020046 & 2025 & 0.07 & 135.747 & 2.012 & 0.05 & 0.85 & 2.12 & 10.34 &  9.87 & 18.5 & 0.45 & 7.36 & -11.14 &-0.79 & 9.53 & 9.74 & 4.71 & 0.08 & 8.11\\
  J085934.1+003629 & 2311* & 0.26 & 134.892 & 0.608 & 0.05 & 0.90 & 2.02 & 11.20 & 10.81 & 16.3 & 0.47 & 8.64 & -11.20 &-0.01 & 9.27 & 9.61 & 4.64 & --   & 2.47\\
  J085842.0+010956 & 2364 & 0.12 & 134.677 & 1.166 & 0.06 & 0.86 & 1.49 & 10.96 & 10.42 & 19.5 & 0.36 & 7.73 & -11.18 &-0.25 & 9.31 & 9.58 & 4.12   & 0.67 & 3.55\\
  J085944.2+011708 & 2702 & 0.16 & 134.933 & 1.285 & 0.05 & 0.94 & 2.18 & 10.88 & 10.43 & 17.2 & 0.52 & 8.11 & -11.73 &-0.86 & 9.37 & 9.66 & $>4.84$ & 0.59 & 1.78\\
  J090634.8+020752 & 2853 & 0.25 & 136.645 & 2.132 & 0.04 & 0.82 & 1.14 & 11.39 & 10.88 & 19.0 & 0.31 & 8.25 & -11.05 & 0.32 & 9.46 & 9.59 & 4.29 &  --  & 2.37\\
  J090210.6+004805 & 2945 & 0.20 & 135.545 & 0.802 & 0.05 & 0.46 & 1.51 & 10.70 & 10.88 & 18.3 & 0.32 & 8.10 &  -9.89 & 0.80 & 9.46 & 9.27 & 2.94 &  --  & 9.97\\
  J085727.4+010847 & 2959 & 0.07 & 134.364 & 1.146 & 0.05 & 0.78 & 1.51 & 10.30 &  9.84 & 20.1 & 0.31 & 7.09 & -10.90 &-0.66 & 9.48 & 9.63 & 4.23 & 2.80 & 5.19\\
  J091359.4+000909 & 3005 & 0.17 & 138.498 & 0.152 & 0.04 & 0.92 & 1.95 & 10.58 & 10.41 & 17.5 & 0.55 & 8.05 & -11.17 &-0.55 & 8.91 & 9.38 & 4.78 & 6.02 & 1.30\\
  J090236.7+011909 & 3252 & 0.09 & 135.653 & 1.320 & 0.06 & 0.95 & 1.58 & 11.03 & 10.14 & 19.7 & 0.19 & 7.49 & -12.37 &-1.37 & 9.47 & 9.77 & 5.36 & 0.17 & 0.38\\
  J090849.5-001846 & 3321 & 0.22 & 137.208 & -0.313& 0.04 & 0.72 & 1.62 & 11.07 & 10.76 & 19.1 & 0.32 & 8.08 & -10.65 & 0.40 & 9.39 & 9.50 & 4.00 &  --  & 2.46\\
  J091435.2-003919 & 3549 & 0.32 & 138.648 & -0.655& 0.04 & 0.54 & 1.72 & 11.19 & 11.24 & 21.5 & 0.43 & 8.17 & -10.09 & 1.09 & 9.41 & 9.36 & 3.00 &  --  & 6.21\\
  J091409.6+000439 & 3702 & 0.16 & 138.541 & 0.078 & 0.04 & 0.74 & 2.20 & 10.69 & 10.49 & 17.4 & 0.51 & 8.06 & -10.55 & 0.11 & 9.15 & 9.45 & 4.15 & 6.99 & 3.99\\
  J090938.9-005753 & 3834 & 0.13 & 137.412 & -0.966& 0.04 & 0.94 & 2.16 & 10.56 & 10.22 & 18.8 & 0.50 & 7.67 & -11.68 &-1.10 & 9.10 & 9.48 & $>4.96$ & 0.04 & 1.30\\
  J091315.8+004445 & 5088 & 0.23 & 138.315 & 0.746 & 0.04 & 0.71 & 1.56 & 11.05 & 10.68 & 16.9 & 0.22 & 8.31 & -10.69 & 0.38 & 8.95 & 9.30 & 4.24 &  --  & 19.50\\
  J090936.0+023324 & 5311 & 0.16 & 137.399 & 2.557 & 0.04 & 0.66 & 1.38 & 10.58 & 10.48 & 19.9 & 0.31 & 7.67 & -10.35 & 0.19 & 9.28 & 9.41 & 3.48 & 7.44 & 15.18\\
  J091054.2+005454 & 5382 & 0.16 & 137.726 & 0.916 & 0.04 & 0.69 & 1.68 & 10.80 & 10.48 & 17.6 & 0.33 & 7.95 & -10.65 & 0.14 & 9.48 & 9.55 & 3.86 & 0.22 & 3.55\\
  J091143.5+012053 & 5489 & 0.07 & 137.932 & 1.349 & 0.04 & 0.82 & 1.81 & 10.49 &  9.83 & 20.9 & 0.26 & 6.98 & -11.20 &-0.70 & 9.38 & 9.62 & 4.77 & 0.04 & 2.70\\
  J090310.3+014233 & 6310 & 0.16 & 135.793 & 1.709 & 0.04 & 0.57 & 1.71 & 10.57 & 10.48 & 17.3 & 0.28 & 7.93 & -10.25 & 0.31 & 9.45 & 9.47 & 3.16 & 0.11 & 8.74\\
  J085916.4+005218 & 6337 & 0.24 & 134.819 & 0.873 & 0.04 & 0.59 & 1.68 & 10.66 & 10.80 & 18.6 & 0.32 & 8.10 & -10.05 & 0.63 & 8.78 & 9.09 & 3.27 &  --  & 8.75\\
  J085947.9-002143 & 6427* & 0.12 & 134.95 & -0.363 & 0.05 & 0.87 & 1.65 & 11.00 & 10.16 & 21.2 & 0.18 & 7.34 & -11.54 &-0.56 & 9.47 & 9.65 & 4.77   & 0.01 & 12.04\\
  J090413.9-004405 & 6640* & 0.20 & 136.058 & -0.734& 0.04 & 0.94 & 1.97 & 11.10 & 10.52 & 20.0 & 0.37 & 7.82 & -12.03 &-0.90 & 9.33 & 9.63 & $>5.05$ & --   & 2.60\\
\hline                                                                                                             
\end{tabular}                                                                                                      
\label{ETG_properties}
\end{center}
\end{table}
\end{landscape}

\begin{figure*}
\begin{minipage}[t]{1.0\textwidth}
    \begin{center}
\includegraphics{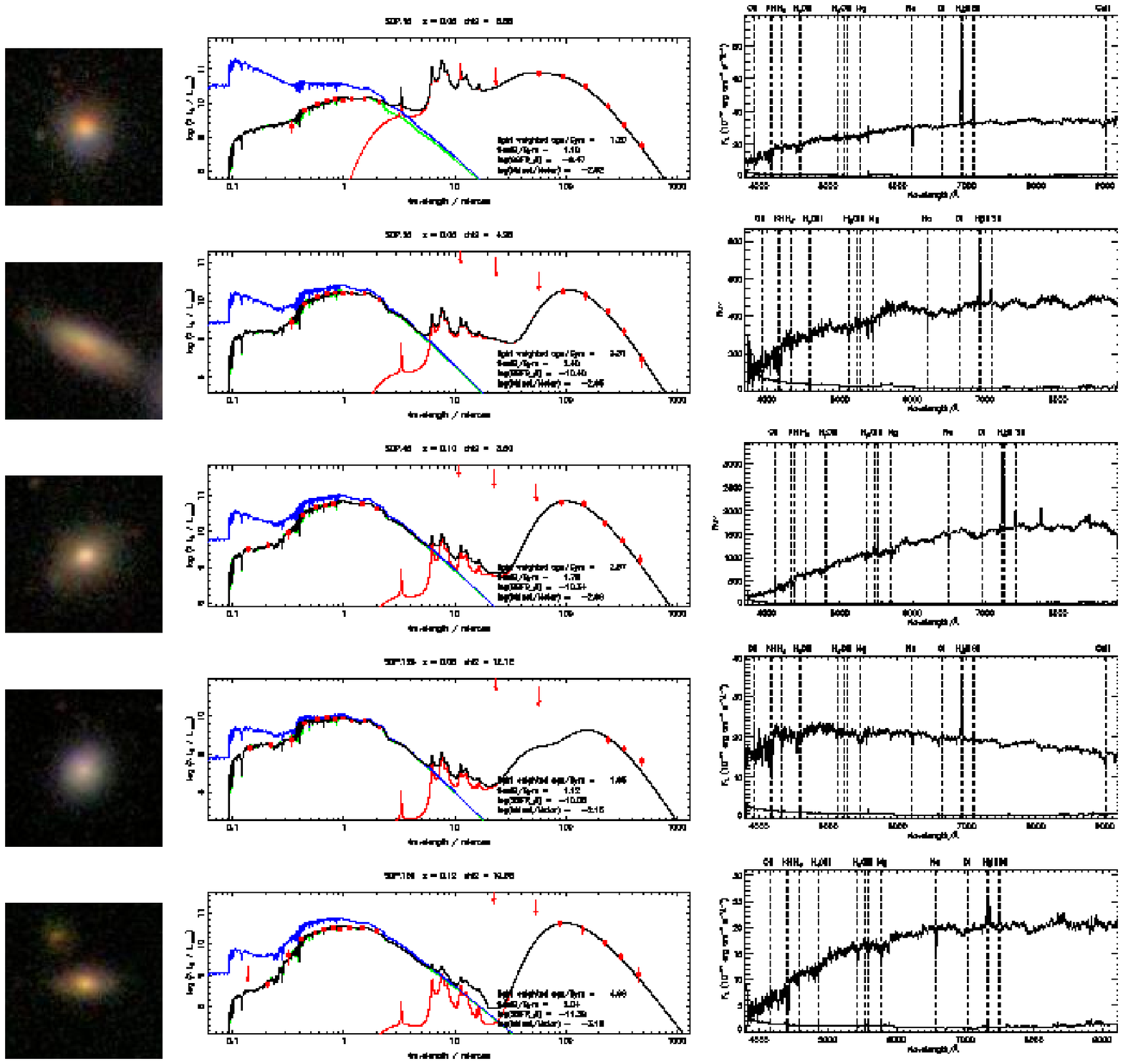}
 \end{center}
\end{minipage}
\caption{Optical images, multiwavelength SEDs and optical spectra of the 42 ETGs in our sample. Images are 30" on a side. The rest-frame SEDs of each ETG are shown, where red points are the observed photometry, with 5$\sigma$ upper limits shown as arrows. Errors on the photometry are described in \citet{Smith11b}. 
The black line is the total best fit model, the green line is the attenuated optical model, the blue line is the unattenuated optical model, the red line is the infrared model. Spectra are from SDSS and GAMA, and the standard deviation in the spectra is also shown. The spectra have been smoothed by a boxcar of 8 pixels.}
\label{fig:ETGs}
\end{figure*}

 \begin{figure*}
 \begin{minipage}[t]{1.0\textwidth}
     \begin{center}  
 \includegraphics{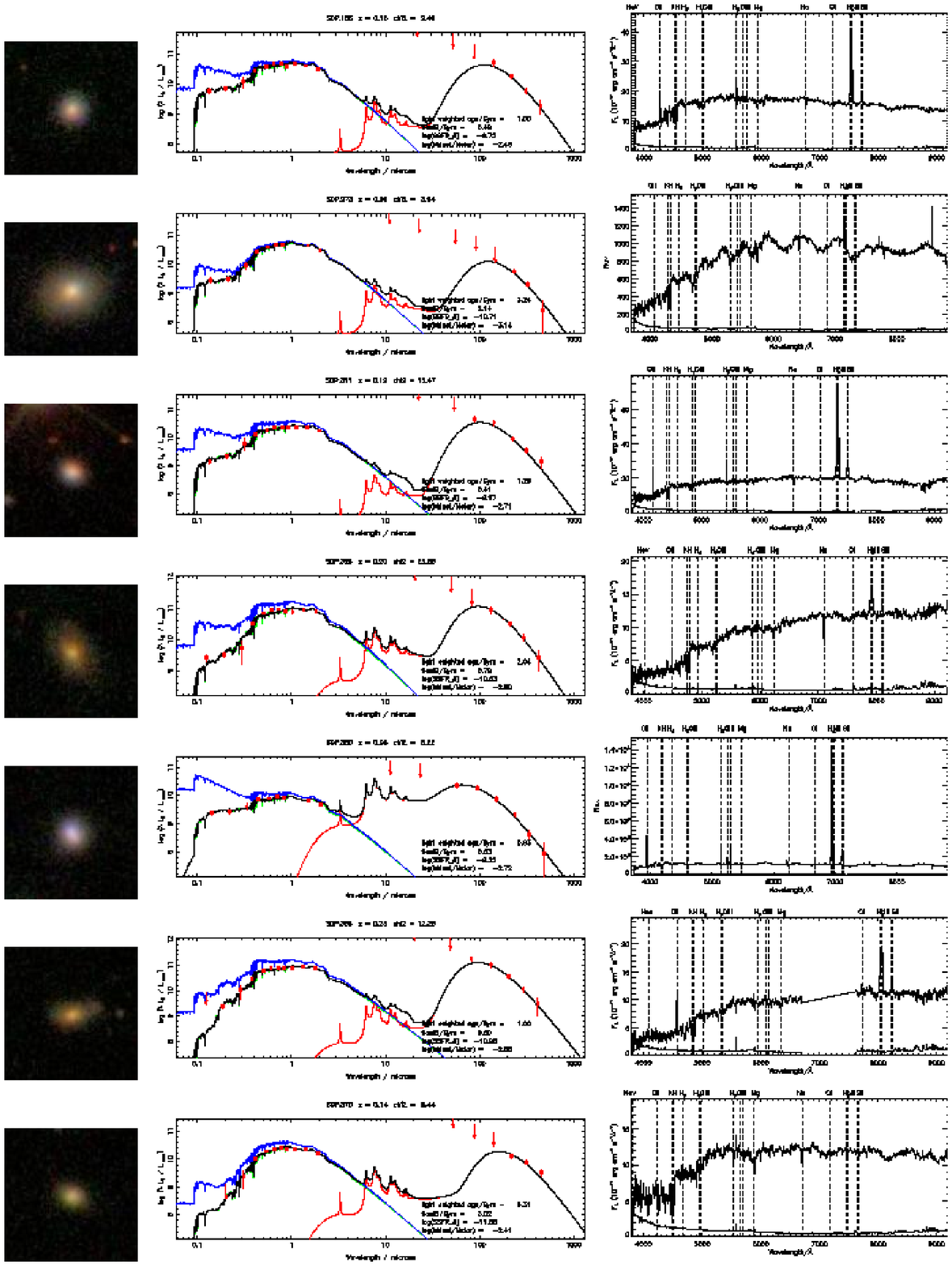}
  \end{center}
 \end{minipage}
 \contcaption{}
 \end{figure*}
 
 \begin{figure*}
 \begin{minipage}[t]{1.0\textwidth}
     \begin{center}  
 \includegraphics{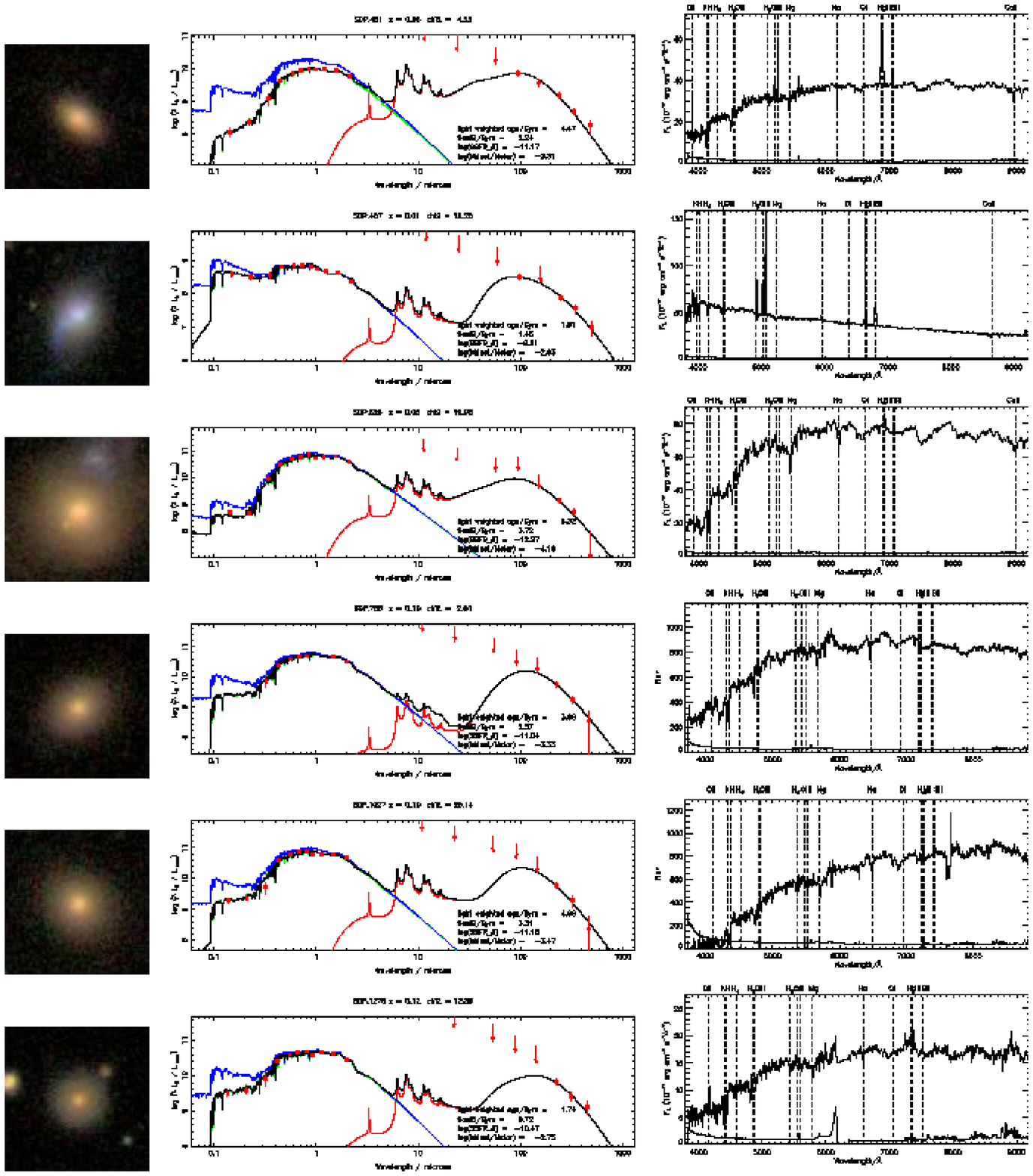}
  \end{center}
 \end{minipage}
 \contcaption{}
 \end{figure*}
 
 \begin{figure*}
 \begin{minipage}[t]{1.0\textwidth}
     \begin{center}  
 \includegraphics{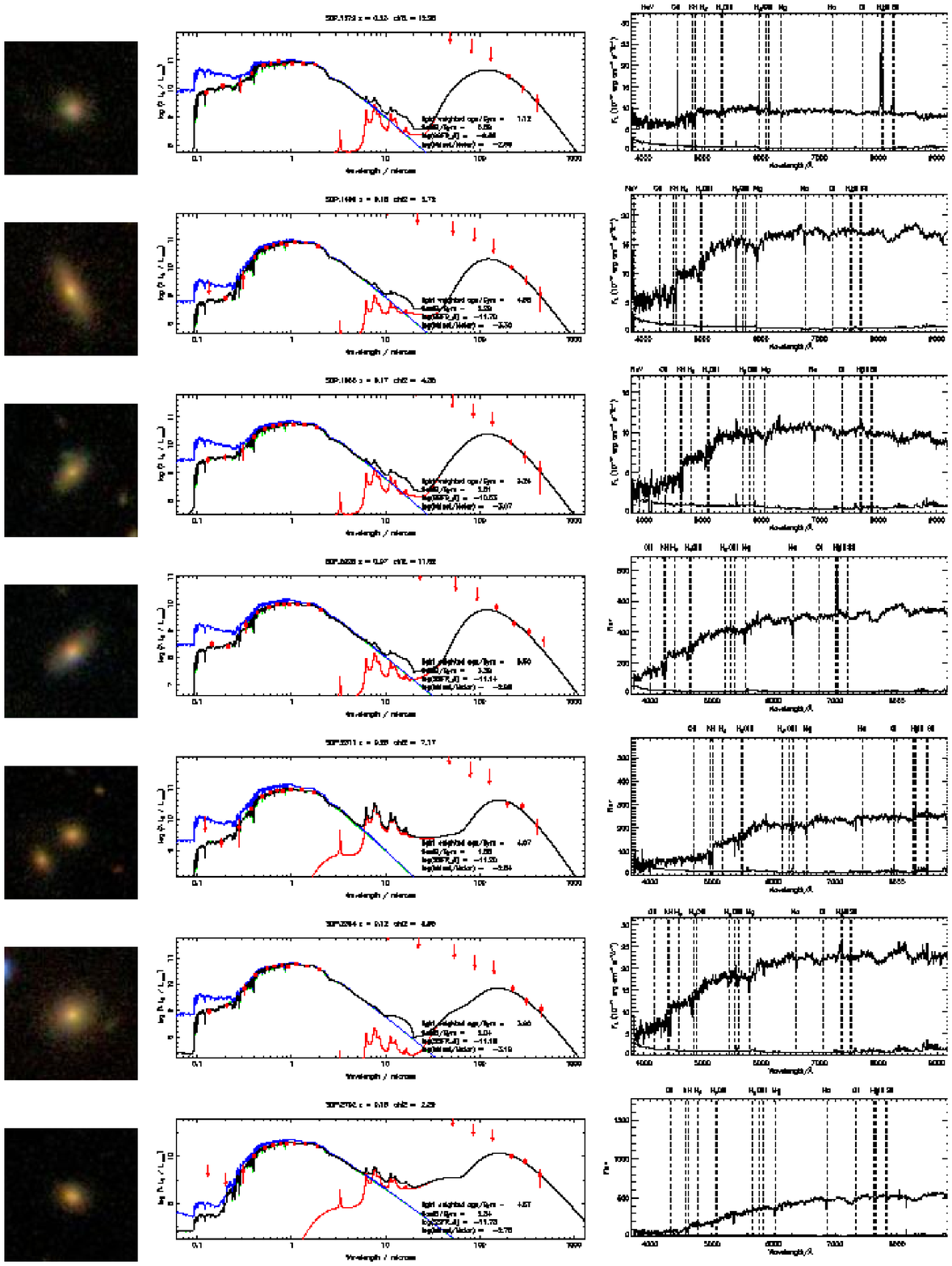}
  \end{center}
 \end{minipage}
 \contcaption{}
 \end{figure*}
 
 \begin{figure*}
 \begin{minipage}[t]{1.0\textwidth}
     \begin{center}  
 \includegraphics{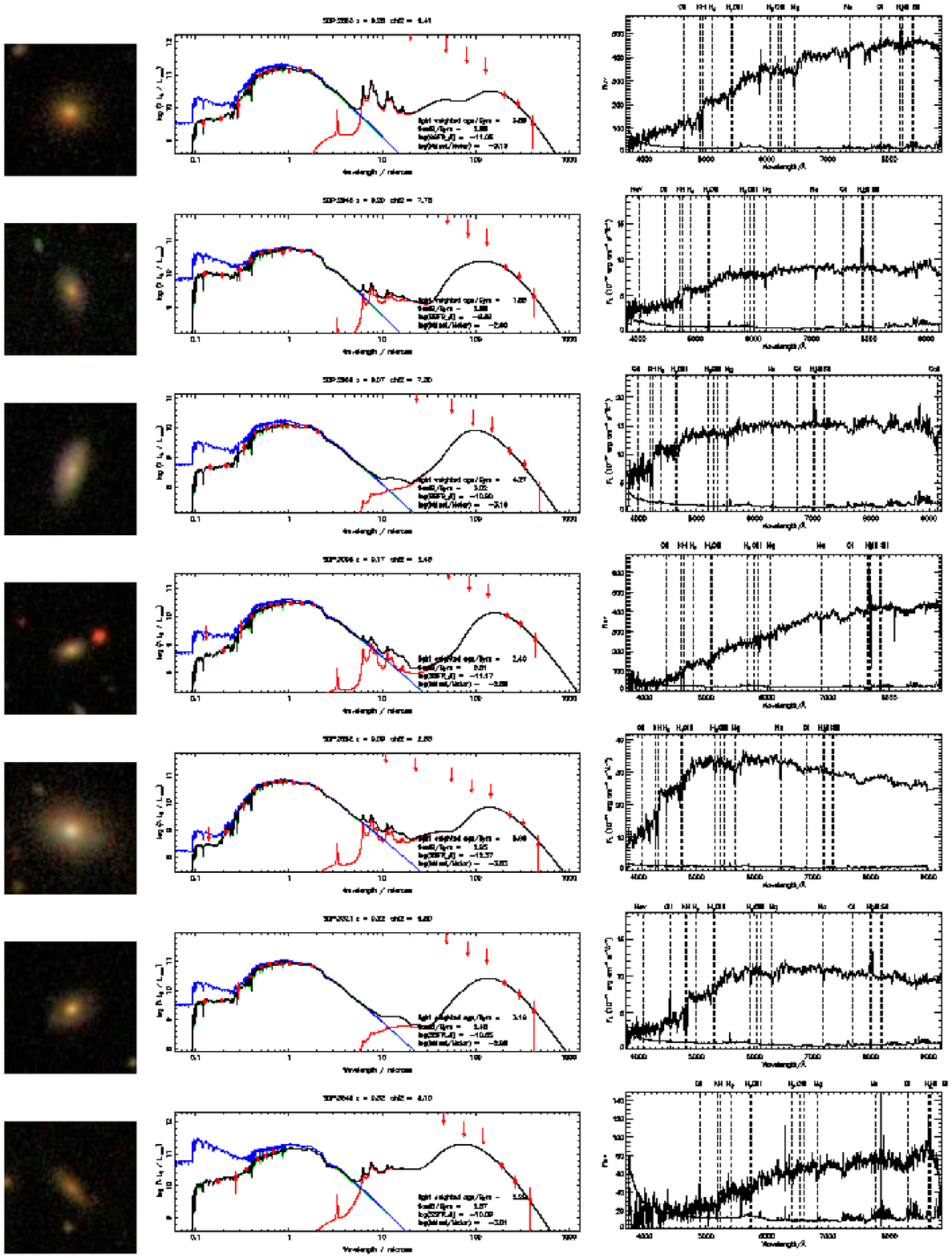}
  \end{center}
 \end{minipage}
 \contcaption{}
 \end{figure*}
 
 \begin{figure*}
 \begin{minipage}[t]{1.0\textwidth}
     \begin{center}  
 \includegraphics{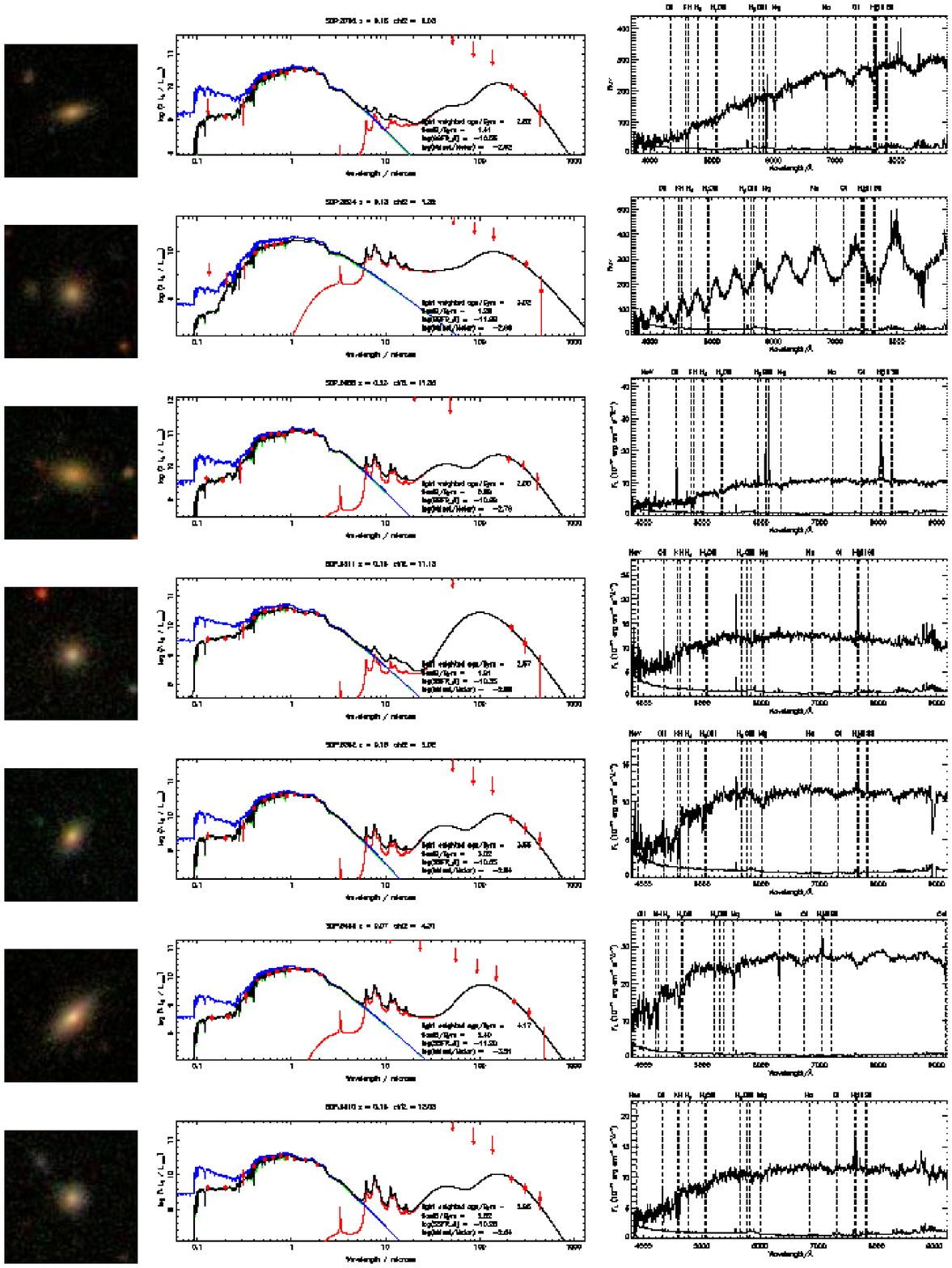}
  \end{center}
 \end{minipage}
 \contcaption{. The wave-like features in the spectrum of SDP.3834 are due to fibre fringing \citep{Colless01_2dF}.}
 \end{figure*}

 \begin{figure*}
 \begin{minipage}[t]{1.0\textwidth}
     \begin{center}  
 \includegraphics{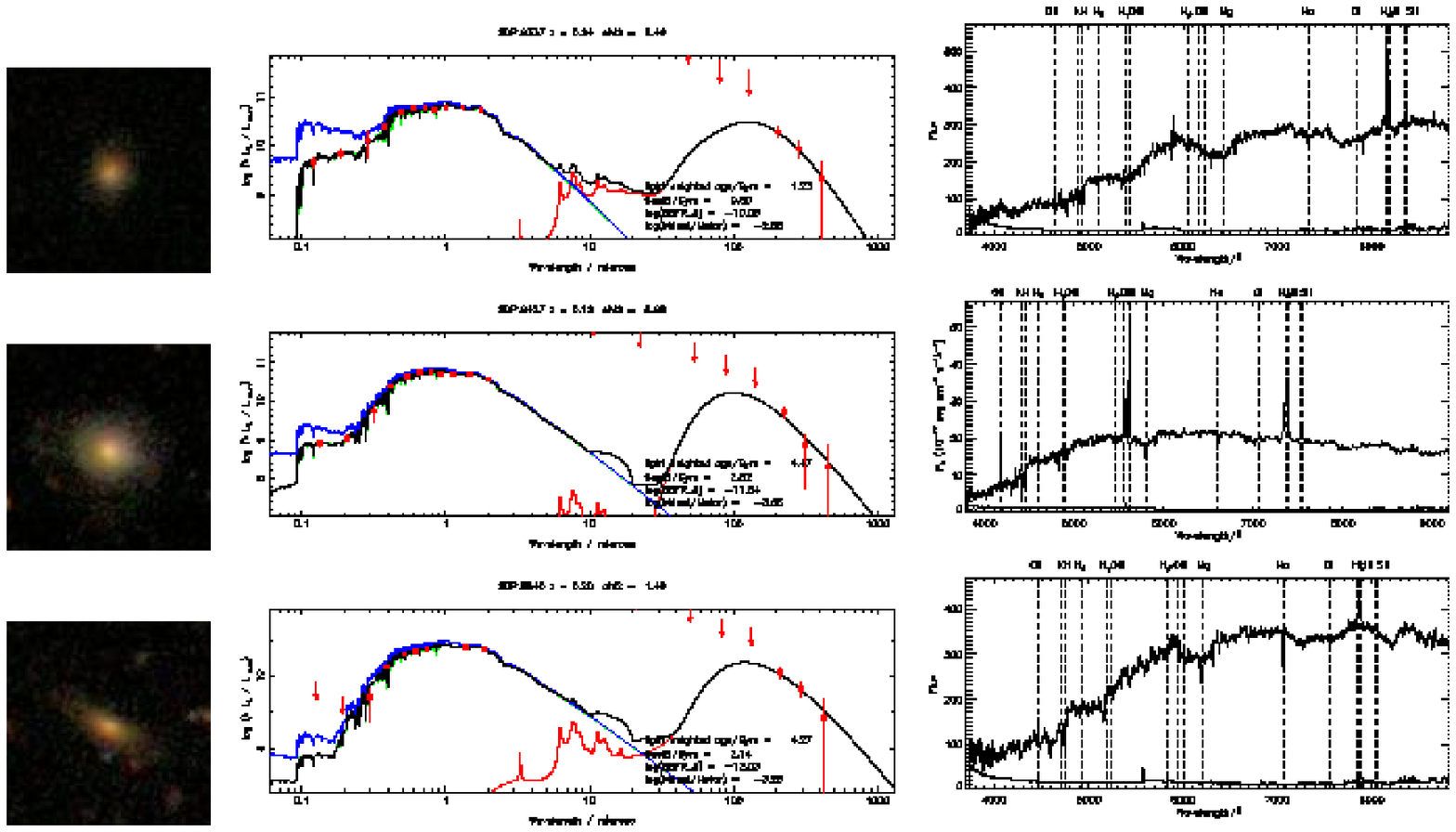}
  \end{center}
 \end{minipage}
 \contcaption{}
 \end{figure*}

\clearpage

\section{Passive spirals}

\begin{landscape}
\begin{table}
\begin{center}
\caption{Properties of passive spirals derived from SED fitting. The columns are (from left to right): ID, SDP ID, redshift, SDSS RA, SDSS DEC, $250\mu$m flux in Jy,
\fmu, the fraction of total dust luminosity contributed by the diffuse ISM; \tauv, total effective V-band optical depth seen by stars in birth clouds; 
$\rm{M}_\ast/\rm{M}_\odot$, log(stellar mass); \ldust/$\rm{L}_\odot$, log(dust luminosity); \tbgscold/K, temperature of the cold ISM 
dust component; \tauvISM, the V-band optical depth in the ambient ISM. $\rm{M}_\mathrm{d}/\rm{M}_\odot$, log(dust mass); \ssfr/$yr^{-1}$, log(SSFR); 
\sfr/$\rm{M}_\odot yr{-1}$, log(SFR), $\rm{t_{LB}}$, log(time of last burst); $\rm{age_r}$, log($r$-band light-weighted age of the stellar population), rest-frame NUV-$r$ colour (Section \ref{Colours}), density ($\Sigma$/galaxies $\rm{Mpc}^{-2}$, see Section \ref{sec:Environment}) H$\alpha$ EW/\AA (corrected for stellar absorption of 1.3\AA\/ if $>3\sigma$ detection).}
\begin{tabular}{|l|l|l|l|l|l|l|l|l|l|l|l|l|l|l|l|l|l|l|l|l|}
\hline
  \multicolumn{1}{|c|}{ID} &
  \multicolumn{1}{|c|}{SDP ID} &
  \multicolumn{1}{c|}{$z$} &
  \multicolumn{1}{|c|}{RA} &
  \multicolumn{1}{|c|}{DEC} &
  \multicolumn{1}{c|}{$\rm{F}_{250}$} &
  \multicolumn{1}{c|}{\fmu} &
  \multicolumn{1}{c|}{\tauv} &
  \multicolumn{1}{c|}{$\rm{M}_\ast$} &
  \multicolumn{1}{c|}{\ldust} &
  \multicolumn{1}{c|}{\tbgscold} &
  \multicolumn{1}{c|}{\tauvISM} &
  \multicolumn{1}{c|}{$\rm{M}_\mathrm{d}$} &
  \multicolumn{1}{c|}{\ssfr} &
  \multicolumn{1}{c|}{\sfr} &
  \multicolumn{1}{c|}{$\rm{t_{LB}}$} &
  \multicolumn{1}{c|}{$\rm{{age}_r}$} &
  \multicolumn{1}{c|}{$NUV-r$} &
  \multicolumn{1}{c|}{$\Sigma$} &
  \multicolumn{1}{c|}{H$\alpha$ EW} \\
\hline
  J085828.5+003814 & 30 & 0.05 & 134.619 & 0.637    & 0.28 & 0.87 & 4.22 & 10.84 & 10.39 & 21.7 & 0.45 & 7.50 & -11.17 & -0.30 & 9.57 & 9.77  & 4.73 & 0.39 & 14.77\\
  J090038.0+012810 & 77 & 0.05 & 135.158 & 1.470    & 0.19 & 0.79 & 1.44 & 10.86 & 10.04 & 17.5 & 0.17 & 7.64 & -11.14 & -0.31 & 9.39 & 9.83  & 4.33 & 0.19 & 1.00\\
  J085946.7-000020 & 143 & 0.05 & 134.945 & -0.006  & 0.15 & 0.85 & 1.80 & 10.78 & 10.15 & 19.5 & 0.26 & 7.45 & -11.22 & -0.48 & 9.41 & 9.65  & 4.69 & 0.39 & 0.57\\
  J090911.8+000029 & 271 & 0.08 & 137.299 & 0.008   & 0.12 & 0.90 & 2.04 & 10.53 & 10.16 & 18.3 & 0.52 & 7.66 & -11.25 & -0.72 & 9.18 & 9.53  & 5.11 & 0.18 & 2.45\\
  J090648.9-005059 & 372 & 0.16 & 136.704 & -0.850  & 0.09 & 0.94 & 2.05 & 11.02 & 10.73 & 19.9 & 0.56 & 8.05 & -11.63 & -0.56 & 9.08 & 9.45  & 5.45 & 0.08 & 6.06\\
  J090312.4-004509 & 1544 & 0.05 & 135.803 & -0.753 & 0.09 & 0.67 & 1.62 & 10.69 & 9.88 & 18.5 & 0.13 & 7.23 & -11.11 & -0.40 & 9.56 & 9.73   & 4.30 & --   & 0.62\\
  J090944.5+022100 & 1773 & 0.05 & 137.435 & 2.350  & 0.06 & 0.94 & 1.57 & 10.50 & 9.78 & 22.7 & 0.25 & 6.76 & -12.10 & -1.57 & 8.93 & 9.62   & 4.70 & 0.56 & 1.63\\
  J090622.3+010014 & 1888 & 0.07 & 136.593 & 1.004  & 0.06 & 0.85 & 1.13 & 10.50 & 9.87 & 19.8 & 0.19 & 7.15 & -11.29 & -0.76 & 9.42 & 9.73   & --   & 0.08 & 1.01\\
  J085827.1+010426 & 2547 & 0.07 & 134.613 & 1.074  & 0.05 & 0.95 & 1.66 & 10.34 & 9.61 & 19.8 & 0.27 & 6.98 & -12.19 & -1.91 & 9.44 & 9.71   & 5.29 & 6.78 & 2.01\\
  J090543.6+010754 & 2612 & 0.05 & 136.432 & 1.132  & 0.05 & 0.92 & 2.00 &  9.94 & 9.45 & 20.3 & 0.40 & 6.74 & -11.53 & -1.56 & 9.19 & 9.56   & 5.08 & 0.22 & 4.25\\
  J090547.8+001136 & 3578 & 0.16 & 136.450 & 0.193  & 0.04 & 0.92 & 2.24 & 10.95 & 10.48 & 20.0 & 0.56 & 7.78 & -11.55 & -0.58 & 9.36 & 9.63  & $>4.57$ & 0.14 & 1.64\\
  J091311.5+001619 & 3935 & 0.17 & 138.299 & 0.274  & 0.04 & 0.94 & 2.44 & 10.71 & 10.27 & 18.9 & 0.46 & 7.72 & -11.72 & -1.07 & 9.36 & 9.69  & $>4.73$ & 0.28 & 69.03\\
  J085738.2+010740 & 4548 & 0.07 & 134.410 & 1.128  & 0.04 & 0.91 & 1.72 & 10.36 & 9.72 & 20.6 & 0.28 & 6.96 & -11.56 & -1.21 & 9.20 & 9.58   & 4.97 & 4.37 & 4.65\\
  J090646.2-004453 & 4639 & 0.16 & 136.693 & -0.749 & 0.04 & 0.85 & 1.51 & 10.99 & 10.48 & 21.2 & 0.30 & 7.57 & -11.21 & -0.17 & 9.48 & 9.73  & --   & 0.17 & 1.30\\
  J091144.5+012952 & 4859 & 0.17 & 137.936 & 1.499  & 0.04 & 0.88 & 2.57 & 10.85 & 10.59 & 21.4 & 0.79 & 7.70 & -11.04 & -0.21 & 9.32 & 9.59  & $>4.29$ & 2.22 & 2.05\\
  J090013.7+004139 & 4964 & 0.24 & 135.057 & 0.693  & 0.04 & 0.94 & 2.33 & 10.49 & 10.48 & 17.6 & 0.48 & 8.15 & -11.41 & -0.93 & 8.70 & 9.05  & 4.26 & --   & 9.60\\
  J090707.3+000805 & 5108 & 0.10 & 136.78 & 0.135   & 0.04 & 0.88 & 2.00 & 10.21 & 9.88 & 18.1 & 0.47 & 7.41 & -11.13 & -0.90 & 9.13 & 9.52   & 4.59 & 0.09 & 17.53\\
  J091230.6-005442 & 5226 & 0.16 & 138.128 & -0.913 & 0.04 & 0.91 & 2.07 & 10.75 & 10.40 & 19.3 & 0.44 & 7.79 & -11.26 & -0.50 & 9.10 & 9.49  & 4.85 & 3.12 & 1.30\\
  J085934.4-000456 & 7324 & 0.17 & 134.895 & -0.082 & 0.03 & 0.92 & 1.97 & 10.73 & 10.28 & 19.6 & 0.43 & 7.65 & -11.50 & -0.75 & 9.24 & 9.56  & $>4.72$ & 2.12 & 2.55\\
\hline
\end{tabular}
\label{Red_spiral_table}
\end{center}
\end{table}
\end{landscape}

 \begin{figure*}
 \begin{minipage}[t]{1.0\textwidth}
     \begin{center}  
 \includegraphics{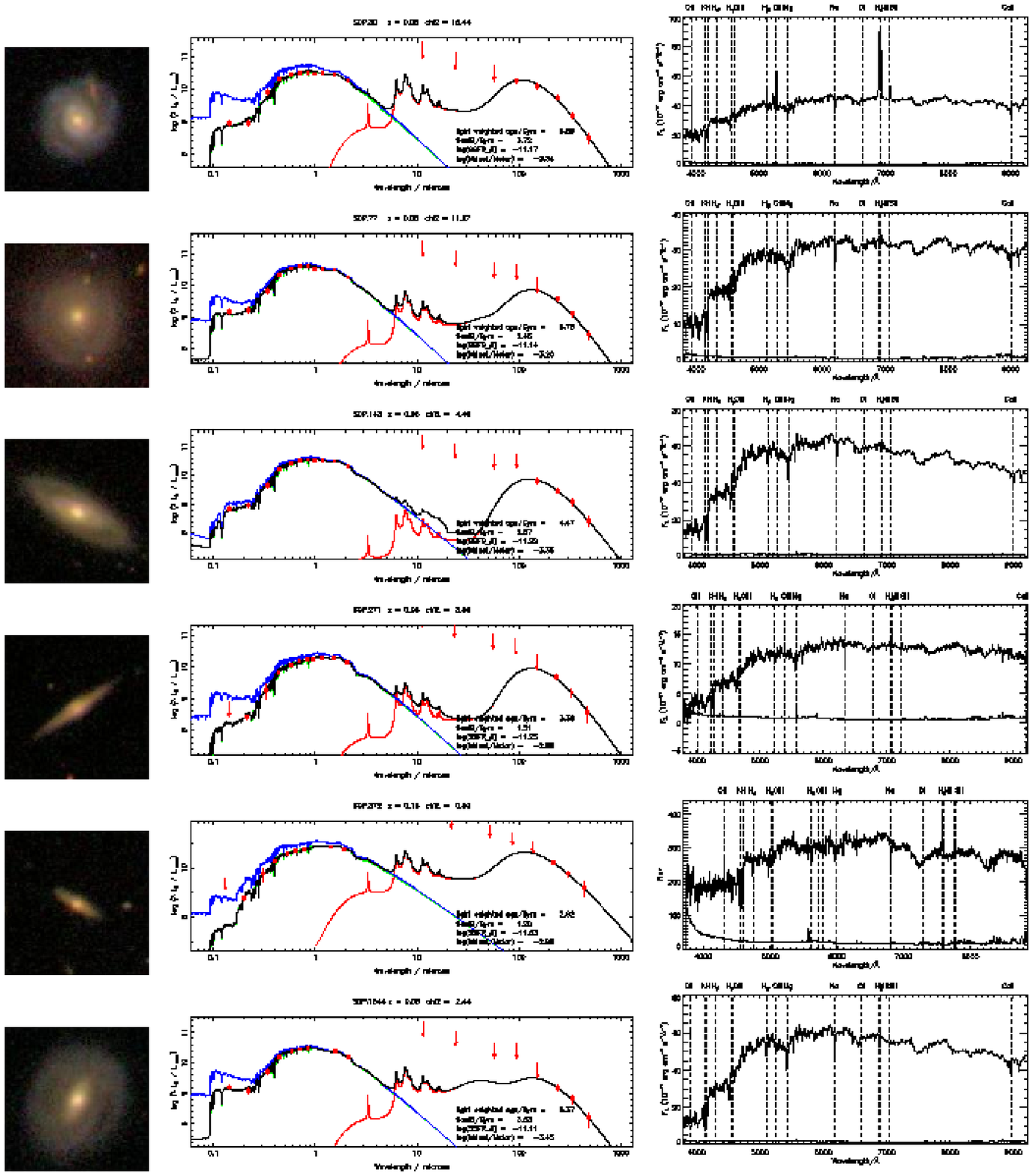}
  \end{center}
 \end{minipage}
\caption{Optical images, multiwavelength SEDs and optical spectra of the 19 passive spirals in our sample. Images are 40" on a side. The rest-frame SEDs of each passive spiral are shown, where red points are the observed photometry, with 5$\sigma$ upper limits shown as arrows. Errors on the photometry are described in \citet{Smith11b}. The black line is the total best fit SED model, the green line is the attenuated optical model, the blue line is the unattenuated optical model, 
the red line is the infrared model. Spectra are from SDSS and GAMA, and the standard deviation in the spectra is also shown. The spectra have been smoothed by a boxcar of 8 pixels.}
 \label{fig:Red_spirals}
 \end{figure*}

 \begin{figure*}
 \begin{minipage}[t]{1.0\textwidth}
     \begin{center}
 \includegraphics{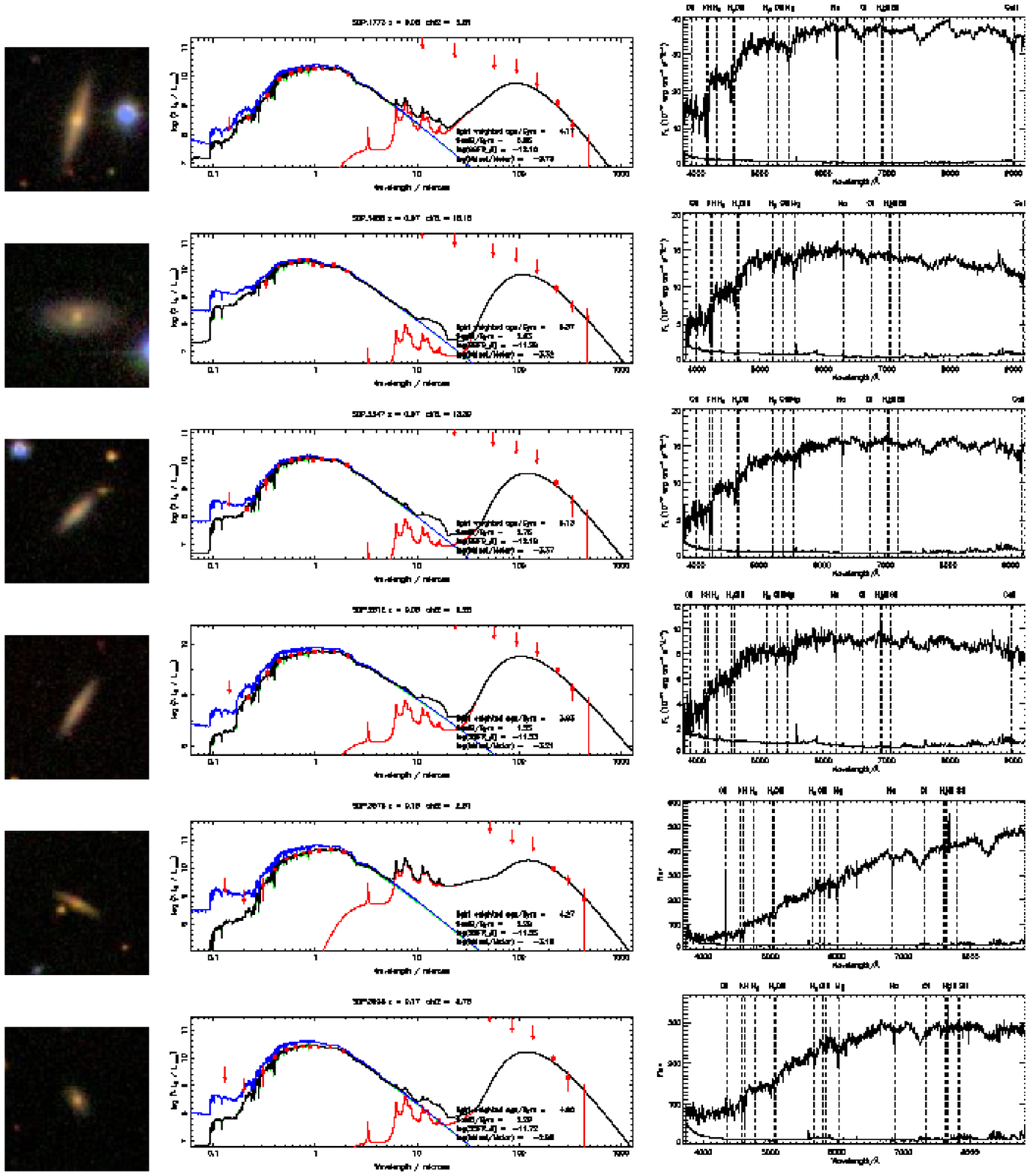}
  \end{center}
 \end{minipage}
 \contcaption{}
 \end{figure*}

 \begin{figure*}
 \begin{minipage}[t]{1.0\textwidth}
     \begin{center}  
 \includegraphics{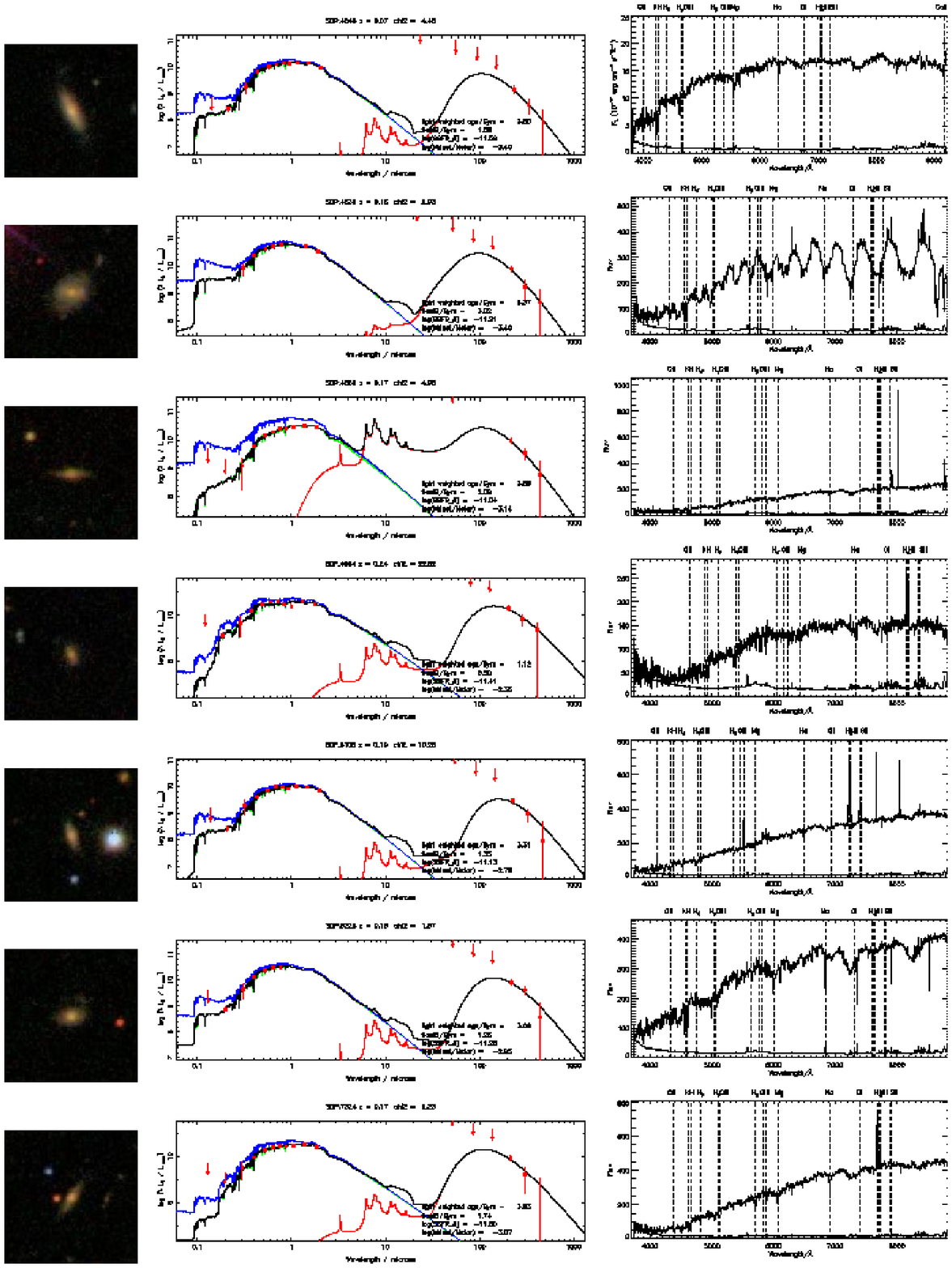}
  \end{center}
 \end{minipage}
 \contcaption{. The wave-like features in the spectrum of SDP.4639 are due to fibre fringing \citep{Colless01_2dF}.}
 \end{figure*}

\clearpage

\section{Summary of mean physical properties}

 \begin{table*}
 \begin{center}
 \caption{Summary of mean physical properties derived from stacking of PDFs for the different populations studied in this paper. The parameters are: \fmu, the fraction of total dust luminosity contributed by the diffuse ISM; $\rm{M}_\ast/\rm{M}_\odot$, log(stellar mass); $\rm{M}_\mathrm{d}/\rm{M}_\odot$, log(dust mass); \mdms, log(dust to stellar mass ratio); \ldust/$\rm{L}_\odot$, log(dust luminosity); \tbgscold/K, temperature of the cold ISM dust component; \tauv, total effective V-band optical depth seen by stars in birth clouds; \tauvISM, the V-band optical depth in the ambient ISM, \ssfr/$\rm{yr}^{-1}$, log(SSFR); \sfr/$\rm{M}_\odot\rm{yr}^{-1}$, log(SFR), $\rm{t_{LB}}$, log(time of last burst); $\rm{age_r}$, log($r$-band light-weighted age of the stellar population).
For each parameter, we use the first moment of the average PDF to estimate the mean for the population.
We can estimate the variance on the population mean as the second
moment of the average PDF minus the mean squared, divided by the number 
of galaxies in the sample. The error on the mean is simply the square
root of the population variance.
The errors for logarithmic parameters are in dex.
The mean parameters from the infrared part of the SED and energy balance parameters are not determined for the control sample, since we only have constraints from upper limits on the FIR-submillimetre flux.}
 \begin{tabular}{ l | c | c | c | c | c | c }
 \hline
  \multicolumn{1}{|c|}{Parameter} &
  \multicolumn{1}{|c|}{H-ATLAS spiral} &
  \multicolumn{1}{c|}{H-ATLAS ETG} &
  \multicolumn{1}{c|}{Normal spiral} &
  \multicolumn{1}{c|}{Passive spiral} &
  \multicolumn{1}{c|}{Control spiral} &
  \multicolumn{1}{c|}{Control ETG} \\
 \hline
   \vspace{0.1cm}
\fmu                & $0.59\pm0.01$ & $0.74\pm0.02$ & $0.58\pm0.01$ & $0.87\pm0.02$ & -  & - \\
$\rm{M}_\ast$       & $10.29\pm0.02$ & $10.69\pm0.08$& $10.27\pm0.02$  & $10.62\pm0.07$  & $10.15\pm0.03$  & $10.77\pm0.03$\\
$\rm{M}_\mathrm{d}$ & $7.72\pm0.02$  & $7.74\pm0.08$ & $7.73\pm0.02$  & $7.47\pm 0.10$ & -                  & - \\
\mdms               & $-2.57\pm0.02$ & $-2.95\pm0.07$& $-2.54\pm0.02$  & $-3.16\pm0.09$ & -                  & - \\
\ldust              & $10.53\pm0.02$ & $10.48\pm0.07$& $10.55\pm0.02$  & $10.14\pm0.09$  & -                  & - \\
\tbgscold           & $19.7\pm0.1$   & $19.8\pm0.5$  & $19.7\pm 0.1$  & $19.8\pm0.6$       & -                  & - \\
\tauv               & $2.28\pm0.07$  & $2.28\pm0.23$ & $2.28\pm0.07$  & $2.34\pm 0.37$ & $1.66\pm0.08$  & $1.61\pm0.10$\\
\tauvISM            & $0.47\pm0.01$  & $0.43\pm0.04$ & $0.48\pm0.02$ & $0.41\pm0.05$ & $0.24\pm0.01$& $0.20\pm0.01$\\
\ssfr               & $-9.99\pm0.03$ & $-10.85\pm0.14$ & $-9.92\pm0.03$ & $-11.59\pm0.18$ & $-10.58\pm0.07$ & $-11.92\pm0.07$\\
\sfr                & $0.30\pm0.03$  & $-0.16\pm0.12$& $0.36\pm0.03$ & $-0.97\pm 0.19$& $-0.43\pm0.05$& $-1.16\pm0.07$\\
$\rm{t_{LB}}$       & $8.70\pm0.07$  & $9.04\pm0.18$ & $8.68\pm0.08$  & $9.26\pm0.10$   & $8.87\pm0.07$  & $9.39\pm0.03$\\
$\rm{{age}_r}$      & $9.21\pm0.02$  & $9.45\pm0.05$ & $9.19\pm0.02$  & $9.59\pm0.05$  & $9.32\pm0.02$  & $ 9.67\pm0.01$\\
\hline                                                                                                             
\end{tabular}                                                                                                      
\label{tab:summary_properties}
\end{center}
\end{table*}

\label{lastpage}

\end{document}